\documentclass[preprint, authoryear]{elsarticle}

\usepackage{natbib}
\usepackage[utf8]{inputenc} 
\usepackage[T1]{fontenc}
\usepackage{lmodern}
\usepackage[fleqn]{mathtools}
\usepackage{booktabs}
\usepackage{rotating}
\usepackage{amssymb}
\usepackage{amsmath}
\usepackage{graphicx}
\usepackage{caption}
\usepackage{epstopdf}
\usepackage{stackrel}
\usepackage{soul}
\usepackage{color}
\usepackage{float}
\usepackage[official]{eurosym}
\usepackage{tabularx}
\usepackage{afterpage}
\usepackage{mathrsfs}
\usepackage{chngcntr}
\usepackage{caption}
\usepackage{subcaption}

\renewcommand*{\underline}{\ul}

\newdefinition{rmk}{Remark}
\newproof{pf}{Proof}
\newproof{pot}{Proof of Theorem \ref{thm2}}

\makeatletter
\def\ps@pprintTitle{%
	\let\@oddhead\@empty
	\let\@evenhead\@empty
	\def\@oddfoot{\centerline{\thepage}}%
	\let\@evenfoot\@oddfoot}
\makeatother

\begin{document}
\title{Re-evaluating cryptocurrencies' contribution to portfolio diversification\\[12pt]
	{\large A portfolio analysis with special focus on German investors}\\{\small(First Draft: June 12, 2020)}}

\author[hhu]{Tim Schmitz\corref{cor1}}
\ead{Tim.Schmitz@hhu.de}
\author[hhu]{Ingo Hoffmann}
\ead{Ingo.Hoffmann@hhu.de}	

\cortext[cor1]{Corresponding author. Tel.: +49 211 81-11515; Fax.: +49 211 81-15316}

\address[hhu]{Financial Services, Faculty of Business Administration and Economics, \\ Heinrich Heine University D\"usseldorf, 40225 D\"usseldorf,
	Germany}	

\begin{abstract}
	 In this paper, we investigate whether mixing cryptocurrencies to a German investor portfolio improves portfolio diversification.
	 We analyse this research question by applying a (mean variance) portfolio analysis using a toolbox consisting of (i) the comparison of descriptive statistics, (ii) graphical methods and (iii) econometric spanning tests. In contrast to most of the former studies we use a (broad) customized, Equally-Weighted Cryptocurrency Index (EWCI) to capture the average development of a whole \textit{ex ante} defined cryptocurrency universe and to mitigate possible survivorship biases in the data. According to \cite{GlasPoddig2018}, this bias could have led to misleading results in some already existing studies. We find that cryptocurrencies can improve portfolio diversification in a few of the analyzed windows from our dataset (consisting of weekly observations from 2014-01-01 to 2019-05-31). However, we cannot confirm this pattern as the normal case. By including cryptocurrencies in their portfolios, investors predominantly cannot reach a significantly higher efficient frontier. These results also hold, if the non-normality of cryptocurrency returns is considered. Moreover, we control for changes of the results, if transaction costs/illiquidities on the cryptocurrency market are additionally considered. 
\end{abstract}

\begin{keyword}
	Capital Allocation \sep Portfolio Diversification \sep Spanning-Tests \\ Stepdown-Tests \sep Cryptocurrencies \sep Transaction Costs\\[10pt]
	\noindent \textit{ORCID IDs:} 0000-0001-9002-5129 (Tim Schmitz), 0000-0001-7575-5537 (Ingo Hoffmann).\\[10pt]
	\noindent \textit{Acknowledgement:} We thank Coinmarketcap.com for generously providing the cryptocurrency time series data for our research. Moreover, very special thanks go to Raymond Kan (University of Toronto) and Christoph J. Börner (Heinrich Heine University Düsseldorf), as well as to participants of the 2019 Unicredit/ HypoVereinsbank (HVB) Doctoral Seminar in Hannover (Germany) for providing helpful advices. 
\end{keyword}
\maketitle
\section*{Management Summary / Key Messages:}
\begin{itemize}
	\item To capture the cryptocurrency market development as unbiased as possible, we construct an Equally-Weighted Cryptocurrency Index (EWCI) of the whole cryptocurrency market at one reference date.
	\item An equal weighting scheme of cryptocurrency returns reduces the expected return of cryptocurrency investments (compared to the usual market cap weighting schemes)
	\item Consequence: in most of the analyzed periods cryptocurrencies do not play a significant role in the optimal portfolio any more
	\item  The (remaining) diversification effects of cryptocurrencies do not necessarily fade away, if transaction cost and illiquidity issues are included in the considerations
\end{itemize}
\section{Introduction}
In the aftermath of the subprime crisis, \cite{Nakamoto2008} developed the technical foundations of Bitcoin as the first cryptocurrency in history. According to the definition of \cite{HeEtAl2016}, cryptocurrencies are virtual currencies, which are additionally characterized by their convertibility to real goods and services, their decentralized transaction networks and the use of cryptographic techniques.\\

After the introduction of the Bitcoin, the cryptocurrency market grew fast, reaching
a number of 2,216 cryptocurrencies (effective date: 2019-06-01) existing on the market and a market capitalization (market cap) of 243.98 billion euro (\cite{Coinmarketcap2019}).\\

Although cryptocurrencies are intended as substitutes for national (fiat) currencies as alternative media of exchange, recent studies in the literature, such as \cite{Yermack2015} or \cite{BaurHongLee2018}, find that cryptocurrencies do not act as a form of money because they do not fulfill the classical three functions of money (store of value, medium of exchange, unit of account). Instead, cryptocurrencies' high attention in the media and social networks in combination with their innovative technology and the expectation of a high return potential make investors rather use them as (speculative) investment vehicles. This development led to a rising speculative value in cryptocurrency prices ($\rightarrow$ bubbles: \cite{CheahFry2015}) -- especially in December 2017 -- and the rise of several indirect forms of cryptocurrency investments (e.g.\ futures, certificates, funds, etc.) besides the traditional direct investments in cryptocurrencies (via specialized, mostly unregulated exchanges). Since december of 2017, where the Chicago Merchantile Exchange (CME) and the Chicago Board Options Exchange (CBOE) as classical (regulated) exchanges started trading the first Bitcoin futures, the cryptocurrency market was also opened for institutional investors (\cite{CorbetEtAl2018}).\\
To put it more precisely, cryptocurrencies can also be classified as a stand-alone asset class, because \cite{KrueckebergScholz2018} find, that they meet the asset class criteria defined by \cite{Sharpe1992}. Because the (emerging) asset class of cryptocurrencies is currently not widely spread in -- institutional and private -- investors' portfolios, it is useful to investigate their (possible) diversification benefits, if cryptocurrencies are added to an already (well) diversified investment portfolio. For the means of our paper, we define diversification as the conscious combination of different assets with varying risk-return-characteristics in the investors portfolio -- aiming at a reduction of portfolio risk (\cite{Markowitz1952}), as it is shown in the Portfolio Selection Theory by \cite{Markowitz1952,Markowitz1959}.\\
Therefore, we analyse, whether there is a measurable improvement of portfolio diversification, if $N$ cryptocurrencies (called test assets hereafter) are mixed into a reference portfolio of $K$ exemplarily selected assets (called benchmark assets hereafter).
\\ 

In the \textit{contentually-related economic literature}, we find different approaches to analyze those possible diversification benefits of cryptocurrencies.\\On the top level, we have to differentiate between two different literature strands: the analysis of diversification benefits on average (means: \textit{in normal times}) and in times of financial distress (means: \textit{in times of crises})\footnote{\cite{BaurLucey2010} use the term \textit{safe haven} to describe assets which are uncorrelated or negatively correlated to other assets or portfolios in times of crises. Building on their quantile regression approach (in combination with a GARCH (1,1) specification), which was originally invented for a save haven analysis in the context of gold investments, or using extensions like dynamic models (e.g.\ DCC-GARCH (1,1) models by \cite{Engle2002}, there is also a growing literature with focus on a possible save haven status of cryptocurrencies (e.g.\ \cite{BouriEtAl2017,Smales2019,Shahzad2019,UrquhartZhang2019}).}. Following the classical Portfolio Selection Theory, where the average correlation is used as an input factor in the portfolio optimization framework (\cite{Markowitz1952,Markowitz1959}), our paper contributes to the analysis of the diversification benefits on average (first literature strand). Within this literature strand, the diversification benefits of the integration of cryptocurrencies in investor portfolios can be analyzed in three different ways
\begin{itemize}
	\item using an approach based on \textit{descriptive statistics},
	\item using a \textit{graphical} approach,
	\item using a \textit{regression-based} approach.
\end{itemize}

The \textit{first approach}, the so-called \textit{descriptive-statistics-based approach}, compares descriptive statistics of portfolios (such as optimal portfolio weights and different return-, risk and performance measures) with and without cryptocurrencies being mixed in the portfolio (e.g.\ used by \cite{EislEtAl2015}).

The \textit{second approach}, the so-called \textit{graphical approach}, covers methods which compare one type of graphic in two situations -- with and without the use of cryptocurrencies in the investment opportunity set of the investors. One possible comparison could be, whether the use of cryptocurrencies in the investors' portfolios leads to an upward shift of the efficient frontier or not (e.g.\ used by \cite{LeeKuoChuenEtAl17}). Such an upward shift of the efficient frontier means reaching a higher return for a given risk level.

The \textit{third approach}, the so-called \textit{regression-based approach}, uses econometric tests to evaluate the contribution to portfolio diversification of different asset classes. Spanning tests are some of the most important econometric tests for diversification issues. Those spanning tests are linked to the graphical approach mentioned above and check, whether possible shifts of the efficient frontier are statistically significant (\cite{KanZou12}).

To create a preferably holistic approach, our analyis combines elements of all the three approaches to analyse the research question from different points of view and to compare the different results. This combination allows us to evaluate, (i) to which extent cryptocurrencies should be included in the optimal portfolios (optimal portfolio weights), (ii) whether these optimal portfolio weights lead to a shift in the efficient frontier and (iii) whether this shift is statistically significant.\\

Independent of the approach used in their analysis, former studies -- except for \cite{GlasPoddig2018} -- consistently find, that considering cryptocurrencies in investors' portfolios leads to a significant upward shift of the efficient frontier, which is (sometimes) confirmed by spanning tests and (sometimes) by descriptive statistics (higher portfolio returns, higher portfolio performance). 
In the methodologically related literature diversification issues are mostly analysed using a traditional Markowitz Mean-Variance-Framework (\cite{Markowitz1952,Markowitz1959}) -- independent of the analyzed asset class. In the context of cryptocurrencies, we identify different approaches for the analysis of diversification effects. First, there are some existing studies, just like {\cite{WuPandey2014} or \cite{EislEtAl2015}, which do not implement such a traditional Mean-Variance-Framework, but a Mean-Conditional-Value-at-Risk-Framework, as it is proposed by \cite{Rockafellar2000} because of the non-normality of cryptocurrency returns (\cite{OsterriederEtAl17}). On the other hand, there is also a huge number of studies, such as \cite{Borri2019}, \cite{GlasPoddig2018}, \cite{Brauneis2019} and \cite{Liu2019}, which nevertheless stick to the (widely spread) Markowitz framework for simplicity reasons, e.g.\ because the scope of their analysis is on additional restrictions (just as the impact of transaction costs or illiquidity issues).

Focusing on the above-mentioned spanning tests, there are a lot of applications in the literature to show the diversification impact of a specific asset classes. A famous application of spanning tests in the context of commodity investments is given by \cite{BelousovaDorfleitner2012}. Moreover, there is already some literature on spanning tests in the cryptocurrency context (e.g.\ \cite{Anyfantaki2018}, \cite{BriereEtAl2015}, \cite{Chowdhury2014}, \cite{GlasPoddig2018}, \cite{LeeKuoChuenEtAl17}). Except for \cite{GlasPoddig2018}, all of these studies also find an additional diversification effect, if cryptocurrencies are considered in an investor's portfolio, which is consistent with the findings of the other approaches mentioned above.\\

With regard to our regional focus, we find that \cite{GlasPoddig2018} is the only study which analyses diversification effects of cryptocurrencies for German investors. Most of the earlier studies dealing with cryptocurrencies' contribution to diversification (explicitly or implicitly) use a global or an US-investor's point of view (e.g.\ \cite{Anyfantaki2018,LeeKuoChuenEtAl17,BriereEtAl2015,EislEtAl2015,WuPandey2014}). But neither the investors nor their financial market behavior are completely comparable in Germany and the US because of cultural differences (e.g.\ different levels of risk aversion)\footnote{A brief look at scores of the cultural dimensions by \cite{Hofstede2001} for Germany and the USA, which were more recently published in \cite{HofstedeEtAl2010}, shows, that there is e.g.\ a higher degree of uncertainty avoidance in Germany than in the USA. Although this term is not completely congruent to the term of risk aversion, it is often interpreted as the approximate inverse of "risk tolerance" (\cite{FrijnsEtAl2013}) or the approximate inverse of "cultural appetite for risk" (\cite{AggarwalEtAl2012}) in the literature, which both come close to the definition of risk aversion.} and structural differences, such as different regulatory frameworks, different status of financial markets in the economy (bank-oriented financial systems in Germany vs. financial-market-oriented financial system in the USA) and -- as a consequence --  different relationships to financial market investments in the context private retirement provisions (\cite{Vitols2005}). Thus, changing the geographical focus of the analysis is interesting and well justified in this context.\\

From a more technical point of view, \cite{GlasPoddig2018} find, that the time series of nearly all existing studies on diversification in the cryptocurrency context are (possibly) biased (e.g.\ survivorship bias) because of some distorting restrictions in their datasets causing misleading results. Consistently, they find that investing in cryptocurrencies -- in reality -- does not lead to measurable diversification advantages for investors. To prevent biased results, we follow their trend-setting approach and use a customized Equally-Weighted Cryptocurrency Index (EWCI) to capture the average development of the cryptocurrency market more accurately.

Moreover, \cite{GlasPoddig2018} state, that those results would become worse, if -- in addition -- the possibility for investors to lose their whole invested amount (because of dying cryptocurrencies) or considering liquidity restrictions (as they are observable on the market) in the optimization framework would be considered. While the latter effect was already analyzed by \cite{Borri2019} from the American point of view, there is no study, which tries to consider temporary inactive or dying cryptocurrencies as long as it is possible for a more accurate measurement of cryptocurrencies' diversification potentials. As a consequence, this study aims at filling this research gap by extending the analysis of \cite{GlasPoddig2018} by considering such (temporary) inactive cryptocurrencies as long as possible in the index calculation and also controlling for liquidity issues and transaction costs in a further analysis.\\

According to \cite{FantazziniZimin2019}, there is no generally accepted definition of such dead cryptocurrencies in the literature. However, the authors state that dying cryptocurrencies suffer significant drops of their market price which is accompanied by the illiquidity of the respective coins. Not least because of those effects, the inclusion of dying cryptocurrencies in their dataset would have worsened the results of \cite{GlasPoddig2018}, as the authors admit in their study.

There are different possibilities, why cryptocurrencies can die, such as the abandoning of coins (e.g.\ due to a lost market traction and/or lost exchange listings), scam issues (e.g.\ Ponzi Schemes or pump and dump patterns), hacks (mostly accompanied by the theft of coins), the identification of coins as (useless) joke coins, and other problems (e.g.\ problems in the programming code, failed forks, failed pre-mining) (\cite{Coinopsy2019}; \cite{Deadcoins2019}; \cite{FantazziniZimin2019}; \cite{GrobysSapkota2019}). In addition, it is noteworthy, that the state of '\textit{death}' can only be a temporary phenomenon for some cryptocurrencies, because they can recover after a substantial revamp (system updates, new programming codes, etc.). As a consequence, even though a certain coin is (temporarily) dead, they can still have a neglegible trading volume on the market. This phenomenon is driven by investors who try to monetarize parts of their initial investments on the one hand, and by investors who buy those coins because they bet on a future revocery (revamp) on the other hand. \cite{FederEtAL2018} try to define a concrete rule for cryptocurrencies, to be classified as '\textit{dead}' and '\textit{recovered}'. Due to their definition, a cryptocurrency can be classified as dead, if its average daily trading volume for a given month is lower or equal to 1\% of its historical peak. On the other hand, a dead cryptocurrency is classified as 'resurrected', if this average daily trading volume reaches a value of more or equal to 10\% of its historical peak again (\cite{FederEtAL2018}).

To capture those effects the best as possible, we adjust the calculation methodology of the cryptocurrency index (in comparison to other studies). Because of the limited data availability, our study does not follow the methodology of \cite{FederEtAL2018}, but uses a more simplified approach to identify temporary inactiveness. Hence, in the following, a cryptocurrency is marked as \textit{temporarily inactive}, if we observe data gaps (with respective to price- and market cap data) in the time series of the respective coins at the rebalancing dates of the index.\\

In a nutshell, this paper analyses, whether considering cryptocurrencies (represented by an Equally-Weighted Cryptocurrency Index (EWCI)) in the investment opportunity set of German investors leads to diversification benefits (and a higher portfolio efficiency) under the assumption of a classical mean-variance framework. The study combines three different analytical (non-econometric and econometric) approaches, two different data aggregation levels (whole sample, subsamples) and two different scenarios (firstly without transaction costs and illiquidity issues and secondly with the consideration of those additional restrictions). 

We find, that considering cryptocurrencies in the investment opportunity set does not lead to (significant) diversification benefits in most of the analysed windows of our dataset (2014-01-01/2019-05-31). However, there were also some observation windows, in which a significant diversification effects were measurable. Moreover, the additional assumption of illiquidity issues and transaction costs (especially on the cryptocurrency market) in the portfolio optimization did not always worsen the diversification benefits of cryptocurrencies, as it might be expected at a first glance (e.g.\ by \cite{GlasPoddig2018}).\\

The remainder of the paper is structured as follows: In section \ref{MethodologicalFramework}, we describe the methodological framework of our study -- with special regard to the underlying portfolio optimization framework and the fundamentals of the implemented spanning tests. In section \ref{Dataset}, we give information about the derivation of the respective dataset and the calculation of our customized cryptocurrency index. Section \ref{Results} shows the results of the different econometric and non-econometric approaches and derives implications for German investors. In section 5, we adjust our portfolio optimization framework to additionally control for the effects of transaction costs and illiquidity considerations on the efficient frontier. The last section discusses the results and summarizes the key points.

\section{Methodological Framework}\label{MethodologicalFramework}
\subsection{Portfolio Optimization Framework and Concept of the Analysis}\label{FrameworkAndConcept}
\subsubsection{Mean-Variance-Portfolio Optimization Framework}\label{Markowitz}
The starting point to understand diversification is the Portfolio Selection Theory by \cite{Markowitz1952,Markowitz1959}. He showed that the mix of assets with a correlation coefficient $\rho < 1$ (= diversification) leads to a reduction of the overall portfolio risk.

Although there are a lot of advanced portfolio optimization frameworks used in the contentually-related literature, just as the mean conditional value at risk (CVaR) optimization framework developed by \cite{Rockafellar2000}, we follow \cite{Brauneis2019} and start with a classical mean-variance-framework as basic optimization model for simplicity and consistency reasons. This simple framework has the advantage, that it is (i) the foundation of the mean-variance spanning tests used afterwards, and (ii) easily allows for additional restrictions (transaction costs, illiquidity) in a further step of the analysis. To prevent the valid objection, that cryptocurrency returns are (mostly) not normally-distributed, so that a mean-variance optimization would underestimate potential losses from cryptocurrencies' additional tail risk (fat tails) and would lead to sub-optimal portfolio allocations (cf.\ e.g.\ \cite{EislEtAl2015}; \cite{Jorion2001}; \cite{McNeilEtAl2005}), we follow \cite{BelousovaDorfleitner2012} and \cite{KanZou12} by applying additional controls, but on the level of specialized spanning tests presented hereafter, whose applicability is not limited to normal distributions.\\

Starting with the construction of the basic (unrestricted) mean-variance optimization framework, we firstly define 
${\boldsymbol R}_{t}=[{\boldsymbol R}_{1t}',{\boldsymbol R}_{2t}']'$ as a transposed return vector of the total returns  of the $K+N$ risky assets, where ${\boldsymbol R}_{1t}'$ is a transposed $K$-vector containing the returns of the $K$ benchmark assets, and ${\boldsymbol R}_{2t}'$ is a transposed $N$-vector containing the returns of the $N$ test assets (\cite{KanZou12}). 

In the next step, we can use this information to calculate vectors of the expected returns $\boldsymbol{\mu}$ and variance-covariance-matrices of this returns $\boldsymbol{V}$. As a result, we have 
\begin{flalign}\label{1}
\boldsymbol{\mu} = \text{E}[{\boldsymbol R}_t] \equiv 
\begin{bmatrix}
\boldsymbol{\mu}_{1}\\
\boldsymbol{\mu}_{2} \\
\end{bmatrix}
\end{flalign}
as an expected return vector of the dimension $(K+N)\times1$, which consists of all expected returns of the $K$ considered benchmark assets $\boldsymbol{\mu}_{1}$ and another vector containing the expected returns of the $N$ test assets $\boldsymbol{\mu}_{2}$. Moreover, we have
\begin{flalign}\label{2}
\boldsymbol{V} = \text{Var}[\boldsymbol{R}_t] \equiv
\begin{bmatrix}
\boldsymbol{V}_{11} & \boldsymbol{V}_{12}\\
\boldsymbol{V}_{21} & \boldsymbol{V}_{22}\\
\end{bmatrix}
\end{flalign}
as the corresponding (non-singular) variance-covariance-matrix, which is segmented in the four sub-matrices $\boldsymbol{V}_{11},\boldsymbol{V}_{12},\boldsymbol{V}_{21},\boldsymbol{V}_{22}$ (block matrix). This block matrix contains the variances and covariances of the benchmark- and test assets -- within ($\boldsymbol{V}_{11}$, $\boldsymbol{V}_{22}$) and between this groups ($\boldsymbol{V}_{21}$, $\boldsymbol{V}_{12}$).\\

In this framework, we calculate two kinds of portfolios -- the Global Minimum Variance Portfolio (GMVP), as it is proposed by \cite{Markowitz1952,Markowitz1959}, and the Tangency Portfolio (TP), which is the result of a maximization of the Sharpe-Ratio by \cite{Sharpe1966,Sharpe1994} (\cite{Chopados2011}). In this first stage of the study, we additionally assume, that there are no transaction costs and there are no liquidity issues for the cryptocurrencies to be considered in the sample of the optimization problem.\\

In the next step, we follow the notations by e.g. \cite{Roncalli2011} and \cite{Chopados2011} and define the investors' optimization problems using the Lagrange formalisms
\begin{flalign}\label{3}
\max_{\boldsymbol{\omega}} \mathcal{L}^{\text{TP}\hphantom{\text{GMVP}}}\! = \max_{\boldsymbol{\omega}}\left[\frac{\boldsymbol{\omega}'\boldsymbol{\mu}-r_F}{\sqrt{\boldsymbol{\omega}'\boldsymbol{V}\boldsymbol{\omega}}}+
\lambda ({ 1}-\boldsymbol{\omega}'{\bf 1})\right]
\end{flalign}
\begin{flalign}\label{4}
\min_{\boldsymbol{\omega}} \mathcal{L}^{\text{GMVP}\hphantom{\text{TP}}} = \, \min_{\boldsymbol{\omega}}\left[\frac{1}{2}\boldsymbol{\omega}'\boldsymbol{V}\boldsymbol{\omega}+\lambda ({1}-\boldsymbol{\omega}'{\bf 1})\right]
\end{flalign}
with $\mathcal{L}$ as Lagrange function, $\boldsymbol{\omega}$ as portfolio weights vector, $\boldsymbol{\mu}$ as expected return vector, $r_F$ as risk-free interest rate (assumption: $r_F = 0\%$), ${\bf 1}$ as vector, wherby every element is equal to 1, and $\lambda$ as the typical Lagrange multiplier. Under the consideration of a budget restriction ($1-\boldsymbol{\omega}'{\bf 1} = 0$) in both cases, we maximize the portfolio's Sharpe Ratio (see Eq.\ \eqref{5}) for the calculation of the tangency portfolio and minimize the variance of the portfolio returns for the calculation of the GMVP (see Eq.\ \eqref{6}). Please note, that we apply the standard version of the optimization model without the introduction of additional short sale constraints (e.g.\ long-only constraints) for consistency reasons, because this unrestricted framework will also be assumed for the spanning tests following in a later part.\footnote{This assumption might not be problematic, because e.g. the introduction of Bitcoin future contracts makes it possible to short sell Bitcoins on the CME and CBOE: e.g. investors can sell the promise to deliver units of Bitcoins at a future date for a previously defined price and speculate, that they can buy those units for an even lower price to make profits (\cite{HaleEtAl2018}). We assume, that those future contracts will be traded on traditional (regulated) exchanges for a growing number of cryptocurrencies in the future. Nevertheless, a parallel application of a (more feasible) long-only portfolio might mitigate possible objections with regard to the final results.} This optimization framework is frequently used in the portfolio optimization and spanning test literature (e.g.\ \cite{GlasPoddig2018}). Nevertheless, as far as it is possible, we additionally calculate the results for the introduction of a long-only assumption ($\omega_{i} \geq 0$  $\forall i$) as a robustness check.\\

Under the assumption of an unconstrained portfolio, it is possible to calculate the optimal portfolio analytically by solving the respective optimization problems (Eq.\ \eqref{3} and \eqref{4}). This calculation generates the (optimal) weights vectors $\boldsymbol{\omega}^{\text{TP}}$ for the tangency portfolio (Eq.\ \eqref{5} with $r_F = 0\% $) and $\boldsymbol{\omega}^{\text{GMVP}}$ for the Global Minimum Variance Portfolio (GMVP) (Eq.\ \eqref{6}):
\begin{flalign}\label{5}
\boldsymbol{\omega}^{\text{TP}\hphantom{\text{GMVP}}} = \frac{\boldsymbol{V}^{-1} \boldsymbol{\mu}}{\boldsymbol{1}'_{K+N}\boldsymbol{V}^{-1}\boldsymbol{\mu}}
\end{flalign}
\begin{flalign}\label{6}
\boldsymbol{\omega}^{\text{GMVP}\hphantom{\text{TP}}} = \frac{\boldsymbol{V}^{-1}\boldsymbol{1}_{K+N}}{\boldsymbol{1}'_{K+N}\boldsymbol{V}^{-1}\boldsymbol{1}_{K+N}}
\end{flalign}
where $\boldsymbol{1}_{K+N}$ is a $K+N$ vector of ones.\\

\subsubsection{Concept of the Portfolio Optimization Analysis}\label{ConceptPortOpt}
For the means of our analysis, we use the abovementioned portfolio optimization framework (cf.\ Sec.\ \ref{Markowitz}) and calculate the optimal portfolio weights for the respective asset classes (with and without the consideration cryptocurrencies) for two cases:
\begin{itemize}
	\item  \textbf{Case A:} The first case should be more general, which means that it contains only one optimization for the whole dataset (Observation period $T_{\text{Obs}} = 65$ months). Subsequently, we compute only one respective (optimal) weight vector for the situations with and without the consideration of cryptocurrencies.
	\item \textbf{Case B:} The second case should control for the changing data base (e.g.\ caused by market trends) more in detail. In this course, we stick to the setting of \cite{EislEtAl2015} and limit the data observation period to the first year of observations in the dataset. In this first year, the investors are assumed to allocate their capital between all benchmark assets in equal portions. After one year of observation time, the portfolio is constructed on the basis of this observed data (Observation period $T_{\text{Obs}} = 12$ months). Hence, the investors have to decide, whether they want to include the test assets in their existing portfolio of benchmark assets. Afterwards, we roll this one-year-window through-out the dataset in monthly steps to assume a monthly rebalancing of the portfolio considering the last year of observations in the optimization. The resulting optimal portfolio allocation (based on the data of the last year) is implemented for the whole month following on that observation window. Because we assume the same initial portfolio with and without cryptocurrencies and no transaction costs in this stage, the initial portfolio allocation is not relevant for the advantageousness of one alternative (with/without cryptocurrencies) until we introduce further assumptions in Sec.\ \ref{ExtendedModel:Framework} concerning transaction costs.
\end{itemize}
After the calculation of the respective optimal portfolios for both cases, we finally focus on Case B more in detail and give the following portfolio metrics for the resulting portfolio returns: mean return, standard deviation, CVaR, maximum drawdown, Sharpe Ratio and -- at last -- the resulting portfolio value at the end of the planning horizon (end value), if $100.00$ EUR are invested in this (frequently rebalanced) portfolio at the portfolio construction date. This backtesting procedure is not applicable to Case A, because the whole dataset is already used as an observation period.

At this stage of our analysis, we do not apply an additional portfolio smoothing technique (such as the usual 3-month EWMA smoothing used by \cite{EislEtAl2015}) for consistency reasons, because we refer to the resulting optimal portfolios later on in the context of the spanning tests. Those techniques are used to prevent extreme changes of the portfolio weights (\cite{EislEtAl2015}), which occur in the course of portfolio rebalancing, because Markowitz portfolios tend to have extreme (positive or negative) portfolio weights (\cite{HaerdleEtAl2018}). Applying such a technique would lead to biased results, so that the results of the portfolio-optmization-approach and the regression-based-approach would not be comparable any more. However, by introducing additional constraints (for the joint capturing of transaction cost and liquidity issues) in Sec.\ \ref{ExtendedModel:Framework}, the shift in the (optimal) portfolio weights will be smoothed indirectly by an adjustment of the respective optimization frameworks in a later stage of the analysis.

\subsection{Econometric Approach: Spanning Tests}
\subsubsection{Mean-Variance-Spanning}\label{SpanningTests}
Besides the comparison of portfolio statistics with and without mixing $N$ cryptocurrencies (test assets) in a portfolio of $K$ benchmark assets, cf.\ Sec.\ \ref{Markowitz}, it is also possible to test, whether the consideration of those $N$ test assets leads to a (significant) shift of the investors' efficient frontier. Therefore, we use the econometric approach of spanning tests as a suitable measure for those (possible) shifts. Note, that the usual spanning tests, which are also selected for this paper, do not allow for long-only restrictions, so that we refer to the unconstrained optimization framework in Sec.\ \ref{Markowitz} in this stage of the analysis.\\

Mathematically, the underlying concept of \textit{spanning}, which is part of the linear algebra, is fulfilled, if a vector can be replicated by a linear combination of other vectors of the same vector space (\cite{LaneBirkhoff1999}). This linear combination is called a \textit{span}. With regard to this mathematical definition, we can use the term \textit{spanning}, if the same efficient frontier (as a linear combination of the single assets' risks and returns) results -- independent of the fact, whether investors include the test assets in their portfolio (\cite{HubermanKandel1987,KanZou12}). Consequently, every efficient portfolio (and also the optimal portfolio) will remain the same, even if there are new investment opportunities available in the investment opportunity set. In other words: the test assets added to the investment opportunity set are not relevant for the construction of an efficient portfolio (means: zero portfolio weights of the test assets), so that they do not have a significant contribution to portfolio diversification. But if we observe a shift in the efficient frontier after the test assets are included in the investment opportunity set, there is no spanning any more.\\

The first group of spanning tests, the so-called mean variance spanning tests building on the work of \cite{HubermanKandel1987}, assume a mean variance portfolio optimization framework, as it is defined in Sec.\ \ref{Markowitz}. Moreover, they are the foundation of different extensions and practical applications in the literature (cf.\ \cite{KanZou12} and others for an exemplary overview). Their theoretical foundation will be presented in the notation of \cite{KanZou12} afterwards.

The foundation of the mean variance spanning tests is the (linear) regression model
\begin{flalign}\label{7}
\boldsymbol{R}_{2t} = \boldsymbol{\alpha} + \boldsymbol{\beta} \boldsymbol{R}_{1t} + \boldsymbol{\epsilon}_t,
\end{flalign}
where the test assets' returns $\boldsymbol{R}_{2t}$ are regressed on the returns of the benchmark assets $\boldsymbol{R}_{1t}$. In this context, we assume that the requirements $\text{E}[\boldsymbol{\epsilon}_t]={\bf 0}_N$ (unbiasedness) and $\text{E}[\boldsymbol{\epsilon}_t\,\boldsymbol{R}'_{1t}]={\bf 0}_{N \times K}$ (no endogeneity) hold. The intuition of the linear regression is, that the test asset returns should be generated by a linear combination of the benchmark asset returns. Moreover, we can rewrite the regression model in Eq.\ \eqref{7} as $\boldsymbol{\alpha} = \boldsymbol{\mu}_{2} - \boldsymbol{\beta}\boldsymbol{\mu}_{1}$ and define the $N\times K$ matrix $\boldsymbol{\beta}$ -- in accordance with its Ordinary Least Squares solution -- as $\boldsymbol{\beta} = \boldsymbol{V}_{21}^{\phantom{-1}}\boldsymbol{V}_{11}^{-1}$. In addition we introduce the vector $\boldsymbol{\delta}$ with $\boldsymbol{\delta} = \boldsymbol{1}_N - \boldsymbol{\beta} \boldsymbol{1}_K$.\\

In sum, by using a linear regression model, we assume a linear relationship between benchmark and test assets in a first step. In other words, we assume that the test assets' returns are replicable by a linear combination of the benchmark assets' returns (spanning). In a second step, we try to falsify this relationship using different hypothesis tests, which check the fit of this assumption. 
Therefore, we formulate the null hypothesis
\begin{flalign}\label{8}
H_0: \hspace{2.5mm}\boldsymbol{\alpha} = {\bf 0}_N, \, \boldsymbol{\delta} = {\bf 0}_N,
\end{flalign}
which simultaneously checks whether the test assets have a zero portfolio-weight in the tangency portfolio (first condition) and in the GMVP (second condition). In other words, the null hypothesis checks whether the test assets are irrelevant in the optimal portfolio. In this context it is important to remind that, if we know two portfolios on the efficient frontier (here: TP, GMVP), we can construct the whole efficient frontier using linear combinations (\cite{GlasPoddig2018}). Hence, if the null  hypothesis holds for both portfolios, we can conclude that it will also hold for the other portfolios on the efficient frontier. \\

For a better understanding of the null hypothesis, it is necessary to derive the null hypothesis from the classical Lagrange optimization formalism for the tangency portfolio in Eq.\ \eqref{3} and for the GMVP in Eq.\ \eqref{4}. Solving these optimization problems generates the (optimal) weights vectors $\boldsymbol{\omega}^{\text{TP}}$ for the tangency portfolio in Eq. \eqref{5} and $\boldsymbol{\omega}^{\text{GMVP}}$ for the GMVP in Eq. \eqref{6}. \\

Inverting the covariance matrix Eq.\ \eqref{2} by using the block matrix inversion formula and substituting
$\boldsymbol{\beta} = \boldsymbol{V}_{21}^{\phantom{-1}}\boldsymbol{V}_{11}^{-1}$ (= Ordinary Least Squares solution) and 
$\boldsymbol{\Sigma} = \boldsymbol{V}_{22}^{\phantom{-1}}-
\boldsymbol{V}_{21}^{\phantom{-1}}\boldsymbol{V}_{11}^{-1}
\boldsymbol{V}_{12}^{\phantom{-1}}$ 
(= $N \times N$ Schur complement) leads to
\begin{flalign}\label{15}
\boldsymbol{V}^{-1} = 
\begin{bmatrix}
\boldsymbol{V}_{11}^{-1} + \boldsymbol{\beta}'\boldsymbol{\Sigma}^{-1} \boldsymbol{\beta} & -\boldsymbol{\beta}' \boldsymbol{\Sigma}^{-1}\\[2pt]
-\boldsymbol{\Sigma}^{-1}\boldsymbol{\beta} & \boldsymbol{\Sigma}^{-1}
\end{bmatrix},
\end{flalign}
which is used in the work of \cite{KanZou12} (Lemma 2 therein).\\

We multiply the weight vectors of Eq.\ \eqref{5} and \eqref{6} by $\boldsymbol{Q} = [{\bf 0}_{N \times K},\boldsymbol{I}_{N\times N}]$ (with $\boldsymbol{I}_{N\times N}$ is the identity matrix) to get a vector only consisting of the test assets' weightings. Simultaneously, we insert $\boldsymbol{V}^{-1}$ as defined in Eq.\ \eqref{15}, so that we have
\begin{flalign}\label{12} \nonumber
\boldsymbol{Q} \boldsymbol{\omega}^{\text{TP}\hphantom{\text{GMVP}}} 
& = 
\frac{{\boldsymbol{Q} \boldsymbol{V}^{-1}} \boldsymbol{\mu}}     {\boldsymbol{1}'_{K+N}\boldsymbol{V}^{-1}\boldsymbol{\mu}} \hspace{6.5mm}
= 
\frac{[-\boldsymbol{\Sigma}^{-1}\boldsymbol{\beta}, 
	\boldsymbol{\Sigma}^{-1}]\,[\boldsymbol{\mu}'_1,\boldsymbol{\mu}'_2]'}
{\boldsymbol{1}'_{K+N}\boldsymbol{V}^{-1}\boldsymbol{\mu}} \\
& = 
\frac{\boldsymbol{\Sigma}^{-1}
	(\boldsymbol{\mu}_{2} - \boldsymbol{\beta}\boldsymbol{\mu}_{1})}
{\boldsymbol{1}'_{K+N}\boldsymbol{V}^{-1}\boldsymbol{\mu}} \hspace{1.5mm}
= 
\frac{\boldsymbol{\Sigma}^{-1}\boldsymbol{\alpha}} {\boldsymbol{1}'_{K+N}\boldsymbol{V}^{-1}\boldsymbol{\mu}} 
\end{flalign}

\begin{flalign}\label{17} \nonumber
\boldsymbol{Q}\boldsymbol{\omega}^{\text{GMVP}\hphantom{\text{TP}}} 
& = 
\frac{{\boldsymbol{Q} \boldsymbol{V}^{-1}}\boldsymbol{1}_{K+N}} {\boldsymbol{1}'_{K+N}\boldsymbol{V}^{-1}\boldsymbol{1}_{K+N}}
= 
\frac{[-\boldsymbol{\Sigma}^{-1}\boldsymbol{\beta}, 
	\boldsymbol{\Sigma}^{-1}]\,[\boldsymbol{1}'_{K},\boldsymbol{1}'_{N}]'} {\boldsymbol{1}'_{K+N}\boldsymbol{V}^{-1}\boldsymbol{1}_{K+N}}\\
& = 
\frac{\boldsymbol{\Sigma}^{-1}(\boldsymbol{1}_{N} - \boldsymbol{\beta} \boldsymbol{1}_{K})}{\boldsymbol{1}'_{K+N}\boldsymbol{V}^{-1}\boldsymbol{1}_{K+N}}
= 
\frac{\boldsymbol{\Sigma}^{-1}\boldsymbol{\delta}}
{\boldsymbol{1}'_{K+N}\boldsymbol{V}^{-1}\boldsymbol{1}_{K+N}} 
\end{flalign}

Now we can show that $\boldsymbol{\alpha} = \boldsymbol{0}_N$ and $\boldsymbol{\delta} = \boldsymbol{0}_N$ is a signal for zero weights in the test assets (spanning), as we assumed in Eq.\ \eqref{8}:
\begin{flalign}\label{21}
\boldsymbol{\alpha} 
= 
\boldsymbol{0}_N \qquad & \rightarrow\qquad \boldsymbol{Q} \boldsymbol{\omega}^{\text{TP}\hphantom{\text{GMVP}}} \hspace{-4mm}
= 
\boldsymbol{0}_N \\[2pt]
\boldsymbol{\delta} 
= 
\boldsymbol{0}_N \qquad & \rightarrow\qquad \boldsymbol{Q}
\boldsymbol{\omega}^{\text{GMVP}\hphantom{\text{TP}}} \hspace{-4mm}
= 
\boldsymbol{0}_N
\end{flalign}
To estimate the regression model, we additionally assume that  $\boldsymbol{\alpha} $ and $\boldsymbol{\beta}$ are constant over time, so that we can re-write Eq.\ \eqref{7}
\begin{flalign}\label{16}
\boldsymbol{R}'_{2t} = \boldsymbol{\alpha}' +  \boldsymbol{R}'_{1t}\boldsymbol{\beta}' + \boldsymbol{\epsilon}'_t,
\hspace{1cm} \textrm{with } t = 1,2,\ldots,T
\end{flalign}
or in a simple matrix notation as
\begin{flalign}\label{17b}
\boldsymbol{Y} = \boldsymbol{XB} + \boldsymbol{E}
\end{flalign}
where $\boldsymbol{Y}$ is a $T \times N$ matrix containing the test asset returns $\boldsymbol{R}'_{2t}$. The $T \times (K+1)$ matrix $\boldsymbol{X}$ contains rows, which are formatted as follows: $[1,\boldsymbol{R}_{1t}']$. The matrix
$\boldsymbol{B} = [\boldsymbol{\alpha}',\boldsymbol{\beta}']'$ pools the regression parameters $\boldsymbol{\alpha}$ and $\boldsymbol{\beta}$ in one  joint matrix with dimension $(K+1) \times N$. The $T \times N$ matrix $\boldsymbol{E}$ contains rows, which contain all the error terms $\boldsymbol{\epsilon}_t'$ of the different time series regressions from Eq.\  \eqref{16} -- one regression for each test asset. At least, it is assumed that the condition $T \geq K+N+1$ is fulfilled and that the matrix $(\boldsymbol{X}'\boldsymbol{X})$ is non-singular. Both conditions are fulfilled in what follows.\\

For the estimation of the regression model, we calculate
\begin{flalign}\label{18} \nonumber
\hat{\boldsymbol B} 
& = 
[\hat{\boldsymbol{\alpha}'},\hat{\boldsymbol{\beta}'}]' \\
& = (\boldsymbol{X}'\boldsymbol{X})^{-1} (\boldsymbol{X}'\boldsymbol{Y})\\[3pt]
\hat{\boldsymbol{\Sigma}} 
& = \frac{1}{T}
(\boldsymbol{Y} - \boldsymbol{X}\hat{\boldsymbol{B}})'
(\boldsymbol{Y} - \boldsymbol{X}\hat{\boldsymbol{B}})
\end{flalign}
as the estimators of the matrix $\boldsymbol{B}$ and the covariance-matrix $\boldsymbol{\Sigma}$, respectively. If there is an additional matrix $\boldsymbol{\Theta}=[\boldsymbol{\alpha},\boldsymbol{\delta}]'$, both requirements ($\boldsymbol{\alpha}={\bf 0}_N$, $\boldsymbol{\delta}={\bf 0}_N$) of the null hypothesis Eq.\ \eqref{8} can be tested simultaneously, if the null hypothesis is rewritten as $H_0: \boldsymbol{\Theta}={\bf 0}_{2 \times N}$. Since the matrix $\boldsymbol{\Theta}$ can be calculated as $\boldsymbol{\Theta}=\boldsymbol{AB}+\boldsymbol{C}$ using the (projection) matrices
\begin{flalign}\label{19}
\boldsymbol{A} = \begin{bmatrix}
1 & {\bf 0}'_K\\
0 & -{\bf 1}'_K\\
\end{bmatrix}  \qquad \text{and} \qquad
\boldsymbol{C} = \begin{bmatrix}
{\bf 0}'_N\\
{\bf 1}'_N\\
\end{bmatrix}
\end{flalign}
the maximum likelihood estimator (MLE) of the matrix $\boldsymbol{\Theta}$ would be $\hat{\boldsymbol{\Theta}}=[\hat{\boldsymbol{\alpha}},\hat{\boldsymbol{\delta}}]'=\boldsymbol{A}\hat{\boldsymbol{B}}+\boldsymbol{C}$. For the calculation of the test statistics, we need the additional $2\times 2$ matrices $\boldsymbol{G}$ and $\boldsymbol{H}$ with
\begin{eqnarray}\label{20}
\hat{\boldsymbol{G}} = & 
T\,\boldsymbol{A}(\boldsymbol{X}'\boldsymbol{X})^{-1}\boldsymbol{A}'   & = 
\begin{bmatrix}
1 + \hat{\boldsymbol{\mu}}_1'\hat{\boldsymbol{V}}^{-1}_{11}\hat{\boldsymbol{\mu}}_1 
& 
\hat{\boldsymbol{\mu}}_1'\hat{\boldsymbol{V}}^{-1}_{11}{\bf 1}_K
\\
\hphantom{1 + \hspace{2.5mm}}
\hat{\boldsymbol{\mu}}_1'\hat{\boldsymbol{V}}^{-1}_{11}{\bf 1}_K 
& 
{\bf 1}'_K \hat{\boldsymbol{V}}^{-1}_{11} {\bf 1}_K
\end{bmatrix}\\[4pt]
\hat{\boldsymbol{H}} = & 
\hat{\boldsymbol{\Theta}}\hat{\boldsymbol{\Sigma}}^{-1}\hat{\boldsymbol{\Theta}}' & =
\begin{bmatrix}
\hat{\boldsymbol{\alpha}}'\hat{\boldsymbol{\Sigma}}^{-1}\hat{\boldsymbol{\alpha}} & 
\hat{\boldsymbol{\alpha}}'\hat{\boldsymbol{\Sigma}}^{-1}\hat{\boldsymbol{\delta}}
\\
\hat{\boldsymbol{\alpha}}'\hat{\boldsymbol{\Sigma}}^{-1}\hat{\boldsymbol{\delta}} 
& 
\hat{\boldsymbol{\delta}}'\hat{\boldsymbol{\Sigma}}^{-1}\hat{\boldsymbol{\delta}}
\end{bmatrix}.
\end{eqnarray}
If $\lambda_1$ and $\lambda_2$ denote the eigenvalues of the matrix $\hat{\boldsymbol{H}}\hat{\boldsymbol{G}}^{-1}$ (with $\lambda_1 \geq \lambda_2 \geq 0$), the test statistics for the Likelihood Ratio Test ($LR$), the Wald Test ($W$) and the Lagrange Multiplier Test ($LM$) are
\begin{eqnarray}
LR = &
T\sum\limits_{i=1}^{2}\ln(1+\lambda_i) 
& \stackrel{A}{\sim}
X^2_{2N}\label{21a}\\[1pt]
W = &
T (\lambda_1 + \lambda_2) \phantom{\sum\limits_{i=1}^{2}\hspace{2mm}} 
& \stackrel{A}{\sim}
X^2_{2N}\label{21b}\\[1pt]
LM = & 
T\sum\limits_{i=1}^{2}\frac{\lambda_i}{1+\lambda_i} \hspace{8mm}
& \stackrel{A}{\sim}
X^2_{2N}.\label{21c}
\end{eqnarray}
Despite the fact of equal asymptotic distributions and the existence of similar test statistics, \cite{BerndtSavin1977} and \cite{Breusch1979} show that there is an order in the result of the different test statistics $W \geq LR \geq LM$, meaning that (under asymptotic distributions) there is the possibility, that those tests provide conflicting results: while the Wald test faster favors a rejection, the LM test longer favors the acceptance of the null hypothesis. Because none of the beforementioned tests is dominant with regard to its predictive power, we decide to follow \cite{BelousovaDorfleitner2012} and perform all three tests simultaneously to achieve more reliable results.

\subsubsection{Additional Tests for non-normally distributed returns}
\subsubsection{GMM-Wald-Test}
The above-mentioned Mean-Variance-Spanning-Tests (as MLE-based tests) assume that the returns of the single assets in the portfolio follow a normal distribution and that the error terms in equation \eqref{16} are homoscedastic (\cite{BelousovaDorfleitner2012}). But, as we will show in the descriptive statistics of our data sample (cf.\ Sec.\ \ref{DescriptiveStatistics}), cryptocurrency returns do not (necessarily) follow such a normal distribution, so that Mean-Variance-Spanning-Tests may be inefficient in that context. Moreover, the multivariate Ljung-Box test\footnote{This test, which is discussed more in detail by \cite{Tsay2014}, is a multivariate version of the univariate version by \cite{LjungBox1978}. and leads to highly significant results ($p \leq 0.01$) for our dataset.} for our dataset indicates, that the error terms ($\epsilon_t$) exhibit indications of conditional heteroscedasticity, which leads to the problem, that the test statistics of the former tests (Eq.\ \eqref{21a}-\eqref{21c}) do not follow a $X^2_{2N}$ distribution any more.\\

But spanning tests are also applicable outside the mean-variance-framework -- especially in the context of non-normal returns and conditional heteroscedasticity. As a first example, we refer to \cite{FersonEtAl1993} (FFE), who propose a specialized spanning test, the so called GMM-based Wald-test (also: GMM-Wald test)\footnote{There is no need to additionally use a GMM-based version of the LR- and LM- test, because \cite{NeweyWest1987} show, that these tests have the same form as the GMM-Wald test (cf.\ \cite{KanZou12} for a more detailed discussion).}, which is not dependent of the underlying distribution (\cite{GlasPoddig2018}). Assuming the stationarity of $\boldsymbol{R}_{1t}$ and $\boldsymbol{R}_{2t}$ (which is fulfilled for our dataset\footnote{We used an Augmented Dickey Fuller (ADF) test by \cite{DickeyFuller1979} to check the stationarity of our return time series. As a result, the $p$-value for every time series was $p \leq 0.01$, so that the assumption of stationarity has not to be rejected in this use case.}), 
\cite{FersonEtAl1993} test $H_0$ by calculating a similar test statistic compared to the Wald test used for the Mean-Variance-framework before (cf.\ Eq.\ \eqref{21b} and \cite{KanZou12}). The main difference is, that the GMM-Wald test builds on a different regression model, a so-called Generalized Method of Moments (GMM) estimation, as it is proposed by \cite{Hansen1982}. Those GMM estimations are dependent of the formulation of the different moment conditions of the regression model (\cite{KanZou12}).\\

Besides this \textit{regression approach}, which is represented by the FFE spanning test, there is also another kind of GMM-Wald test used in the literature, which was (among others) proposed by \cite{BekaertUrias1996} (BU). In contrast to the FFE spanning test introduced before, the BU spanning test does not use regression coefficients (like FFE), but statistical discount factors (SDF) based on the work of \cite{HansenJagannathan1991} instead. Thus, the BU approach is refered to the literature strand of the so-called \textit{SDF approaches} (e.g.\ like \cite{Ferson1995,DeSantis1993}). The BU spanning test is available in two different versions (i) with adjustment of \textit{Errors in Variables} (EIV) and (ii) without this adjustment. For a more precise estimation of the model parameters, we rely on the alternative with EIV adjustment in addition to the FFE spanning test in our study.\\

The resulting test statistics of both GMM-Wald test approaches 	(FFE, BU with EIV adjustment), which we use but are not introduced here in detail, are (asymptotically) $\chi^2_{2N}$ distributed. We do not focus on a deeper methodological foundation here, but refer to \cite{FersonEtAl1993}, \cite{Hansen1982}, \cite{BekaertUrias1996}, \cite{HansenJagannathan1991} and especially \cite{KanZou12} for a more detailed description of the necessary fundamentals of both approaches and the necessary theoretical fundamentals.\\

According to a comparison of the performance of both approaches by \cite{KanZou12}, the regression approach is more powerful than the SDF-approach. Nevertheless, we decide to capture both approaches to apply an additioanl robustness check for our results, like it is also done by \cite{KanZou12} and \cite{GlasPoddig2018}. In the light of the consistency of all the different results, we can evaluate the underlying model risk.

\subsubsection{Stepdown-Test}
All the formerly introduced spanning tests are joint test, which test both conditions of the null hypothesis Eq.\ \eqref{8} simultaneously. An important shortcoming of these tests is that the estimation $\hat{\boldsymbol{\delta}}$ can be estimated more precisely than $\hat{\boldsymbol{\alpha}}$ which influences the weighting of both conditions of the null hypothesis in the test results (\cite{KanZou12}). As a consequence, it might be possible, that the spanning tests' results indicate significant changes in the tangency portfolio, even though there are no significant changes in reality. On the other hand, it is possible that the spanning tests do not lead to significant results with regard to the GMVP, even though, there are significant changes in reality.\\

Subsequently, we follow \cite{KanZou12}, implementing an additional stepdown procedure as an additional robustness check, which allow us to control for the acceptance or rejection of both conditions seperately. In other words, this procedure allows us to account for different diversification impacts for investors depending on their optimization strategy (the global minimum variance portfolio, tangency portfolio).\\

From the methodological point of view, those stepdown tests apply two stages of $F$-tests: In the \textit{first stage} we use an $F$-test (so called $F_1$-test) to check, whether the condition $\boldsymbol{\alpha} = {\bf 0}_N$ can be accepted. Thus, we compute the test statistics 
\begin{flalign}\label{F1}
F_1 = \Bigg(\frac{T-K-N}{N}\Bigg)\Bigg(\frac{\lvert\bar{\boldsymbol{\Sigma}}\rvert}{\lvert\hat{\boldsymbol{\Sigma}}\rvert}-1\Bigg) \sim F_{N,T-K-N}
\end{flalign}
with $\hat{\boldsymbol{\Sigma}}$ denoting the unconstrained estimate of $\boldsymbol{\Sigma}$ and $\bar{\boldsymbol{\Sigma}}$
denoting the respective constrained estimate of $\boldsymbol{\Sigma}$, where the constraint is defined as $\boldsymbol{\alpha} = {\bf 0}_N$.\\

In the \textit{second stage}, we run an additional $F$-test (so called $F_2$-test) to check, whether the condition $\boldsymbol{\delta} = {\bf 0}_N$ can be accepted, if the first condition $\boldsymbol{\alpha} = {\bf 0}_N$ holds. In this context significant results for the $F_1$-test indicate, that we have a significant change in the tangency portfolio, whereas significant result in the $F_2$ test indicate a significant change in the GMVP. The respective test statistics is
\begin{flalign}\label{F2}
F_2 = \Bigg(\frac{T-K-N+1}{N}\Bigg)\Bigg(\frac{\lvert\tilde{\boldsymbol{\Sigma}}\rvert}{\lvert\bar{\boldsymbol{\Sigma}}\rvert}-1\Bigg) \sim F_{N,T-K-N+1}
\end{flalign}
with $\tilde{\boldsymbol{\Sigma}}$ as the constrained estimate of $\boldsymbol{\Sigma}$ by imposing both constraints
of (i) $\boldsymbol{\alpha} = {\bf 0}_N$ and (ii) $\boldsymbol{\delta} = {\bf 0}_N$. In what follows, we denote with $\xi_1$ the significance level of the $F_1$-test and $\xi_2$ denotes the respective significance level of the $F_2$-test. Under the step-down procedure, the spanning hypothesis has to be accepted, if both tests lead to significant results. Therefore, the significance level of this step-down test (in general) can be calculated as $\xi_{\text{joint}} = 1 - (1-\xi_1)(1-\xi_2) = \xi_1 + \xi_2 - \xi_1\xi_2$.\\

In the further analyzes, the results of this stepdown-tests give us a deeper insight in the reasons, why a joint spanning test leads to significant or insignificant results. Moreover, it is possible to use different significance levels ($\xi_1 \neq \xi_2$) for the different stages, if the investors put more focus on a certain optimization strategy (compared to the other). \cite{KanZou12} propose to use this instrument to choose a $\xi_1$ higher than $\xi_2$ to capture, that it is more difficult to detect the significance of $F_1$. Therefore, we set the a significance level of $\xi_1 = 0.10$ for $F_1$ and $\xi_2 = 0.05$ for $F_2$ (and all the other spanning tests mentioned before) as suitable significance levels.

\subsubsection{Concept of the Spanning Test Analysis}\label{ConceptSpann}
The conception of the spanning test analysis follows the conception of the portfolio optimization analysis. This means: before we can run those beforementioned spanning tests, we need to differentiate two different cases (consistent to the portfolio optimization approach), which are both relevant for our methodology:
\begin{itemize}
	\item  \textbf{Case A:} The first case should be -- again -- more general, which means that it contains only one optimization for the whole dataset (Observation period $T_{\text{Obs}} = 65$ months). Subsequently, we compute one efficient frontier for each case -- one with cryptocurrencies, one without -- and also one regression result for each applied spanning test. In this case, the results can be easily reported in detail afterwards.
	\item \textbf{Case B:} The analysis should control for the changing data base (e.g.\ caused by market trends), as it was already done in the non-econometric analysis above. Thus, the spanning tests need to be re-run every month based on the updated observation window (Observation period $T_{\text{Obs}} = 12$ months), which is consistent to the setting in Sec.\ \ref{Markowitz}. In the latter case, we have $54$ efficient frontiers and $54$ corresponding regression results, so that -- for simplicity -- we only give the share of the significant results compared all results afterwards. This kind of subsample analysis was already implemented beyond the specialized cryptocurrency literature, e.g.\ by \cite{KanZou12}, but in a less granular way.
\end{itemize}
For the means of our study, we decide to compute the results for both of the beforementioned cases (A and B). We use case A to firstly demonstrate the functioning of the applied spanning tests. By additionally focusing on a more granular analysis (as in case B), we will get a more detailed insight in the diversification effects of crpyoturrencies than former studies (such as \cite{GlasPoddig2018}), which only report aggregated results (as in case A). Furthermore, case B allows for rebalancing, as we did in the non-econometric analysis before.

\section{Dataset}\label{Dataset}
\subsection{Identification of typical Asset Classes for a German Investor's Portfolio}
\subsubsection{Benchmark Assets}
The dataset used in this study consists of historical weekly total return index data of different asset classes, which are relevant from a German investors' perspective, from 2014-01-01 to 2019-05-31. A weekly data frequency, like e.g.\ \cite{WuPandey2014} already used for their analysis, has the advantage, not to be prone to weekday effects (less noisy) and to exclude weekends, because only cryptocurrency data is available for weekends (\cite{GibbonsHess1981,Yermack2015}).\\

Therefore, as a first step, we need to define an exemplary benchmark portfolio for German investors without cryptocurrencies (\textit{benchmark assets} only).

On the superordinate level, we start with the asset class categories \cite{HornOehler2019} and \cite{OehlerHorn2019} identified in their analysis of German households' balances: stocks, cash equivalents, bonds, real estate and luxury goods. Because of the heterogeneity of these different categories, we use a more granular classification. We subsume money market investments and international currencies under the cash equivalents. The category bonds is divided in the asset classes sovereign bonds, covered bonds and corporate bonds. Finally, for the definition of the category of luxury goods, we do not only refer to luxury goods in the narrow sense (such as jewellery, paintings, oldtimer-cars, etc.), but also to commodities (especially precious metals) because of their use in jewellery, fashion (haute couture), for collectors' coins, etc. 

We identify benchmark indices for all the benchmark asset classes, which are shown in Table \ref{tab:Dataset} below. All the relevant indices are provided by Bloomberg, Thomson Reuters Eikon and the Frankfurt Stock Exchange (see Appendix for a detailed line-up of the indices and the related data sources). Besides the already existing benchmark indices, we also use an own (customized) currency basket index based on exchange rate data provided by Bloomberg.\footnote{The customized equally-weighted currency basket is an index, which tracks the development of the Euro exchange rates of important international currencies: US-Dollar (USD), Swiss Franc (CHF), Japanese Yen (JPY), Australian Dollar (AUD), New Zealand Dollar (NZD), Canadian Dollar (CAD), Norwegian Krone (NOK), Danish Krone (DKK), Swedish Krona (SEK), British Pound (GBP), Turkish Lira (TRY), South African Rand (ZAR), Russian Ruble (RUB), Polish Zloty (PLN), Mexican Peso (MXN), Indian Rupee (INR) and Chinese Yuan (CNY).}
\begin{table}[H]
	\footnotesize \noindent
	\begin{tabularx}{\textwidth}{X l}
		\toprule
		Asset Class & Index\\
		\midrule
		Stocks & Stoxx Europe 600 TR \\
		Money Market & iBoxx EUR Jumbo TR 1-3 \\
		Currencies (FX) & Equally-weighted Currency Index (customized)\\
		Sovereign Bonds & iBoxx Euro Eurozone Sovereign Overall TR\\	
		Covered Bonds & iBoxx Euro Covered TR\\	
		Corporate Bonds & iBoxx Euro Liquid Corporates Diversified TR\\			
		Real Estate & RX REIT Performance Index\\	
		Luxury Goods & Solactive Luxury and Lifestyle Index TR  \\		
		Commodities / Gold & GSCI Commodity Index TR \\
		\bottomrule
	\end{tabularx}
\caption{: Dataset (Benchmark-Assets).}
\label{tab:Dataset}
\end{table}
This index selection deliberately ignores, that a global diversification could be more rational in the context of asset allocation. To capture the investment behavior of German investors more realisticly, the chosen indices mostly have a European focus, which is a slightly different perspective compared to other studies in the literature. This index selection does not only reflect the typical, significant home bias of German investors' portfolios, as it is reported in the related literature (e.g.\ by \cite{BaltzerEtAl2015,HornOehler2019,OehlerEtAl2007,OehlerEtAl2008}). Instead, we refer to \cite{BalliEtAl2010}, who identify, that this original home bias already became a euro bias in the course of (European) financial integration. Thus, we follow this approach and implement mostly European indices -- except for commodities, luxury goods, currencies and real estate. While the first three asset classes are traded on global markets, we propose that for real estate investments, we should restrict the focus on Germany. Most of the households own real estate properties in their own environment. As an example, consider owner-occupied real estate, (rented-out) heritages or memberships in German housing cooperatives.\\

For the further analysis, we follow \cite{HornOehler2019} and assume, that investors use exchange-traded funds (ETFs) or certificates to easily and cheaply trace the development of those selected indices without the necessity to buy all the index constituents as individual direct investments.

\subsubsection{Test Assets}
Besides those benchmark assets, we also need to define the cryptocurrencies under study as test assets. In the related literature, there are three different approaches to capture the cryptocurrency market:
\begin{itemize}
	\item \textbf{First Approach:  Individually-Selected Cryptocurrency Portfolios}\\
	The first approach to capture the asset class of cryptocurrencies is the individual definition of a customized cryptocurrency portfolio. Especially early studies on diversification effects of cryptocurrencies used a special case of this approach -- a Bitcoin-only Approach (e.g.\ \cite{WuPandey2014} or \cite{EislEtAl2015}). The idea of those studies is the use of Bitcoin as a representative cryptocurrency for the whole market because of its dominant market position, its superior level of awareness and the biggest data history available for all cryptocurrencies.\\
	Besides the Bitcoin-only approach, there are also studies, which slightly expand the size of the cryptocurrency universe under study, but still select those analyzed cryptocurrencies individually, e.g.\ a certain number of the most popular ones such as \cite{Borri2019}. But we interpret the results of \cite{ElBahrawyEtAl2018}, \cite{GlasPoddig2018}, \cite{Glas2019} and \cite{Sovbetov2018} in a way, that there are remarkable differences of the cryptocurrency performance -- probably driven by their position in the marketcap ranking (e.g.\ because of their higher visibility, a higher trading volume, the access of institutional investors, etc.).\\
	Therefore, the individual selection of a customized cryptocurrency portfolio would lead to a selection bias in the data in both cases, so that we need to reject this approach.	 
	\item \textbf{Second Approach: Market-Cap-Weighted Cryptocurrency Index}\\ 
	More recent studies, such as \cite{TrimbornEtAl2018b}, use Market-Cap-weighted cryptocurrency indices to capture a broader, more representative sample of the cryptocurrency market. One index, which is mostly used in this context, is the so called CRIX index developed by \cite{TrimbornEtAl2018a}, covering a group of the most important cryptocurrencies with regard to their market cap. Market-cap-weighting is a usual weighting scheme which is also implemented in other well-established indices, such as the stock indices DAX, EUROSTOXX50, S\&P500 and NASDAQ.
	
	\item \textbf{Third Approach: Equally-Weighted Cryptocurrency Index}\\
	\cite{GlasPoddig2018} criticize former approaches and add a third approach using a customized equally-weighted cryptocurrency index. They argue, that former indices -- related to the first approaches -- do not use data on the whole cryptocurrency market, but too small samples of the crypto-universe only including the biggest (= large-cap) cryptocurrencies with regard to their marketcap. But this crypto-universe is changing fast, so that there is still no consistent data base existing. 
	In this context, the development of small (possibly dying) cryptocurrencies is ignored by the development of the index, which leads to a survivorship bias in the data. Moreover, the index development is driven by the cryptocurrencies with the biggest market capitalization -- especially by the Bitcoin, which has shares of around $50-90\%$ of the total market cap during the observation window. But if we assume, that the development of the cryptocurrencies returns depends on their market cap, there would be another bias in the index data, if this problematic weighting scheme would still be used. More general, the European regulations of UCITs (containing: exchange-traded funds (ETFs)) (cf.\ Sec.\ XIII. No.\ 49 in \cite{ESMA2014}, e.g.\ §§ 207-209 KAGB\footnote{The Kapitalanlagegesetzbuch (KAGB) is a German capital investment law implementing European directives for both UCITS (as codified in Directive 2009/65/EC) and AIF (as codified in the Directive 2011/61/EU).}) and Alternative Investment Funds (AIF) (e.g.\ § 221 KAGB) consistently forbid such high weightings of individual assets to secure diversification within the portfolios of those investment products. This fact would make those indices not be suitable for e.g.\ cryptocurrency ETFs.
	In contrast, the equal weighting scheme of \cite{GlasPoddig2018} captures the average development of all included cryptocurrencies.\footnote{In this context \cite{ElBahrawyEtAl2018} use a Wright-Fisher model of neutral evolution to analyse the patterns of the cryptocurrency market. They state, that the (technical) similarity of the cryptocurrencies under study and the randomness of the next generation of deadcoins leads to the absence of a logical order of all the existing cryptocurrencies (no \textit{"selective advantage"}) to some extent, because there is no consensus between the investors, which cryptocurrency could survive the cryptocurrency competition. If investors are not able to bring the investment alternatives into a logical order, an 1/N weighting would be obvious because of Laplace criterion \cite{BoernerEtAl2020}). We do not follow this argumentation in total, but rather partly, because we are of the opinion, that there are different criteria, such as the quality of the whitepaper, or the degree of innovation (compared to existing cryptocurrencies), which make investors at least be able to differentiate between serious (emerging) cryptocurrencies and e.g.\ junk coins or parody coins. Our motivation to use the equal weighting scheme is to capture the average market development, because the research question is not only limited to large-cap cryptocurrencies, but to cryptocurrencies in general.} On the other hand, this weighting scheme ignores the smaller market dephts of those cryptocurrencies with smaller market caps. As a consequence, (especially institutional) investors could have problems to realize those equally-weighted cryptocurrency portfolios -- especially with rising investment volumes. 
\end{itemize}
In this study, we follow \cite{GlasPoddig2018} (Third Approach) and implement a (customized) Equally-Weighted Cryptocurency Index (EWCI), which tries to capture the average development the whole cryptocurrency market at the reference date 2014-01-01. To mitigate those liquidity issues mentioned above, we will add an adjusted portfolio optimization approach after the regular analysis, which will also capture transaction costs and market liquidity in a combined approach (see Sec.\ \ref{ExtendedModel:Framework}). For index creation, we now need to define a suitable cryptocurrency universe.\\

Our cryptocurrency universe consists of $N = 66$ cryptocurrencies listed in the CoinMarketCap Cryptocurrency Market Cap Ranking as at 2014-01-01 (cf.\ Tab. \ref{tab:DatasetTestAssetNutshell}). For consistency reasons, we consider weekly closing prices for each cryptocurrency ($T = 283$ weeks), as we also did for the benchmark assets. 

Contrary to most of the other existing approaches, such as the CRIX by \cite{TrimbornEtAl2018a}, the EWCI index tries to capture the market as broad as possible to prevent survivorship biases (\cite{GlasPoddig2018}), not just a small (allegedly) representative segment. On the other hand the use of a closed cryptocurrency universe has the disadvantage, that some of the included cryptocurrencies become temporarily inactive or even so-called dead coins, which leads to the necessity of additional index creation rules in this context to challenge those additional hurdles. 
As a reaction, the created EWCI index uses a flexible index size (such as \cite{TrimbornEtAl2018a}, which is adjusted at the beginning of each month in steps of five (or even multiples). 
\begin{table}[H]
	\footnotesize \noindent
	\begin{tabularx}{\textwidth}{l  l  X  l  l  X}
		\toprule
		\# & Cryptocurrency & ID & \# & Cryptocurrency & ID\\
		\midrule
		1 & Bitcoin & BTC & 34 & MinCoin & MNC\\
		2 & Litecoin & LTC & 35 & I0Coin & I0C\\
		3 & Ripple & XRP & 36 & Deutsche e-Mark & DEM\\
		4 & Peercoin & PPC & 37 & TagCoin & TAG\\
		5 & Omni & OMNI & 38 & SexCoin & SXC\\
		6 & Nxt & NXT & 39 & StableCoin & SBC\\
		7 & Namecoin & NMC & 40 & FLO & FLO\\
		8 & BitShares PTS & PTS & 41 & Junkcoin & JKC \\
		9 & Quark & QRK & 42 & Datacoin & DTC\\
		10 & Megacoin & MEC & 43 & LottoCoin & LOT\\
		11 & WorldCoin & WDC & 44 & BitBar & BTB\\
		12 & Primecoin & XPM & 45 & CatCoin & CAT\\
		13 & Feathercoin & FTC & 46 & GrandCoin & GDC\\
		14 & Novacoin & NVC & 47 & TigerCoin & TGC\\
		15 & Infinitecoin & IFC & 48 & ByteCoin & BTE\\
		16 & Dogecoin & DOGE & 49 & Philosopher Stones & PHS\\
		17 & NetCoin & NET & 50 & EZCoin & EZC\\
		18 & Zetacoin & ZET & 51 & Luckycoin & LKY\\
		19 & Devcoin & DVC & 52 & Globalcoin & GLC\\
		20 & Digitalcoin & DGC & 53 & Diamond & DMD\\
		21 & Tickets & TIX & 54 & BetaCoin & BET\\
		22 & Anoncoin & ANC & 55 & CasinoCoin & CSC\\
		23 & Terracoin & TRC & 56 & Phoenixcoin & PXC\\
		24 & Freicoin & FRC & 57 & Orbitcoin & ORB\\
		25 & Copperlark & CLR & 58 & Franko & FRK\\
		26 & Ixcoin & IXC & 59 & Argentum & ARG\\
		27 & Earthcoin & EAC & 60 & Joulecoin & XJO\\
		28 & Bullion & CBX & 61 & HoboNickels & HBN\\
		29 & BBQCoin & BQC & 62 & Noirbits & NRB\\
		30 & GoldCoin & GLC & 63 & CraftCoin & CRC\\
		31 & FastCoin & FST & 64 & Elacoin & ELC\\
		32 & AsicCoin & ASC & 65 & Spots & SPT\\
		33 & MemoryCoin & MMC & 66 & Fedoracoin & TIPS\\	
		\bottomrule
	\end{tabularx}
	\caption{: Creation of the Dataset (Test-Assets), Data Source: CoinMarketCap.}
	\label{tab:DatasetTestAssetNutshell}
\end{table}
At a rebalancing date, we count the members of the active cryptocurrency universe in this point of time to compute the new index size ($N_{\text{Ind}}$), which is the closest multiple of 5 and not bigger than the size of the active cryptocurrency universe ($N_{\text{Act}}$). To mitigate liquidity issues, we rank all the active cryptocurrencies by their market capitalization and choose those {$N_{\text{Ind}}$} cryptocurrencies with the largest market capitalization. Now, if a current index constituent becomes inactive during the month, we follow \cite{TrimbornEtAl2018a} and use the Last-Observation-Carried-Forward (LOCF) method until the active cryptocurrency universe is updated again. The LOCF method is a frequently used technique for index calculation in practise, especially with regard to illiquid assets or the consideration of holidays (\cite{Brown1994,MSCI2018,Stoxx2018}). 

At those rebalancing dates, we use additional normalization factors to ensure, that the index value at this date is the same for the old and the new index composition. This rebalancing technique is frequently used in index calculation (cf.\ exemplary \cite{TrimbornEtAl2018a}  and \cite{CCI302020}).\\

Note, we do not use further data cleaning techniques in the sense of \cite{IncePorter2006}, because there are (still) no such techniques available for cryptocurrencies (as already concluded by \cite{GlasPoddig2018}). 

\subsection{Descriptive Statistics of the Dataset}{\label{DescriptiveStatistics}}
After we defined both benchmark assets and test assets as subsets of the final dataset, we now calculate log returns $r_{i,t} = \log\left[\frac{P_{i,t}}{P_{i,t-1}}\right]$ for this whole dataset. As a first impression of the result, we report descriptive statistics (cf.\ Tab.\ \ref{tab:DescriptiveStatistics}).\\

\begin{table}[htbp]
	\footnotesize \noindent
	\begin{tabularx}{\textwidth}{ l  X  X  X X  X }
		\toprule
		Data & ID & Min & Mean & Median & Max \\
		\midrule
		\multicolumn{6}{l}{Test-Assets: Cryptocurrencies}\\
		\midrule         
		EWCI & EWCI & -0.3884 & 0.0024 & -0.0141 & 0.7708 \\
		\midrule
		\multicolumn{6}{l}{Benchmark-Assets:}\\
		\midrule
		Money Market & MON & -0.0022 & -0.00006 & 0.0000 & 0.0014\\
		Currencies & CUR & -0.0243 & -0.0002 & 0.000002 & 0.0296\\
		Sovereign Bonds & SOV & -0.0222 & 0.0008 & 0.0013 & 0.0157\\ 
		Covered Bonds & COV & -0.0083 & 0.0005 & 0.0007 & 0.0058\\ 
		Corporate Bonds & COR & -0.0137 & 0.0005 & 0.0010 & 0.0105\\ 
		Stocks & STO & -0.0691 & 0.0011 & 0.0044 & 0.0499\\ 
		Real Estate & RES & -0.0508 & 0.0027 & 0.0030 & 0.0584\\
		Luxury Goods & LUX & -0.0879 & 0.0022 & 0.0038 & 0.0625\\
		Commodities / Gold & COM & -0.0899 & -0.0017 & 0.0005 & 0.0574\\
		\bottomrule
	\end{tabularx}
	\begin{tabularx}{\textwidth}{ l  X  X  X  X  X  X}
		\toprule
		ID & t-Test & B \& H- & St.-Dev. & Maximum & Sharpe & Sortino\\
		& ($p$-value)  & Return &  & Drawdown & Ratio & Ratio \\
		\midrule
		\multicolumn{7}{l}{Test-Assets: Cryptocurrencies}\\
		\midrule           
		EWCI &  0.2669 & 0.1319 & 0.1498 & 0.9630 & 0.0159 & 0.0253\\
		 &  (0.7898) &  &  &  &  & \\		
		\midrule
		\multicolumn{7}{l}{Benchmark-Assets}\\
		\midrule
		MON &  2.2405 & 0.0029 & 0.0004 & 0.0046 & 0.1334 & 0.1955\\
		 & (0.0258)  &  &  &  &  & \\
		CUR &  -0.5319 & -0.0125 & 0.0076 & 0.1791 & -0.0317 & -0.0428\\
			& (0.5952)  &  &  &  &  & \\
		SOV  & 2.5392 & 0.0402	& 0.0050 &  0.0585 & 0.1512 & 0.2156\\
			& (0.0117)  &  &  &  &  & \\
		COV & 3.6380 & 0.0243 & 0.0021 & 0.0237 & 0.2166 & 0.3278\\
			& (0.0003)  &  &  &  &  & \\
		COR &  2.8247 & 0.0285 & 0.0032 & 0.0364 & 0.1682 & 0.2458\\
			& (0.0051)  &  &  &  &  & \\
		STO &  0.9221 & 0.0589 & 0.0200 & 0.2385 & 0.0549 & 0.0752\\ 
			& (0.3572)  &  &  &  &  & \\
		RES &  2.5786 & 0.1509 & 0.0176 & 0.1766 &	0.1536 & 0.2457\\
			& (0.0104)  &  &  &  &  & \\
		LUX & 1.6614 & 0.1201 &	0.0220 & 0.2561 & 0.0989 & 0.1406\\
			& (0.0978)  &  &  &  &  & \\
		COM & -1.0872 & -0.0838 & 0.0260 & 0.5631 & -0.0647 & -0.0815\\
			& (0.2779)  &  &  &  &  & \\
		\bottomrule
\end{tabularx}
\begin{tabularx}{\textwidth}{ l  X  X  X  X  X  X}
	\toprule
	ID & Skewness & Kurtosis & Jarque- & Shapiro- & Anderson- & Cramer- \\
	& & & Bera & Wilk & Darling & von Mises \\
	& & & ($p$-value) & ($p$-value) & ($p$-value) & ($p$-value)\\	
	\midrule
	\multicolumn{7}{l}{Test-Assets: Cryptocurrencies}\\
	\midrule           
	EWCI &  0.9172 & 3.5608 & 188.520 & 0.9448 & 250.00 & 0.644\\
		 & - & - & (< 0,001) & (< 0,001) & (< 0,001) & (< 0,001)\\
	\bottomrule
\end{tabularx}
	\caption{: Descriptive Statistics, Source: Own Calculations.}
	\label{tab:DescriptiveStatistics}
\end{table}
According to these descriptive statistics (cf.\ Tab.\ \ref{tab:DescriptiveStatistics}), the sampling and weighting of cryptocurrencies affects the measured performance of the cryptocurrency market, as it was expected before (as in \cite{GlasPoddig2018}). This means: In this sample, cryptocurrency returns are not extraordinary high compared to both cryptocurrency returns in other studies (with marketcap weighted cryptocurrency indices) and to the other asset classes' returns in this study. Although cryptocurrencies have the second highest buy-and-hold (B\&H) return \textit{per annum} for the whole planning horizon, the expected returns (on a weekly basis) do not significantly differ from 0, as an additional $t$-test indicates. In addition to this (quasi non-existent) expected returns, cryptocurrency returns are much more volatile (riskier) than returns of the other assets under study, which is also reflected by the extreme values of the respective returns. In contrast to the results of most of the existing studies before, cryptocurrency returns are now not high enough to offset their large volatility, leading to smaller performance measures (Sharpe Ratio, Sortino Ratio) compared to most of the other asset classes.\\

Skewness and Kurtosis indicate that cryptocurrencies are not normally-distributed as already shown by \cite{KrueckebergScholz2018}. This indication is confirmed by additional empirical normal distribution tests (Jarque-Bera-Test, Shapiro-Wilk-Test, Anderson-Darling-Test, Cramer-von-Mises-Test), which all led to significant results (cf.\ Tab.\ \ref{tab:DescriptiveStatistics}), so that the normality assumption (null hypothesis) must be rejected. 

\section{Results of the (Standard) Portfolio Analysis}\label{Results}
\subsection{First Indication: Correlation Analysis}
As a first indication of cryptocurrencies' contribution to portfolio diversification it is necessary to focus on the (static) correlations of cryptocurrencies and the other portfolio assets.\\ 

Therefore, we calculate the Pearson correlation matrix for all the assets under study (Tab. \ref{tab:CorrMatrix}). 
\begin{table}[htbp]
	\footnotesize \noindent
	\begin{tabularx}{\textwidth}{ l  X  X  X  X  X  X  X  X  X  X}
		\toprule
		 & MON & CUR & SOV & COV & COR & STO & RES & LUX & COM & EWCI  \\
		\midrule
		MON & 1.00 &  &  &  &  & & & & &\\
			& (***) & & & & & & & & &\\
		CUR & 0.29 & 1.00 &  &  &  & & & & &\\
			& (***) & (***) & & & & & & & &\\
		SOV & 0.55 & 0.17 & 1.00 &  &  & & & & &\\
			& (***) & (***) & (***) & & & & & & &\\	
		COV & 0.78 & 0.16 & 0.73 & 1.00 &  & & & & &\\ 
			& (***) & (***) & (***) & (***) &  &  &  &  & &\\		
		COR & 0.56 & 0.25 & 0.65 & 0.74 & 1.00 & & & & &\\
			& (***) & (***) & (***) & (***) & (***) &  &  &  & &\\
		STO & 0.01 & 0.51 & 0.02 & -0.11 & 0.17 & 1.00 & & & &\\
			& & (***) & & & (**) & (***) &  &  &  &\\		 
		RES & 0.09 & 0.29 & 0.11 & 0.10 & 0.26 & 0.48 & 1.00 & & & \\ 
			& & (***) & (***) & (**) & (***) & (***) & (***) &  &  &\\		
		LUX & 0.05 & 0.49 & 0.02 & -0.08 & 0.18 & 0.80 & 0.45 & 1.00 & &\\ 
			& & (***) & & & (**) & (***) & (***) & (***) &  &\\		
		COM & -0.04 & 0.33 & -0.10 & -0.11 & 0.04 & 0.39 & 0.13 & 0.29 & 1.00 & \\ 
			& & (***) & & & & (***) & (**) & (***) & (***) &\\		
		EWCI & 0.06 & -0.02 & 0.03 & 0.02 & 0.06 & 0.03 & 0.07 & 0.04 & 0.02 & 1.00\\
			&  &  &  &  &  &  &  &  &  & (***)\\
		\midrule
		\multicolumn{11}{l}{\textit{Significance-Levels: } *** $\leq$ 0.001 < ** $\leq$ 0.01 < * $\leq$ 0.05}\\
		\bottomrule
	\end{tabularx}
	\caption{: Pearson Correlation Matrix, Source: Own Calculations.}
	\label{tab:CorrMatrix}
\end{table}
The Pearson correlation matrix indicates, that cryptocurrencies have an insignificant correlation with the other asset classes, which means their returns develop independently of the other asset classes. According to \cite{Markowitz1952,Markowitz1959} and \cite{Malkiel2019}, independent assets have a medium diversification potential.\footnote{Because of the non-normality of cryptocurrency returns, we follow \cite{KrueckebergScholz2018} and re-run our calculations using the two non-parametric correlation measures Spearman's rho and Kendall's tau. Because the results of the different correlation measures remain stable (only incremental changes in the coefficients), we do not report those results seperately.} 

In the following, we want to further examine the results shown in the last line of Tab.\ \ref{tab:CorrMatrix} (cryptocurrency correlations) by considering the changes of the correlation coefficients $\rho$ over time. Fig.\ \ref{fig:RollingCorrelations} shows the results of this analysis: we calculate 10-week rolling-window-correlations between cryptocurrencies and other exemplary assets (here: stocks, corporate bonds and commodities) to check the stability of the correlation coefficient close to $ \rho = 0 $. With the probability density $p(r;n,\rho)$ of the correlation coefficient (Eq.\ 2.1 in \cite{OlkinPratt1958} for the case of unknown parameters) and the assumed correlation ($\rho \approx 0 $), the variance of the correlation coefficient can be determined. This variance occurs when evaluating finite time series (number of data points $ n < \infty $) due to statistical errors. The calculation of the integral leads to the following equation for the variance: $ \text {Var} [r] = \frac {1}{n-1} $. Since the cryptocurrencies do not have normally distributed returns, this calculation is an indicative estimate in the present case. For the 10-week rolling window ($n = 10$), the estimate of the standard deviation, as the square root of the variance $\sigma = \sqrt{\text{Var}[r]} = \sqrt{\frac{1}{10-1}} = \frac{1}{3}$, is visualized by an error band (black) in Fig.\ \ref{fig:RollingCorrelations}. It becomes obvious, that the majority of the correlation coefficients lies within the error band. In addition, no clear trends can be identified that permanently lead the correlation coefficient out of the band. This results lead to the strong assumption that the correlation coefficient between cryptocurrencies and other asset classes is stable over time in the observation period, without trends and close to zero.
\begin{figure}[H] 
	\includegraphics[width=\textwidth]{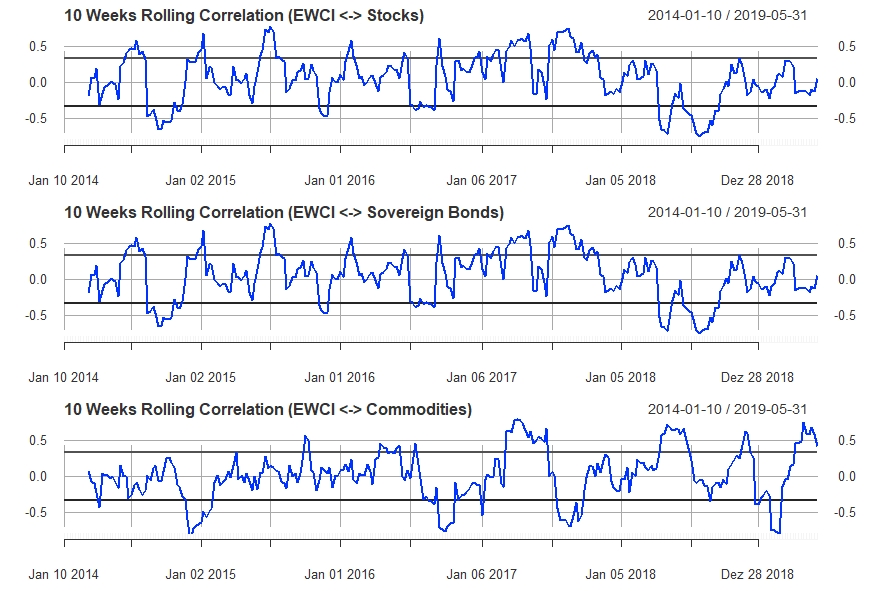}
	\caption{: 10-Week Rolling Correlations (EWCI to selected Asset Classes), Source: Own Calculations.}
	\label{fig:RollingCorrelations}
\end{figure}

\subsection{Portfolio Optimization Analysis: Results}\label{OptimizationResults}
\subsubsection{(Optimal) Portfolio Weights}\label{pfweights}
After the correlation analysis gave a first indication of cryptocurrencies' diversification potential, we now show the results for the graphical and descriptive statistics based approaches of the portfolio analysis. Therefore, we need to differentiate the results for the Cases A and B (Sec.\ \ref{Markowitz}):\\

\noindent \textit{Results of Case A:}\\
As a first step, we focus on the results of the portfolio optimization, if the whole dataset is used as an observation period (here: in the unrestricted case). From now on, we differentiate between the results for 
\begin{itemize}
	\item the (Global) Minimum Variance Portfolio (GMVP) and
	\item the Tangency Portfolio (TP)
\end{itemize}
for both cases -- with and without the consideration of cryptocurrencies. It becomes obvious, that cryptocurrencies only have a small (to neglegible) portfolio weight (close to $0.00\%$) in the optimization results (cf.\ Tab.\ \ref{tab:PortfolioAllocationA}). Presumably neglegible portfolio weights of cryptocurrencies are also observable for the respective portfolios considering additional long-only restrictions (cf.\ Tab.\ \ref{tab:PortfolioAllocationLongA} in the Appendix). Beyond the cryptocurrency weights, it becomes obvious, that (especially) the unconstrained TPs are dominated by extreme long- and short positions. However, this typical drawback\footnote{Although previous studies did not necessarily disclose all of their optimal portfolio weights, we expect similar patterns for the unconstrained optimization results in at least some of those studies due to the comparability of the applied optimization approaches. According to \cite{GreenHollifield1992} such effects can be driven by a single dominant factor in the covariance matrix. Because of the high dimensionality and the limited data availability, we checked all of our covariance matrices, whether they are well-conditioned. As a result, we do not see substantial problems with regard to a suitable condition of the covariance matrices used in this paper. However, if a substantial problem with a suitable condition of the covariance matrices arises in practise, it is proposed to select an alternative covariance matrix estimator e.g.\ by \cite{LedoitWolf2004}.} of unconstrained Mean-Variance optimizations (\cite{HaerdleEtAl2018}) fades away after the introduction of the portfolio weight restrictions (\cite{FanEtAl2012,JagannathanMa2003}), as the results of the long-only portfolios confirm.
\begin{table}[H]
	\footnotesize \noindent
	\begin{tabularx}{\textwidth}{ X  c  c c c}
		\toprule
		Asset & \multicolumn{4}{c}{Portfolio Weights}\\
		&\multicolumn{2}{c}{\textit{Minimum-Variance-Portfolio}} & \multicolumn{2}{c}{\textit{Tangency-Portfolio}} \\	
		\midrule
		& No EWCI & With EWCI & No EWCI & With EWCI\\
		\midrule
		\multicolumn{5}{l}{Unconstrained Portfolio: No Transaction Costs (Full Sample: 2014-01/2019-05)}\\
		\midrule
		MON & 1.156794 & 1.156753 & -17.085677 & -15.793628\\
		CUR & -0.008207 & -0.008307 & -3.130671 & -2.934682\\
		SOV & 0.003906 & 0.003891 & -2.524664 & -2.348019\\
		COV & -0.172364 & -0.017240 & 25.849404 & 23.980406\\
		COR & 0.019226 & 0.019418 & -3.832146 & -3.504868\\
		STO & -0.002055 & -0.002065 & 0.228977 & 0.209941\\
		RES & 0.002053 & 0.002067 & 0.645085 & 0.603059\\
		LUX & -0.000234 & -0.000217 & 1.078560 & 1.006219\\
		COM & 0.000883 & 0.000895 & -0.228868 & -0.209167\\
		EWCI & 0.000000 & -0.000034 & 0.000000 & -0.009261\\
		\bottomrule
	\end{tabularx}
	\caption{: (Optimal) unconstrained Portfolio for Case A (with and without cryptocurrencies), Source: Own Calculations.}
\label{tab:PortfolioAllocationA}
\end{table}
\noindent \textit{Results of Case B:}\\
As a second step, we re-run the analysis under the assumption of monthly rebalancing. A brief look at the optimal portfolio weights for the unconstrained model indicates that cryptocurrencies still play a minor role in the portfolio composition, because their optimal portfolio weights are consistently close to $0.00\%$ -- except for some still small, but slightly more extreme outliers in the TP (Min: 0.029514; Max: 0.098583) (cf.\ Fig.\ \ref{fig:HeatmapUnconst}). An introduction of an additional long-only constraint ($\omega_i \geq 0$ $\forall i$) leads to comparably small cryptocurrency weightings and to less extreme outliers (Min: 0.000000; Max: 0.064390) in the TP (cf.\ Fig.\ \ref{fig:HeatmapLongOnly} in the Appendix.).\\

This first indication is confirmed by an additional look at the average portfolio weights of the respective asset classes (cf.\ Tab.\ \ref{tab:AvgPortfolioWeights}, Tab.\ \ref{tab:PortfolioStatisticsLongOnly}), where the average portfolio weight of cryptocurrencies is -- again -- close to $0.00\%$ with the most extreme cryptocurrency weight in the tangency portfolio of the long-only model ($1.07\%$). This result is consistent with the fact, that cryptocurrencies were usually not included (or only included with a very small weight) in most of the observation windows.
\begin{figure}[H]
	\begin{subfigure}{.5\textwidth}
		\centering
		\includegraphics[width=\linewidth]{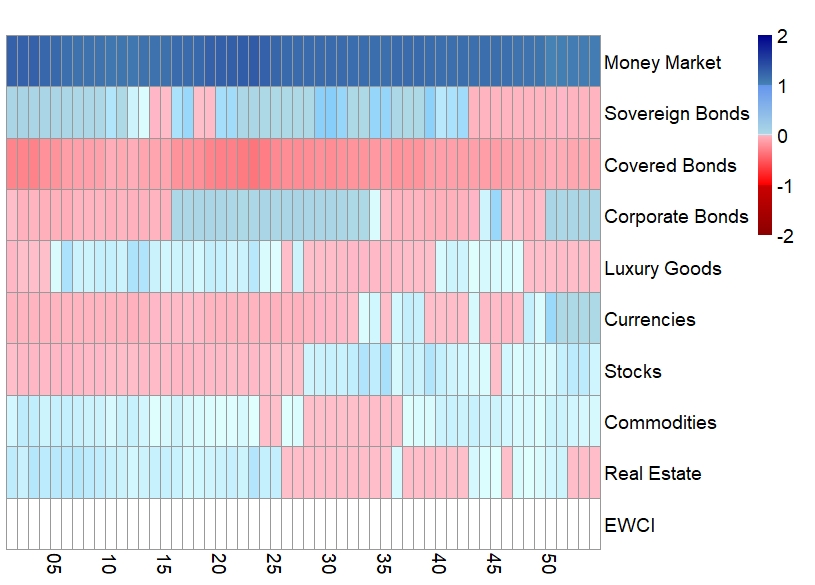}
		\caption{GMVP: Benchmark assets only}
		\label{fig:sub1a}
		\includegraphics[width=\linewidth]{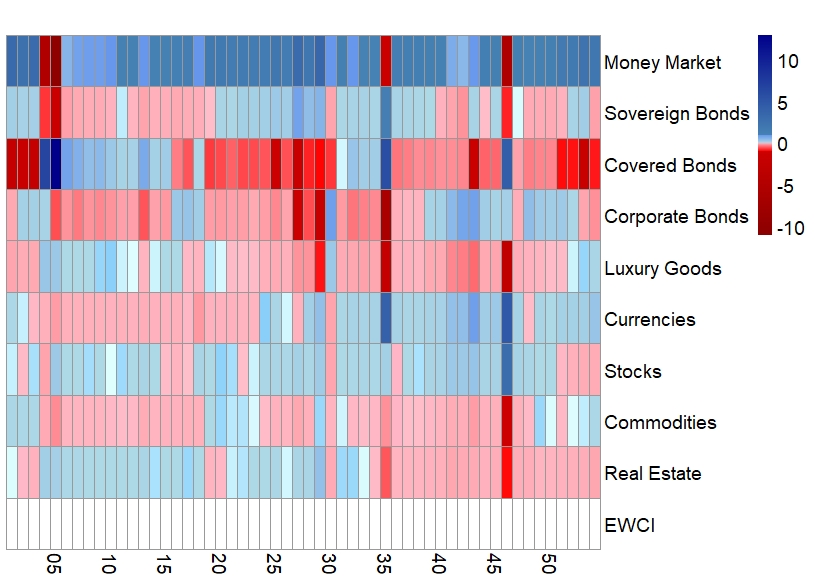}
		\caption{TP: Benchmark assets only}
		\label{fig:sub1b}
	\end{subfigure}%
	\begin{subfigure}{0.5\textwidth}
		\centering
		\includegraphics[width=\linewidth]{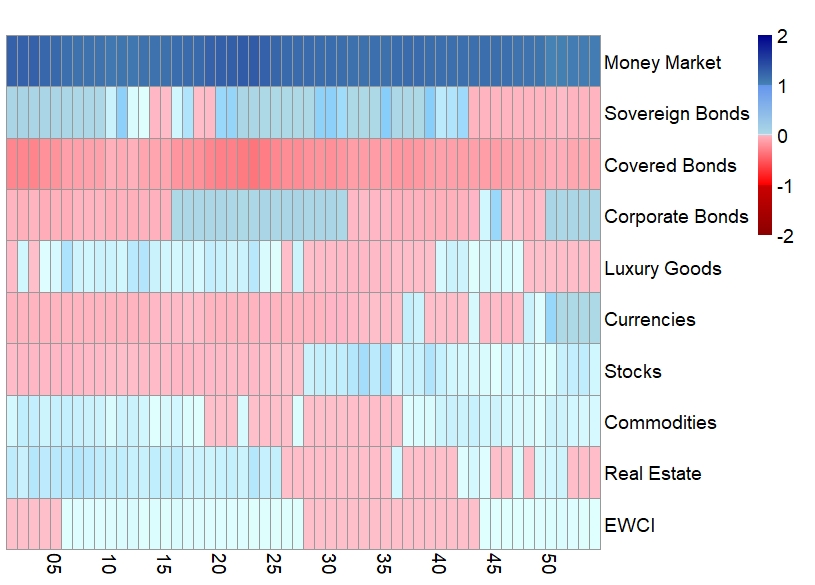}
		\caption{GMVP: Considering Cryptocurrencies}
		\label{fig:sub1c}
		\includegraphics[width=\linewidth]{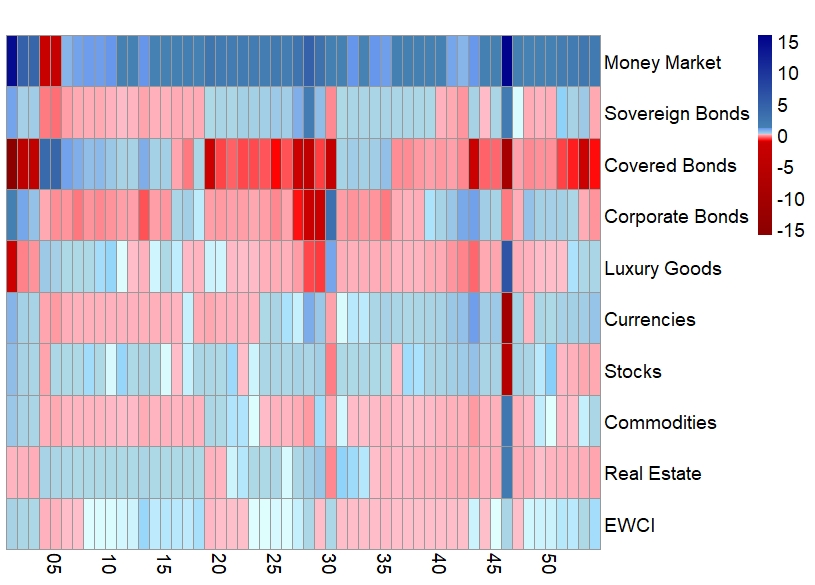}
		\caption{TP: Considering Cryptocurrencies}
		\label{fig:sub1d}
	\end{subfigure}
	\caption{: Portfolio Weight Heatmap for Case B ($T = 54$, 1-Year Rolling Windows): Unconstrained Portfolios (No Transaction Costs), Source: Own Calculations.}
	\label{fig:HeatmapUnconst}
\end{figure}
\begin{table}[H]
	\footnotesize \noindent
	\begin{tabularx}{\textwidth}{ X  c  c c c}
		\toprule
		Asset & \multicolumn{4}{c}{Average Portfolio Weights}\\
		&\multicolumn{2}{c}{\textit{Minimum-Variance-Portfolio}} & \multicolumn{2}{c}{\textit{Tangency-Portfolio}} \\	
		\midrule
		& No EWCI & With EWCI & No EWCI & With EWCI\\
		\midrule
		\multicolumn{5}{l}{Unconstrained Portfolio: No Transaction Costs (1-Year Rolling Windows)}\\
		\midrule
		MON & 1.160374 & 1.157118 & 0.891922 & 1.823775\\
		CUR & -0.007018 & -0.007123 & 0.220982 & -0.124588\\
		SOV & 0.009944 & 0.010162 & 0.013924 & 0.092893\\
		COV & -0.168913 & -0.162026 & 0.164702 & -0.810179\\
		COR & 0.004965 & 0.000874 & -0.179664 & -0.019372\\
		STO & -0.001195 & -0.001144 & 0.091388 & -0.081516\\
		RES & 0.000871 & 0.001178 & -0.023147 & 0.033939\\
		LUX & -0.000067 & 0.000002 & -0.138160 & 0.051076\\
		COM & 0.001039 & 0.000937 & -0.041947 & 0.029609\\
		EWCI & 0.000000 & 0.000022 & 0.000000 & 0.004364\\
		\bottomrule
	\end{tabularx}
	\caption{: Unconstrained Portfolio: Average Portfolio Weights for Case B (with and without cryptocurrencies), Source: Own Calculations.}
	\label{tab:AvgPortfolioWeights}
\end{table}
As a result of the Cases A and B, these (mostly) low portfolio weights of cryptocurrencies lead to the following first indications:
\begin{itemize}
	\item[(i)] The benchmark assets might span the test assets in most of the analyzed windows, so that there would be no significant diversification potentials in most of the analysed windows.
	\item[(ii)] The differentiation of the Cases A and B discloses, that a too aggregated analysis of the diversification potentials of cryptocurrencies (case A) can lead to a (to some extent misleading) conclusion, that there are no diversification potentials of cryptocurrencies at all (like \cite{GlasPoddig2018} concluded), whereas a more granular analysis (Case B) indicates, that there are also a few important exceptions from this conclusion.
	\item[(iii)] The diversification effects seem to depend on the respective optimization framework (unconstrained, long-only) in detail -- even though most of the results of both frameworks seem to be comparable. Moreover, the advantageousness of a consideration of cryptocurrencies is also dependent of the investors' risk attitude (means: preference of the GMVP over the TP or vice versa).
\end{itemize}
These first conclusions might be surprising at a first glance, because the independency of cryptocurrency returns (as shown in the correlation analysis) and former results in the literature may drive the expectation of existing diversification potentials. A comprehensible reason for that largely contradictory result could be the high volatility and comparably low expected returns of cryptocurrencies in the observation window, which is notoriously also driven by the equal weighting scheme of the EWCI (cf.\ \cite{GlasPoddig2018} for similar results). Thus, it is interesting to run the additional spanning tests in a later step to check the significance of the test assets' potential diversification effects more precisely.

\subsubsection{Portfolio Metrics and Wealth Development}
In the former analysis we found that the optimal portfolio weights of cryptocurrencies (and therefore their diversification potentials) for a certain observation window also depend on the assumed optimization model (e.g. unconstrained or long-only). Besides those optimal portfolio weights, which gave interesting results as preliminary works for the following spanning tests, it is also interesting to stress this differences by comparing divergent metrics of the resulting portfolios and the development of how the wealth invested in these optimal portfolios develops in time.\\

Rebalancing the portfolios in accordance to the abovementioned optimal portfolio weights of Case B leads to the following portfolio metrics for the unconstrained portfolio (cf.\ Tab.\ \ref{tab:AvgPortfolioWeights}):
\begin{table}[H]
	\footnotesize \noindent
\begin{tabularx}{\textwidth}{X  c  c  c  c}
	\toprule
	Measure & \multicolumn{4}{c}{Portfolio Statistics}\\
	&\multicolumn{2}{c}{\textit{Minimum-Variance-Portfolio}}&\multicolumn{2}{c}{\textit{Tangency-Portfolio}}\\
	\midrule
	& No EWCI & With EWCI & No EWCI & With EWCI\\
	\midrule
	Minimum Return & -0.002498 & -0.002504 & -0.057127 & -0.249672\\
	Mean Return & -0.000003 & -0.000004 & 0.000048 & -0.003051\\	
	Maximum Return & 0.001036 & 0.001061 & 0.111843 & 0.017646\\	
	\midrule		
	Standard Deviation & 0.000353 & 0.000358 & 0.011902 & 0.021174\\
	Max Drawdown & 0.005660 & 0.006424 & 0.214632 & 0.536954\\
	\midrule
	Sharpe Ratio & -0.007686 & -0.011475 & 0.003994 & -0.144067\\
	Sortino Ratio & -0.009840 & -0.014770 & 0.007129 & -0.144041\\
	\midrule
	End Value ($C_T$ in EUR) & 99.94 & 99.90 & 99.51 & 46.50\\	
	\bottomrule
\end{tabularx}
	\caption{: Key Statistics of the Unconstrained Portfolio for Case B (with and without Cryptocurrencies, 1-Year Rolling Window), Source: Own Calculations.}
	\label{tab:PortfolioStatistics}
\end{table}
In this context, we find that mixing cryptocurrencies in the benchmark portfolios leads to lower mean returns, higher risks and lower performance measures (cf.\ Tab.\ \ref{tab:PortfolioStatistics}). Moreover, an initial investment of $C_0 = 100.00$ EUR in the (continuously rebalanced) portfolios leads to lower end values $C_T$ at the end of the planning horizon, if cryptocurrencies are included in the investment opportunity set of the investors. This results hold for both -- the GMVP and the TP. In contrast to the unconstrained case, the consideration of cryptocurrencies in the investment opportunity set of the long-only model now leads to a remarkably higher return, (again) to a higher risk and -- as a consequence -- to a higher performance. Moreover, the consideration of cryptocurrencies now leads to (remarkably) higher end values $C_T$ for both portfolios (GMVP, TP; cf.\ Tab.\ \ref{tab:PortfolioStatisticsLongOnly} in the Appendix).\\

These two examples demonstrate, that the consideration of cryptocurrencies cannot be called '\textit{efficient}' or '\textit{inefficient}' in general, but the results depend on the optimization framework and the current market trends, instead. In our example, we can observe, that the disadvantegeousness of considering cryptocurrencies for the TP in the unconstrained framework is mainly driven by window 46 with an extremely negative portfolio return (caused by extreme short sales and extreme returns). As a consequence, we can justify our granular subsample analysis by the obvious impact of a few windows on the overall portfolio development and therefore the evaluation of the efficiency of cryptocurrencies in general. 

\subsection{Econometric Analysis: Spanning-Tests}\label{SpanningTestResults}
After this typical portfolio optimization analysis, we now change the point of view and repeat the analysis for the unconstrained portfolio framework using the spanning tests introduced above (as a regression-based approach). Again, we differentiate two different cases (in analogy to the portfolio optimization analysis).\\

\noindent \textit{Results of Case A:}\\
To demonstrate the (general) functioning of the introduced spanning tests, we start with a graphical analysis of the efficient frontier in the (unconstrained) mean-variance-model (cf.\ Fig.\ \ref{fig:EfficientFrontierShort} in Sec.\ \ref{ExtendedModelResults}), where the observation period contains the whole dataset (Case A).

As we can see, the efficient frontier is only shifted minimally, if cryptocurrencies (test assets) are considered in the investment opportunity set.\footnote{The key results also hold, if we do not assume an unconstrained portfolio, but a long-only portfolio instead (cf.\ Fig.\ \ref{fig:EfficientFrontierLongOnly}). But the spanning tests only refer to the unconstrained portfolios.} This can be interpreted as the next indication, that the diversification effect of cryptocurrencies is small to neglegible in this context. To analyze, whether this graphical impression is a significant effect, we firstly focus on the regression results of the mean-variance-spanning tests summarized in Tab.\ \ref{tab:SpanningTestResults}. In addition to this mean-variance-model, we also run the further spanning-tests introduced above (GMM-Wald-Tests, Stepdown Tests) to gain deeper insights, if these results remain robust, if the methodology is changed (robustness checks).
\begin{table}[H]
		\footnotesize \noindent
			\begin{tabularx}{\textwidth}{ l  l  l  l  ||l l ||l  l }
			\toprule
			Test- & \multicolumn{3}{c}{MV-Spanning-Tests} & \multicolumn{2}{c}{GMM-Wald-Tests}&\multicolumn{2}{c}{Stepdown-Test} \\
			Assets    & W & LR & LM & FFK & BU (EIV Adj.) & F1 & F2\\
			&($p$-val.)&($p$-val.)&($p$-val.)&($p$-val.)&($p$-val.)&($p$-val.)&($p$-val.)\\
			\midrule
			EWCI   & 0.1138 & 0.1138 & 0.1138 & 0.0937 & 0.0899 & 0.0385 & 0.0715\\
			        & (0.945) & (0.945) & (0.945) & (0.954) & (0.956) & (0.845) & (0.789)\\    
			\bottomrule
		\end{tabularx}
		\caption{: Results of the Spanning-Tests (Case A), Source: Own Calculations.}
		\label{tab:SpanningTestResults}
\end{table}
The econometric results confirm the former conjecture, that the implementation of cryptocurrencies (test assets) in German investors' portfolios does not lead to a (significant) shift of the efficient frontier -- and therefore significant diversification effects. This result does not only hold for the mean variance spanning tests but also for the GMM-Wald spanning test (which accounts for non-normal and heteroscedatstic returns). Besides those joint tests, the stepdown procedure makes it possible to differentiate between changes in the tangency portfolio and in the GMVP: here we see, that for both optimization strategies, including cryptocurrencies in the investment universe does not lead to significant diversification effects in this (broad) observation window. Hence, because cryptocurrencies (test assets) obviously seem to have a nearly zero weight in both of the portfolios (which is consistent to the results in Sec.\ \ref{OptimizationResults}), we can conclude, that they also have it in all of the other possible (efficient) portfolios (linear combinations). In other words: cryptocurrencies are not part of all the portfolios on the efficient frontier (\cite{GlasPoddig2018}). This result could be shown before in the graphical analysis of this section and the portfolio optimization analysis applied before (Sec.\ \ref{OptimizationResults}).\\

\noindent \textit{Results of Case B:}\\
If this spanning test analysis is now repeated for smaller subsamples of a 1-year rolling-window (Case B), we can observe results as in Tab.\ \ref{tab:SpanningTestResultsRollingWindow}. For a better overview, the results are aggregated, so that we count all regression results, in which we find significant diversification effects for a given significance level $\xi$ (with $\xi = (0.10,0.05,0.01,0.001)$).\footnote{The significance level of $\alpha = 0.10$ is only used to detect significant results in the course stepdown procedure ($F_1$), as it is described above. All other tests only use significance levels up to $\alpha = 0.05$ for this purpose.}Afterwards, we compute their shares to the total number of regression results ($54$) for each test (cf.\ Tab.\ \ref{tab:SpanningTestResultsRollingWindow}). 
\begin{table}[H]
	\footnotesize \noindent
\begin{tabularx}{\textwidth}{ l  l  l  l  ||l l ||l  l }
	\toprule
	Test- & \multicolumn{3}{c}{MV-Spanning-Tests} & \multicolumn{2}{c}{GMM-Wald-Tests}&\multicolumn{2}{c}{Stepdown-Test} \\		
	Assets    & W & LR & LM & FFK & BU (EIV Adj.) & F1 & F2\\
	&&&&&&&\\	
	\midrule
		Significant & 0.204 & 0.185 & 0.148 & 0.222 & 0.167 & 0.204 & 0.074\\
		\midrule 
		Thereof: &&&&&&&\\
		$10\% \leq \xi < 5\%$ & --- & --- & --- & --- & --- & 0.074 & ---\\
		$5\% \leq \xi < 1\%$ & 0.111 & 0.093 & 0.074 & 0.093 & 0.093 & 0.111 & 0.074\\
		$1\% \leq \xi < 0.1\%$ & 0.037 & 0.093 & 0.074 & 0.074 & 0.056 & 0.019 & 0.000\\  
		$0.1\% \leq \xi$ & 0.056 & 0.000 & 0.000 & 0.056 & 0.019 & 0.000 & 0.000\\
		\midrule
		Non-Significant & 0.796 & 0.815 & 0.852 & 0.778 & 0.833 & 0.796 & 0.926\\ 
		\midrule
		Sum of shares & 1.000 & 1.000 & 1.000 & 1.000 & 1.000 & 1.000 & 1.000\\  
		\bottomrule
	\end{tabularx}
	\caption{: Results of the 1-Year Rolling-Window-Spanning-Tests (Case B), Source: Own Calculations.}
	\label{tab:SpanningTestResultsRollingWindow}
\end{table}
We find that the question, whether considering cryptocurrencies in German investors' portfolios leads to a (significant) diversification effect or not, must be answered more differentiated than it was done in case A (and the former studies in the literature). Our results reflect, that the (joint) spanning tests' do not reject the null hypothesis most of the time (in around $80-85\%$ of the cases) because of insignificant results. On the other hand, it means, that there are also around $15-20\%$ significant results, anyway. This shows, that neglecting the diversification effects of cryptocurrencies in general would be misleading.  
\begin{figure}[H]
	\includegraphics[width=\linewidth]{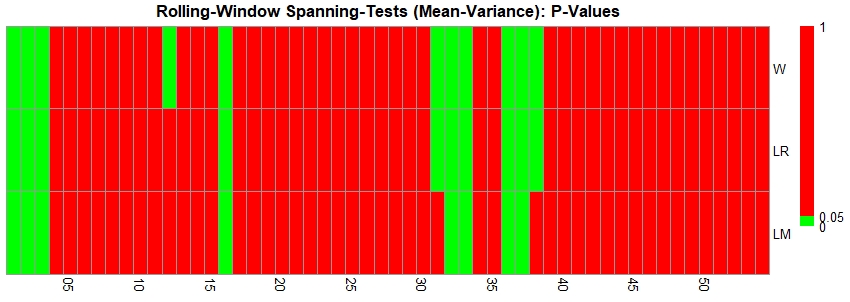}
	\includegraphics[width=\linewidth]{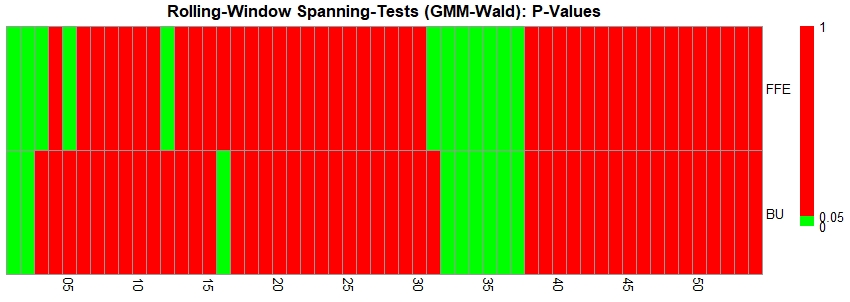}
	\includegraphics[width=\linewidth]{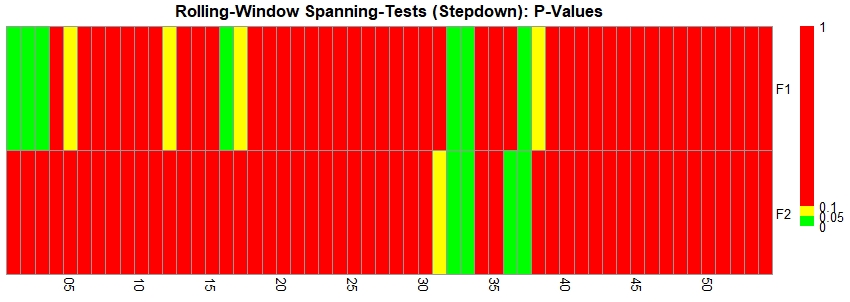}
	\label{fig:HeatmapSpanning}
	\caption{: Rolling-Window-Spanning-Tests (Case B): $p$-values, Source: Own Calculations.}
\end{figure}
As a next step, we present the results of the Rolling-Window-Spanning-Tests more in detail (Fig.\ 3). These disclose three interesting perspectives: 
\begin{itemize}
	\item[(i)]The observation windows with significant diversification effects do not appear randomly, but rather in clusters instead, as the green (or yellow) blocks in the heatmaps signalize. Moreover we find widely consistent clusters and only small differences in the results of all the tests applied in this context. If we now refer to Fig.\ 4, it is obvious, that the first bigger cluster (windows 1-5) contains a remarkable market downturn of the cryptocurrency market, while the second bigger cluster (windows 31-38) contains the cryptocurrency boom in 2017 and the bust in January 2018, which was a consequence of the introduction of Bitcoin futures. But this result should not lead to the hasty conclusion, that extreme movements of the cryptocurrency market are the only premise for significant diversification benefits. The -- in some parts unexpected -- negative portfolio weights for the windows 31-38 (cf. Fig.\ \ref{fig:HeatmapUnconst}) point out, that other factors beyond the (expected) cryptocurrency returns, such as the (co-)variances of the different benchmark- and test assets also seem to be relevant drivers in this context. Moreover, the respective portfolio weights for the windows with significant spanning tests confirm, that even in those times with extreme market movements on the cryptocurrency market the optimal cryptocurrency weights still remain small.
	\item[(ii)]The relevance of cryptocurrencies for the optimal portfolio mostly depends on the optimization strategy. This means: if we measure a significant diversification effect of cryptocurrencies for the GMVP at a certain rebalancing date, we do not necessarily find significant results for the TP (and vice versa). This pattern indicates, that the efficient frontier is not really shifted, but rather rotated in such cases. Therefore, it is theoretically possible, that the joint test still accept the null hypothesis, whereas the stepdown procedure indicates significant diversification effects (\cite{KanZou12}).
	\item[(iii)]Comparing the share of significant results of both $F_1$ and $F_2$, we find that $F_1$ has a remarkably higher share of significant results -- which also holds if we (temporarily) assume the same significance levels for both tests ($\xi_1 = \xi_2 = 0.05$) against the recommendation of \cite{KanZou12}. Thus, cryptocurrencies are more relevant in the tangency portfolio. Following the interpretation of the $p$-values in the stepdown procedure by \cite{BelousovaDorfleitner2012}, cryptocurrencies are then more able to improve the return of a portfolio than reducing its level of risk, which is an expectable result with regard to their high level of price volatility. Then it is likely, that  e.g. extreme expected cryptocurrency returns during remarkable market movements outweigh their high volatility, which was identified as at least one possible driver of significant cryptocurrency shares (besides the covariance matrix) in the investors' tangency portfolios in (i). However, according to Fig.\ 4, cryptocurrencies predominantly do not lead to significant diversification benefits in normal times, when their returns are presumably not able to outweigh the respective volatilites and/or the (co-)variances do not lead to remarkable risk reduction potentials. This obvious difference to most of the existing cryptocurrency literature, which observes a more general diversification benefit of cryptocurrencies, can be explained by the different cryptocurrency weighting scheme and (possibly) by different observation windows. 
\end{itemize}
\begin{figure}[H]
	\centering
	\includegraphics[width=\linewidth]{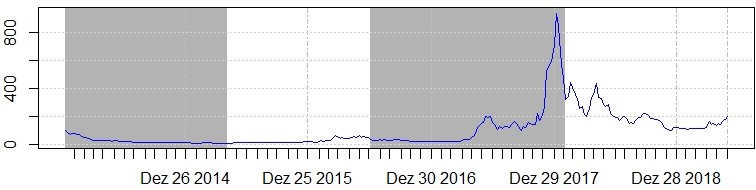}
	\label{fig:SpanWindows}
	\caption{: Comparison of the EWCI index development and obvious clusters of observation windows with significant cryptocurrency weights (grey), Source: Own Calculations.}
\end{figure}
 
\section{Effects of Additionally Accounting for Transaction Costs and a Reduced Market Liquidity on the Efficient Frontier}
\subsection{Methodological Framework}\label{ExtendedModel:Framework}
The analysis above assumed that there are no transaction costs for investors when (re-)allocating their capital between the given investment alternatives -- in analogy to the optimization framework by \cite{Markowitz1952,Markowitz1959}. Moreover, the optimization
framework assumed, that the market depth is high enough so that no illiquidities must be considered.\\

But according to \cite{Borri2019} and \cite{BorriShaknov2018}, investors definitively suffer transaction costs and illiquidity issues on the cryptocurrency market. A short look on the development of Bitcoin's historical transaction costs confirms that point, because it shows, that there seems to be a positive relation between investment demand and transaction costs (\cite{BitcoinVisuals2020a,BitcoinVisuals2020b}). The main reason for that positive relationship is, that investors can influence the probability of their transactions to be added to the blockchain faster by bidding higher transaction fees to the miners, which becomes more important, if the trading volume rises and more transactions compete with each other (\cite{Choi2018,Dwyer2015}). Looking back to the cryptocurrency boom in 2017, the Bitcoins' transaction fees  exploded up to over 40.00 EUR per trade, as the development of the daily mean of all median transaction fees per trade and block (cf.\ Fig.\ \ref{fig:TransactionCost}; blue) indicates.\footnote{We used the daily median transaction fee data (per transaction and block, converted in USD and excluding coinbase transactions) by \cite{BitcoinVisuals2020a} and converted it to EUR by using the daily USD-EUR exchange rate from Thomson Reuters Eikon.} In contrast, the average of these daily average transaction fee data for this observation window (1.29 EUR; black) indicates, that the transaction fees are much lower or even neglegible (as the respective minimum of 0.02 EUR shows) in \textit{normal times}.
\begin{figure}[H]
	\centering
	\includegraphics[width=\linewidth]{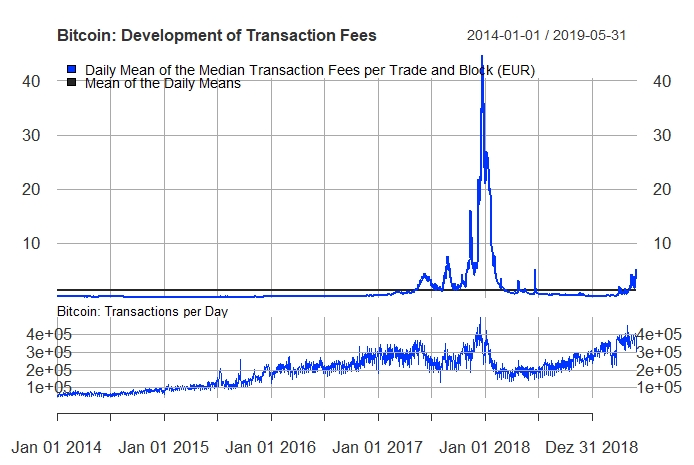}
	\caption{: Bitcoin: Historical Daily Median Transaction Fees per Trade versus Daily Transactions. Data Sources: \cite{BitcoinVisuals2020a,BitcoinVisuals2020b} and Thomson Reuters Eikon.}
	\label{fig:TransactionCost}
\end{figure}

The beforementioned liquidity issues can be confirmed by a closer look on the market depths of those cryptocurrencies: if a cryptocurrency has a higher market cap (and therefore a higher ranking in the CoinMarketCap Cryptocurrency Ranking) than other cryptocurrencies, it tends to have more attention (visibility) and is traded more frequently, which leads to a higher market depth.

The fact, that we use an equally-weighted cryptocurrency index to capture the average development of an \textit{ex-ante} defined cryptocurrency market, might lead to a problem: For high portfolio weights of the cryptocurrency index or very huge investment amounts, it is possible that we are forced to buy more coins of a single cryptocurrency than it is offered on the cryptocurrency market (\cite{TrimbornEtAl2018b}). In this context, beeing forced to buy (nearly) all of the available coins on the market would lead to extreme market movements at first. On the other hand, if more investors would use this strategy, the smaller cryptocurrencies' prices, trading volumes and market depth would rise, so that this problem would probably fade in time.\footnote{On the other hand, those smaller cryptocurrencies are prone to so-called pump and dump schemes, where investors appoint to buy a certain cryptocurrency to boost its market price (because of its small market capitalization and trading volume), which is followed by a wave of selling transactions to realize profits from speculation or to stop potential losses (\cite{LiEtAl2019}). But this phenomenon is distinctive for the cryptocurrency market and can be scheduled for a (random) cryptocurrency. It is useless (and nearly impossible) to identify all historical attempts of pump and dump schemes and to exclude all cryptocurrencies from the dataset, which were related to this attempt. But it is noteworthy, that an exclusion is not necessary, because those pump and dump scenarios only last just a few seconds, which could (more likely) bias intra-day data, but not our weekly observation windows.}\\

To account for those liquidity constraints mentioned above, \cite{TrimbornEtAl2018b} propose a Liquidity Bounded Risk Return Optimization (LIBRO) approach based on the methodological foundation of \cite{DarollesEtAl2015}. In this course, they add a liquidity constraint for the individual portfolio weights on the cryptocurrency level (based on their respective turnover values) as a further equation in the portfolio optimization framework. But this approach is harder to apply, if not only single cryptucurrencies, but (aggregated) indices representing dozens of single titles of different asset classes are used in the optimization.\\

\cite{Borri2019} and \cite{BorriShaknov2018} use an alternative approach to capture both transaction costs and illiquidity issues. They define a portfolio optimization framework assuming unilateral (or: asymmatric) transaction costs. This means: transaction costs are assumably not equal for all asset classes and -- at least in their study -- limited to cryptocurrency trades, which can be reasoned by the above-mentioned explosion of cryptocurrency transaction costs in booms. As a (positive) side-effect, those transaction costs on cryptocurrency trades cause, that investors are penalized for extreme shifts in their cryptocurrency positions (leading to consistently small portfolio shares of cryptocurrencies in their framework). This effect also reduces the probability, that there are less cryptocurrency units available on the exchanges than the number of units demanded (illiquidity issues). 

Other works, just as \cite{Anyfantaki2018} and \cite{TopaloglouTsomidis2018}, also assume  asymmetric transaction costs (means: penalizing transactions of a certain asset class more than the trades of other asset classes) to the disadvantage of cryptocurrencies in another methodological context, but in contrast to \cite{Borri2019} and \cite{BorriShaknov2018} they do not set the benchmark assets' transaction costs to zero. Moreover, they do not distinguish different transaction cost levels for taker- and maker- transactions. Instead, they assume transaction costs of $35$ BP for all benchmark asset transactions and $50$ BP for all cryptocurrency transactions. However, for their final asset allocation, investors primarily focus on the difference between both transaction cost levels, so that there is no big difference between these approaches mentioned before.\\

In order to control for (possible) changes of the results, if transaction costs are implemented and illiquidity issues are reduced, we refine the general ideas of the beforementioned studies and build a customized alternative optimization framework, which also takes a more realistic (asymmetric) transaction cost scheme into consideration.

For the benchmark assets, we assume a linear transaction cost function
\begin{flalign}\label{22d}
\tau_{i}^{\text{Benchmark}} = c_{i}^{\text{Benchmark}} \text{ } |\Delta\omega_{i}| \text{ } V_{0},
\end{flalign}
where $c_{i}^{\text{Benchmark}}$ is a fixed transaction cost factor for asset $i$, $\Delta\omega_{i}$ is the change in the portfolio weights of asset $i$ and $V_{0}$ is the wealth invested in the initial portfolio. We set $c_{i}^{\text{Benchmark}} = 35$ BP,  as it is proposed by \cite{Anyfantaki2018} and \cite{TopaloglouTsomidis2018}.  

But with regard to the cryptocurrencies (still our test assets), we use a divergent approach. In the light of demand-dependent transaction cost factors, a (linear) transaction cost function $\tau_{i}^{\text{Test (Lin.)}}$ including a fixed transaction cost factor $c_{i}^{\text{Test}}$, such as
\begin{flalign}\label{22a}
\tau_{i}^{\text{Test (Lin.)}} = c_{i}^{\text{Test}} \text{ } |\Delta\omega_{i}| \text{ } V_{0},
\end{flalign}
which is (implicitly) proposed in the former studies in the literature (e.g.\ \cite{Borri2019,BorriShaknov2018,Anyfantaki2018,TopaloglouTsomidis2018}), would not be suitable, because of the overproportional rise of the transaction costs, which was observed in Fig.\ \ref{fig:TransactionCost}. We now implement, that the transaction cost factor $c_i^{\text{Test}}$ is not fixed any more, but now depends on the shifts $|\Delta\omega_{i}|$ in the portfolio allocation and a marginal cost contribution factor $\tilde{c}_{i}^{\text{Test}}$, which consists of the fixed transaction cost factor $c_{i}^{\text{Test}}$ (from the linear case) and a special scalability factor $\Psi$. The latter has to be chosen in a further step. Consequently, these changes lead to 
\begin{flalign}\label{22c}
c_{i}^{\text{Test (new)}} = \frac{1}{2} \underbrace{\vphantom{\frac{1}{2}}\Psi c_{i}^{\text{Test}}}_{\substack{\tilde{c}_{i}^{\text{Test}}}}\text{ }|\Delta\omega_{i}|.
\end{flalign}
If Eq.\ \eqref{22c} is inserted in Eq.\ \eqref{22a}, we have a quadratic transaction cost function, so that we can simply write
\begin{flalign}\label{22b}
\tau_{i}^{\text{Test (Quad.)}} = \frac{1}{2} \tilde{c}_{i}^{\text{Test}} \text{ } |\Delta\omega_{i}|^2 \text{ } V_{0}.
\end{flalign}
This beforementioned form of the transaction cost function can also be derived from the Taylor series (see Appendix A.3). The function now captures rising transaction cost factors $c_{i}^{\text{Test (new)}}$, if investors buy/sell high shares of cryptocurrencies.  Furthermore, it penalizes those great shifts of the portfolio allocation in favor of cryptocurrencies, which is desirable with regard to the illiquidity issues on the cryptocurrency market. A more practical rationale for this formula could be: if investors face extreme shifts of their optimal portfolio to the (dis)advantage of cryptocurrencies, this is probably caused by a significant market upturn (downturn). In both situations investors are assumed to bid higher transaction fees to accelerate the current transactions. 

On the other hand, for small changes of the respective portfolio weight, we have smaller transaction costs -- sometimes even smaller than the transaction costs of other asset classes (calculated with the linear case formula), which is consistent with the low (average) transaction cost level on the cryptocurrency market in normal times.

As a generalization of the cryptocurrency case, the comparison of both transaction cost functions (linear/quadratic) discloses, that we need to seperate three different cases: 
\begin{itemize}
	\item[(i)] $\tau_{i}^{\text{Test (Quad.)}} < \tau_{i}^{\text{Test (Lin.)}}$ 
	($\forall$ $|\Delta$$\omega_{i}| < |\Delta\omega_{i}^{*}|$)
	\item[(ii)] $\tau_{i}^{\text{Test (Quad.)}} = \tau_{i}^{\text{Test (Lin.)}}$ 
	($\forall$ |$\Delta$$\omega_{i}| = |\Delta\omega_{i}^{*}|$)
	\item[(iii)] $\tau_{i}^{\text{Test (Quad.)}} > \tau_{i}^{\text{Test (Lin.)}}$ 
	($\forall$ $|\Delta$$\omega_{i}| > |\Delta\omega_{i}^{*}|$).
\end{itemize}
The calculation of the intersection point $|\Delta\omega_{i}^{*}|$ is demonstrated in the Appendix in detail. But this calculation depends on the individual parameterization of the transaction cost function (with regard to $\Psi$). In this study, we aim at a parametrization, which ensures, that we have the same expected transaction cost factor as \cite{Anyfantaki2018} and \cite{TopaloglouTsomidis2018}: $\text{E}[\tau_{i}^{\text{Test (Quad.)}}] = \text{E}[\tau_{i}^{\text{Test (Lin.)}}] = 50$ BP. This result is reached, if the integral between both cost functions is zero\footnote{This relationship only holds, if the distribution of the $\Delta\omega_{i}$ is an equal distribution in the interval $[\min,\max]$. This assumption is based on the Laplace critereon, because we do not know the empirical distribution of this variable. If further research has a bigger data base, which makes estimations of this distributions more reliable, our approach can easily be adjusted.}:
\begin{flalign}\label{22}
\int_{\text{min}}^{\text{max}} d(|\Delta\omega_{i}|)\left[\frac{1}{2}\tilde{c}_{i}^{\text{Test}} \text{ } |\Delta\omega_{i}|^2 \text{ } V_{0} - c_{i}^{\text{Test}} \text{ } |\Delta\omega_{i}| \text{ } V_{0}\right]\overset{!}{=} 0.
\end{flalign} 
Note, that the interval $[\text{min},\text{max}]$ can be set exogeneously depending on the optimization strategy, e.g.\ $[0,1]$ for a long-only strategy. In this case, the parameter $ \tilde{c}_{i}^{\text{Test}}$ must be set to  $\tilde{c}_{i}^{\text{Test}}= \Psi c_{i}^{\text{Test}} = 3 c_{i}^{\text{Test}}$ to fulfill this condition in the long-only case.\footnote{See Appendix for a detailed derivation of this conditions.} For the unconstrained portfolio, it is important to mention, that max is not allowed to reach infinity for mathematical reasons. However, this restriction does not mean, that it is not possible to calculate the unconstrained portfolio: we do not restrict the portfolio weights to a maximum value, but only changes in the portfolio weights, so that building up a high position of a certain asset with gradual changes in the portfolio would still be allowed. Moreover, in the further analysis we introduce a transaction cost budget $\tilde{K}_0$, which limits the maximum amount of transactions in any case, so that the restriction of max will not be binding any more. Thus, we can consistently assume $[\min,\max] = [0,1]$ also for the unconstrained portfolios.\\

We are now able to adjust our previous Lagrange functions (Eq.\ \eqref{3} and \eqref{4}) by an additional transaction costs restriction, which leads to
\begin{flalign}\label{22}
\max_{\boldsymbol{\omega}} \mathcal{L}^{TP} = \max_{\boldsymbol{\omega}}\left[\frac{\boldsymbol{\omega}'\boldsymbol{r}-r_F}{\sqrt{\boldsymbol{\omega}'\boldsymbol{V}\boldsymbol{\omega}}}+\lambda_1 (1-\boldsymbol{\omega}'\boldsymbol{1})+\lambda_2\left(\tilde{K}_0 \geq C\right)\right]
\end{flalign}
\begin{flalign}\label{23}
\min_{\boldsymbol{\omega}} \mathcal{L}^{GMVP} = \min_{\boldsymbol{\omega}}\left[ \frac{1}{2}\boldsymbol{\omega}'\boldsymbol{V}\boldsymbol{\omega}+\lambda_1(1-\boldsymbol{\omega}'\boldsymbol{1})+\lambda_2\left(\tilde{K}_0 \geq C\right)\right]
\end{flalign}
in the unconstrained case with $\tilde{K}_0$ as a transaction cost budget (for the current portfolio shift) and $C = \boldsymbol{{\mathscr{T}}}'\boldsymbol{1}$  as the cumulated transaction costs, where $\boldsymbol{{\mathscr{T}}}$ is a ($K+N$)-vector containing all the individual transaction costs $\boldsymbol{{\mathscr{T}}} = \left[\left(\boldsymbol{\tau}_{K \times 1}^{\text{Benchmark}}\right)',\left(\boldsymbol{\tau}_{N \times 1}^{\text{Test (Quad.)}}\right)'\right]'$ caused by trades of the test assets $\boldsymbol{\tau}_{N \times 1}^{\text{Test (Quad.)}}$ and benchmark assets $\boldsymbol{\tau}_{K \times 1}^{\text{Benchmark}}$. According to the additional transaction cost restriction, the (cumulated) transaction costs caused by the shift of the assets' weightings is not allowed to exceed the exogeneously given transaction cost budget. This budget is here set to $\tilde{K}_0 = 10$ BP at first (and varied afterwards as different budget levels). In this context, it is important to mention, that our optimization model does not aim at a transaction cost optimal portfolio (such as a transaction cost minimal portfolio; cf.\ e.g. \cite{LoboEtAl2007} for examples), but uses the same optimization strategies as introduced in the sections before -- under the additional restriction, that portfolio shifts are only allowed to a certain extent. Like in Sec.\ \ref{Markowitz}, it is also possible to introduce an additional long-only constraint ($\omega_{i} \geq 0$ $\forall i$) as a robustness check, if the possibility of short sales should be excluded from the analysis.\\

We run this optimization model using a dataset covering the whole range of the sample (2014-01-01 to 2019-05-31) as an observation period, like we did in Sec.\ \ref{SpanningTestResults} in the course of the spanning tests. As a starting point for the optimization, we stick to the assumption an (equally-weighted) initial portfolio only containing the benchmark assets (like in Sec.\ \ref{Markowitz}), which is shifted to the calculated (optimal) portfolio after a certain observation period. The results give some indication, whether cryptocurrencies should considered in the optimal portfolio, if liquidity- and transaction cost considerations restrict the optimal portfolio choice. This composition of the initial portfolio is feasible in the light of the fact, that investors first have to get in touch (and gain experiences) with the emerging asset class of cryptocurrencies, while they are more experienced with the other traditional asset classes. Moreover, this composition directly translates the usual research question \textit{"How does the possibility of mixing cryptocurrencies into an investor's portfolio of benchmark assets affect the efficiency and diversification potentials of this portfolio?"} into an optimization model. If cryptocurrencies would be already included in the initial portfolio, there is no need to add them to the portfolio in a further step.\\

In this context, \cite{GlasPoddig2018} expected that the results generated by the spanning tests become even worse, if transaction costs would be introduced, but do not analyse the effects on the efficient frontier in detail. \cite{Borri2019}) consistently showed a negative shift of the efficient frontier after the introduction of transaction costs, but did not seperate between the efficient frontiers with and without the consideration of cryptocurrencies. Other works, just as \cite{Anyfantaki2018} and \cite{TopaloglouTsomidis2018} directly included transaction costs in their spanning tests, but their results (significant portfolio diversification potentials) are presumably driven by their limited, individually selected cryptocurrency portfolio, which might lead to biased results (\cite{GlasPoddig2018}). As a consequence, it is interesting to observe, whether the expectations of \cite{GlasPoddig2018} can be verified in this analysis.
 
\subsection{Results of the Extended Model}\label{ExtendedModelResults}
Applying the portfolio optimization under transaction costs for the given observation period, we find that the portfolio weights of cryptocurrencies remain at a small level at the first glance, which is not a surprising result: the cryptocurrency weightings were already small before the trading of huge amounts of cryptocurrencies was penalized with high transaction costs. On the other hand, low shifts in the cryptocurrency weightings cause lower transaction costs than shifts of the other asset classes' weightings, which might have favorable effects on the cryptocurrency weightings. A closer look on the optimization results confirms this prediction: the introduction of transaction costs leads changing signs in the TP (before: -0.009261; after: 0.006391) and to (slightly) more extreme (but still very small) cryptocurrency weights in the GMVP (before: -0.000034; after: 0.000640) in the unconstrained framework. These shifts in the cryptocurrency weights are also observable for the TP (before: 0.000000; after: 0.004317) and the GMVP (before: 0.000103; after: 0.000544) in the long-only framework ($\omega_i \geq 0$). Comparing the different results, the most extreme (positive or negative) cryptocurrency results (under the consideration of transaction costs) are consistently measured for the tangency portfolio, which is not surprising in the light of the spanning tests' results.\\

Normally, compared to the situation without transaction costs, the expected return of the portfolio for a given risk level shrinks because of the transaction cost constraints in the calculation of the optimal portfolio, which is tantamount with a (negative) shift of the efficient frontier (cf.\ Fig.\ \ref{fig:EfficientFrontierShort})\footnote{Note, that the underlying transaction cost function remains the same for both optimization frameworks (unconstrained, long-only, so that the resulting curve is only printed once in Fig.\ \ref{fig:EfficientFrontierShort} as representative example for all the other graphical analyses hereafter.}). Consistent to the small weights of cryptocurrencies in all cases (with and without transaction costs), we can conclude, that this shift is not only due to the consideration of cryptocurrencies. Instead, the main part of the shift is driven by the transaction costs caused by trades of the benchmark assets. 

Nevertheless, we can observe, that the consideration of cryptocurrencies leads to small shares of cryptocurrencies in the GMVP and TP (and therefore to a slightly higher efficient frontier) with and without transaction costs. But in this case, the beforementioned shifts are so extremely small, that it is not really observable in Fig.\ \ref{fig:EfficientFrontierShort}. This result was expected for the global analysis (with an observation period of 65 months), because the spanning test results already indicated non-significant effects for the situation without transaction costs. On the other hand, we cannot evaluate the significancy of the shift of the efficient frontier for the situation with transaction costs because of the lack of a suitable spanning test for the optimization model used in this study. 
 
However, these minimal shifts of the efficient frontiers, including the different composition of the GMVP and TP, are also observable in the long-only framework (cf.\ Fig.\ \ref{fig:EfficientFrontierLongOnly}). 
\begin{figure}[H] 
	\begin{subfigure}{.5\textwidth}
		\centering
		\includegraphics[width=\textwidth]{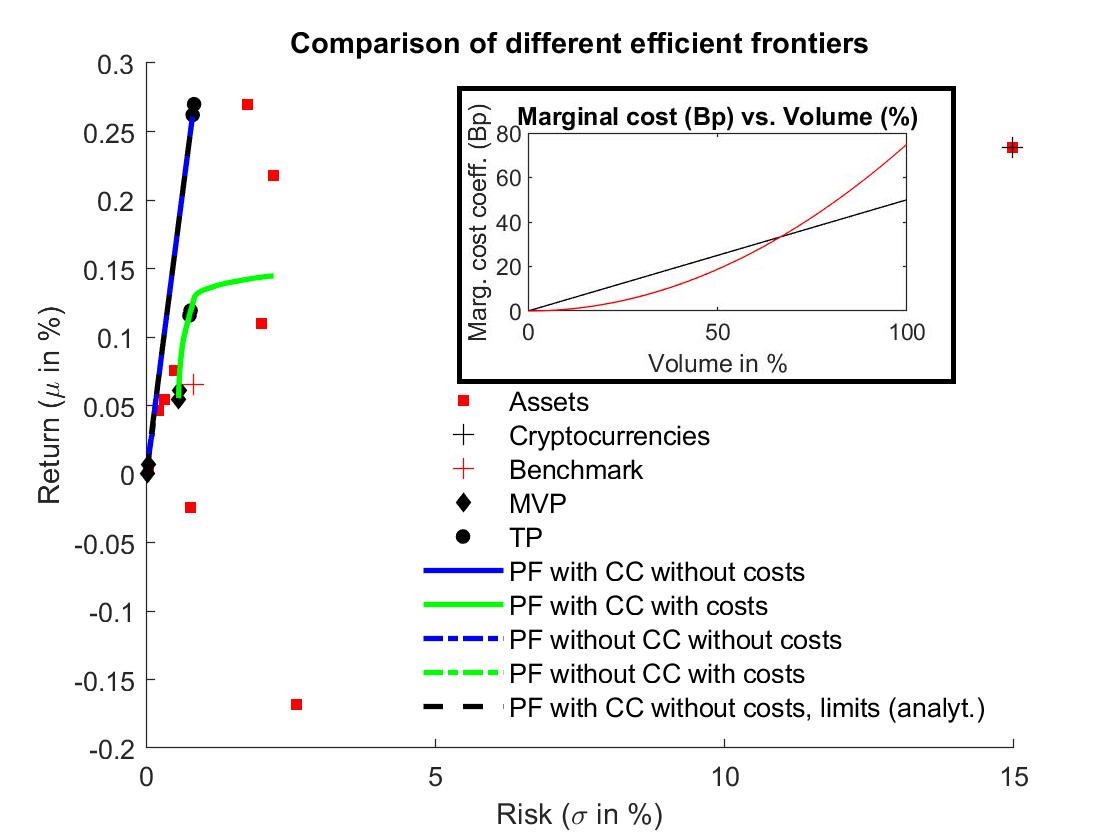}
		\caption{Base Case: Unconstrained Portfolio.}
		\label{fig:EfficientFrontierShort}
	\end{subfigure}%
	\begin{subfigure}{.5\textwidth}
	\centering
	\includegraphics[width=\textwidth]{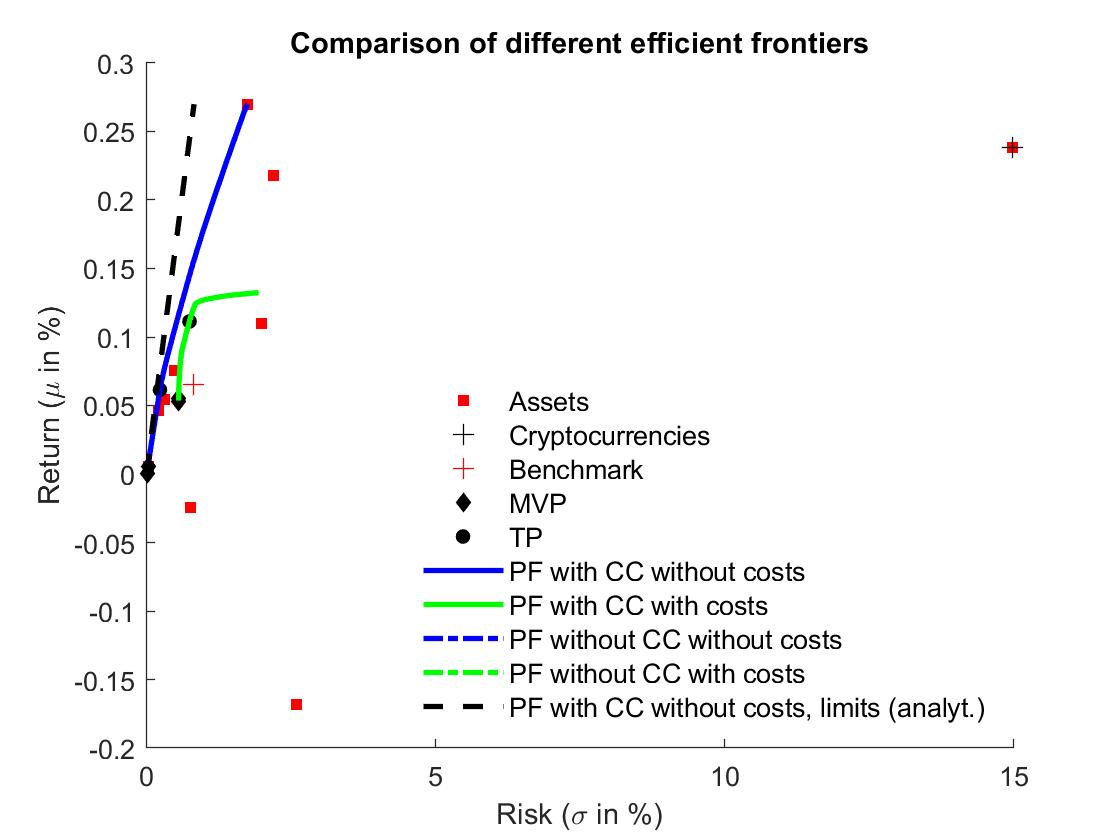}
	\caption{Robustness Check: Long-Only Portfolio.}
	\label{fig:EfficientFrontierLongOnly}
	\end{subfigure}%
	\caption{: Efficient frontiers with and without the consideration of cryptocurrencies and transaction costs (Full Sample: 2014-01/2019-05), Source: Own Calculations.}
	\label{fig:EfficientFrontier}
\end{figure}
Although we did not detect significant effects in the spanning tests' results  of the global analysis (Case A) before transaction costs, it does not mean, that cryptocurrencies did not play a significant role in smaller subsamples. In the results of the rolling-window spanning tests before transaction costs (cf.\ Fig.\ A.3 in the Appendix), we identified smaller subsamples, in which cryptocurrencies played a significant role in the unconstrained portfolio. A view on the portfolio weight heatmaps (cf.\ Fig.\ \ref{fig:HeatmapUnconst} and \ref{fig:HeatmapLongOnly}) will strenghten this point, because this significant role also becomes manifest in high cryptocurrency weights. Finally a more visible positive shift of the efficient frontier has to be expected for these subsamples (without transaction costs). In this situation, it would be interesting to observe, how the efficient frontier and the portfolio allocation is affected by the introduction of transaction cost. In other words, the question arises: does the introduction of transaction costs lead to smaller cryptocurrency weights (and therefore smaller diversification potentials) of cryptocurrencies in those periods, where we actually identified significant results in the spanning test framework?

To answer this research question, we choose a subsample from 2014-01 to 2014-12 and repeat the calculation of the same efficient frontier figure under the consideration of transaction costs. This subsample has the advantage, that it already led to significant results in the spanning tests (especially for the tangency portfolio) and is also the first possible window for the optimization. Thus, we can -- again -- observe the first shift from the initial portfolio (see above) to the first optimal portfolio.

In this course, considering cryptocurrencies in the investment opportunity set has a (again small, but now more visible) positive impact on the efficient frontier at first, if no transaction costs are assumed (gross efficient frontier), which also manifests in slightly different TPs (cf.\ Fig.\ \ref{fig:EfficientFrontierSubsample}). After the introduction of transaction costs, the efficient frontier for the situation with cryptocurrencies is -- again -- shifted downwards (net efficient frontier), but still remarkably above the net efficient frontier without cryptocurrencies, which can be interpreted as a remaining diversification effect. However, because of the lack of a suitable spanning test for our customized optimization model, it could be an interesting future research project to check, whether this remaining effect stays significant (as in the traditional spanning tests before). On the other hand, a closer look on the portfolio weights shows some considerable cryptocurrency weights in the TP, as it was expected. Moreover, it can be observed, that the cryptocurrency weights become less extreme (with a changing sign) after the additional consideration of transaction costs for the TP (before: 0.098583; after: -0.047093) and more extreme (with a changing sign) the GMVP (before: -0.000797; after: 0.012052). As a consequence, if cryptocurrencies have a (too) high weight in the situation without cryptocurrencies, the higher transaction costs lead to a lower portfolio weight and therefore a less importance with regard to portfolio diversification, just as \cite{GlasPoddig2018} expected.

The contrary effect of more extreme cryptocurrency weights (and sign changes) for small portions of cryptocurrencies and more restrictive transaction cost budgets can also confirmed -- at least by tendency -- by Tab.\ \ref{tab:VariationBudget}, which compares the (optimal) cryptocurrency weights for the global analysis (whole sample) and different transaction cost budgets $\tilde{K}_0 = 2,4,6,8,10,12,15,20,30,100$ BPs. However, despite this obvious effect, the cryptocurrency weightings remain comparatively small (as it was expected for this restrictive weighting scheme).
\begin{table}[H]
	\footnotesize \noindent
	\begin{tabularx}{\textwidth}{ X  X  X}
		\toprule
		$\tilde{K}_0$ & EWCI Weight (GMVP) & EWCI Weight (TP)\\
		\midrule
		100 BP & 0.000001 & -0.000794\\
		30 BP & -0.000033 & -0.000338\\
		20 BP & -0.000039 & 0.001950\\
		15 BP & -0.000957 & 0.003453\\
		12 BP & 0.000541 & 0.003382\\
		10 BP & 0.000640 & 0.006391\\
		8 BP & 0.000708 & 0.007592\\
		6 BP & 0.000810 & 0.008989\\
		4 BP & 0.000888 & 0.010396\\
		2 BP & 0.001192 & 0.011813\\
		\bottomrule
	\end{tabularx}
	\caption{: Cryptocurrency Portfolio Weights for varying transaction cost budgets (Unconstrained Scenario, Full Sample: 2014-01/2019-05), Source: Own Calculations.}
	\label{tab:VariationBudget}
\end{table}
Moreover, it becomes obvious, that the beforementioned variations of the transaction cost budgets lead to variations of the efficient frontier under the consideration of transaction costs (cf.\ Fig.\ \ref{fig:sub3b}). Those transaction cost budgets restrain the amount of possible changes of the optimal portfolio compared to the initial equally-weighted portfolio without cryptocurrencies (before the optimization). In this course, higher transaction cost budgets allow for more extreme shifts in the optimal portfolio (from the initial portfolio) and therefore for a closer convergence to the unconstrained efficient frontier without transaction costs. On the other hand, smaller transaction cost budgets lead to more extreme downward shifts of the efficient frontier. But, it is noteworthy, that this effect seems to be increasing with the extent of the reduction of the transaction cost budget. The first reduction from $100$ BPs to $30$ BPs seems to have comparable effects than the following reductions with smaller decrements in the budget. 
\begin{figure}[H] 
	\begin{subfigure}{.5\textwidth}
		\centering
		\includegraphics[width=\textwidth]{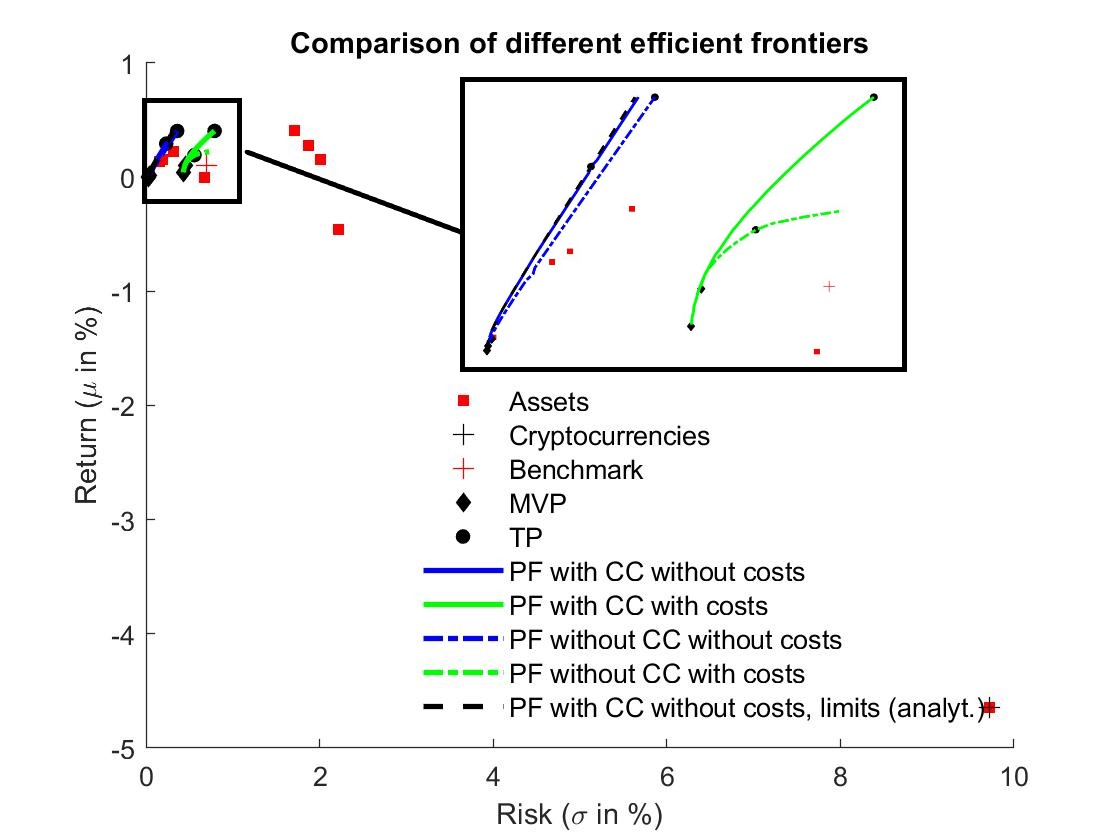}
		\caption{Variation of the Observation Period:\\Subset (2014-01/2014-12), Unconstrained.}
		\label{fig:sub3a}
	\end{subfigure}%
	\begin{subfigure}{0.5\textwidth}
		\centering
		\includegraphics[width=\textwidth]{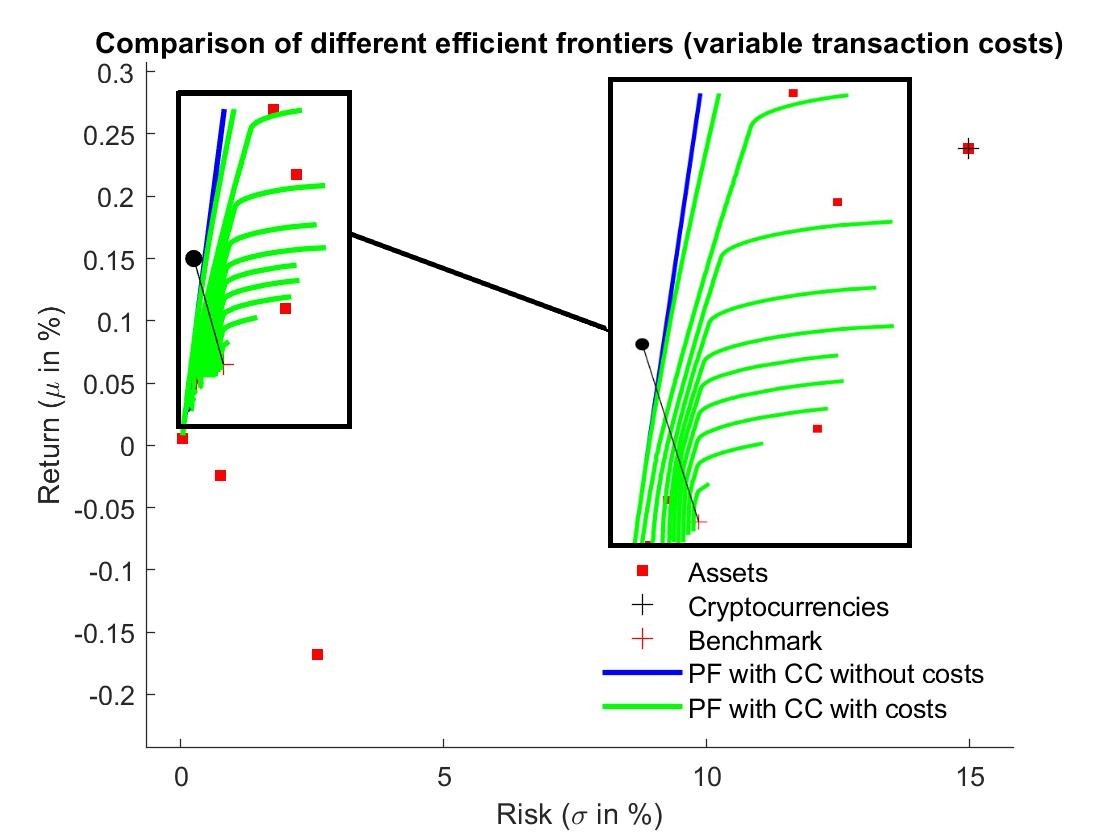}
		\caption{Variation of the Transaction Cost Budget: Full Sample (2014-01/2019-05), Unconstrained.}
		\label{fig:sub3b}
	\end{subfigure}
	\caption{: Robustness Checks: Effects of a changing optimization framework (Subsample Analysis, Variation of Transaction Costs) on the (unconstrained) efficient frontier, Source: Own Calculations.}
	\label{fig:EfficientFrontierSubsample}
\end{figure}
In toto, the beforementioned results are an indication, that parts of the recognized diversification effects seem to remain even after an introduction of transaction costs (and illiquidity reflections). The results of the Markowitz framework and the respective spanning tests are therefore biased to some extent -- just as \cite{GlasPoddig2018} expected before. But in contrast to their expectations, it could be demonstrated, that it is also possible, that the cryptocurrency weightings become more extreme after the introduction of transaction costs by tendency.

\section{Conclusion}\label{DiscussionConclusion}
In this paper, we analyzed, whether cryptocurrencies are able to improve the diversification of German investors' portfolios.\\

In contrast to most of the studies in the literature, we used a customized Equally-Weighted Cryptocurrency Index (EWCI) as a measure for the development of the cryptocurrency market to reduce the usual survivorship bias in cryptocurrency datasets and to measure the (average) development of the cryptocurrency market more accurately than already existing cryptocurrency indices (CRIX, CCI30).\\

The analysis combined three different methodological approaches, which were all applied in previous studies in the related literature: (i) an approach based on \textit{descriptive statistics}, (ii) a \textit{graphical} approach, (iii) an \textit{econometric} approach based on spanning tests.\\

We have shown, that in the absence of transaction costs cryptocurrencies do not improve portfolio diversification for German investors significantly in most of the analysed cases for our dataset including weekly observations from the window 2014-01-01 to 2019-05-31. By including cryptocurrencies in their portfolios, investors predominantly cannot reach a (significantly) higher efficient frontier, which means that they cannot realize (significantly) higher returns at a given risk level. However, the results of the rolling-window spanning tests showed, that there were also a few periods, in which the consideration of cryptocurrencies improved portfolio diversification, anyway.

By additionally demonstrating the effects of introducing liquidity- and transaction cost considerations, we show that the the beforementioned results of a traditional Mean-Variance-Optimization framework seem to be biased to some extent. In contrast to the expectations of e.g.\ \cite{GlasPoddig2018}, we demonstrated, that cryptocurrencies do not necessarily become less attractive for investors after the introduction of transaction cost budgets, which would manifest in shrinking portfolio weights. Instead, the comparatively low transaction costs of smaller cryptocurrency trades in our model led to a higher attractiveness in some of the analyzed contexts. Only in the case of high portfolio weights in the scenario without cryptocurrencies the introduction of (demand-dependent) transaction costs led to a reduction of the portfolio weight.

Because of the lack of a suitable spanning test for our customized optimization model, our research could be a starting point for future research to check, whether the introduction of transaction costs leads to more periods with significant shifts of the (net) efficient frontier.\\ 

In toto, cryptocurrencies must be characterized as extremely speculative and risky assets with many hazards (hacker attacks, fraud, etc.), which may not be suitable for inexperienced private investors and do not really fit to their risk attitude, too. Moreover, even if they would fit to the risk attitude an (experienced) investor, they are often not relevant for portfolio diversification, as the optimization results indicated. Nevertheless, we also gave some examples of periods, where cryptocurrencies had at least a small (but significant) portfolio weight, so that a complete exclusion from the investors' investment opportunity set could be misleading to some extent.
\section*{A. Appendix}
\appendix
\subsection*{A.1 Tabular Appendix}
\begin{table}[H]
	\footnotesize \noindent
	\rotatebox{90}{
		\begin{tabular}{l  l  l  l  l}
			\toprule
			Data & Index & Frequency & From ... to ... & Source\\
			
			\midrule
			\textit{Stocks:} & & & & \\ 
			\hspace*{5mm}Stoxx Europe 600 TR & Index & weekly & 2014-01-01 / 2019-06-01 & Bloomberg\\
			\textit{Money Market:} & & & &\\ 
			\hspace*{5mm}iBoxx EUR Jumbo TR 1-3 & Index & weekly & 2014-01-01 / 2019-01-01 & Bloomberg\\
			\textit{Sovereign Bonds:} & & & & \\ 
			\hspace*{5mm}iBoxx Euro Eurozone Sovereign Overall TR & Index & weekly & 2014-01-01 / 2019-01-01 & Bloomberg\\	
			\textit{Covered Bonds:} & & & & \\ 
			\hspace*{5mm}iBoxx Euro Covered TR & Index & weekly & 2014-01-01 / 2019-01-01 & Bloomberg\\	
			\textit{Corporate Bonds:} & & & & \\ 
			\hspace*{5mm}iBoxx Euro Liquid Corporates Diversified TR & Index & weekly & 2014-01-01 / 2019-01-01 & Bloomberg\\
			\textit{Real Estate:} & & & & \\		
			\hspace*{5mm}RX REIT Performance Index & Index & weekly & 2014-01-01 / 2019-01-01 & Frankfurt Stock Exchange\\
			\textit{Luxury Goods:} & & & & \\
			\hspace*{5mm}Solactive Luxury and Lifestyle Index TR & Index & weekly & 2014-01-01 / 2019-01-01 & Thomson Reuters Eikon\\
			\textit{Commodities / Gold:} & & & & \\
			\hspace*{5mm}GSCI Commodity Index TR & Index & weekly & 2014-01-01 / 2019-01-01 & Thomson Reuters Eikon\\				
			\bottomrule
		\end{tabular}
	}
	\renewcommand{\thetable}{A.\arabic{table}}
	\setcounter{table}{0}
	\caption{: Dataset (Benchmark-Assets except for Currencies).}
	\label{tab:DatasetAppendix}
\end{table}

\begin{table}[H]
	\footnotesize \noindent
		\begin{tabular}{l  l  l  l  l}
			\toprule
			Currency & ID & Frequency & From ... to ... & Source\\
			\midrule
			US-Dollar & USD & weekly & 2014-01-01 / 2019-06-01 & Bloomberg\\
			Swiss Franc & CHF & weekly & 2014-01-01 / 2019-06-01 & Bloomberg\\
			Japanese Yen & JPY & weekly & 2014-01-01 / 2019-06-01 & Bloomberg\\
			Australian Dollar & AUD & weekly & 2014-01-01 / 2019-06-01 & Bloomberg\\
			New Zealand Dollar & NZD & weekly & 2014-01-01 / 2019-06-01 & Bloomberg\\
			Canadian Dollar & CAD & weekly & 2014-01-01 / 2019-06-01 & Bloomberg\\			
			Norwegian Krone & NOK & weekly & 2014-01-01 / 2019-06-01 & Bloomberg\\
			Danish Krone & DKK & weekly & 2014-01-01 / 2019-06-01 & Bloomberg\\
			Swedish Krona & SEK & weekly & 2014-01-01 / 2019-06-01 & Bloomberg\\
			British Pound & GBP & weekly & 2014-01-01 / 2019-06-01 & Bloomberg\\			
			Turkish Lira & TRY & weekly & 2014-01-01 / 2019-06-01 & Bloomberg\\	
			South African Rand & ZAR & weekly & 2014-01-01 / 2019-06-01 & Bloomberg\\
			Russian Ruble & RUB & weekly & 2014-01-01 / 2019-06-01 & Bloomberg\\
			Polish Zloty & PLN & weekly & 2014-01-01 / 2019-06-01 & Bloomberg\\
			Mexican Peso & MXN & weekly & 2014-01-01 / 2019-06-01 & Bloomberg\\
			Indian Rupee & INR & weekly & 2014-01-01 / 2019-06-01 & Bloomberg\\
			Chinese Yuan & CNY & weekly & 2014-01-01 / 2019-06-01 & Bloomberg\\			
			\bottomrule
		\end{tabular}
	\renewcommand{\thetable}{A.\arabic{table}}
	\caption{: Dataset (Benchmark-Assets: Currencies).}
	\label{tab:DatasetAppendix2}
\end{table}
\newpage
\begin{table}[H]
	\footnotesize \noindent
	\begin{tabularx}{\textwidth}{ X  c  c c c}
		\toprule
		Asset & \multicolumn{4}{c}{Portfolio Weights}\\
		&\multicolumn{2}{c}{\textit{Minimum-Variance-Portfolio}} & \multicolumn{2}{c}{\textit{Tangency-Portfolio}} \\	
		\midrule
		& No EWCI & With EWCI & No EWCI & With EWCI\\
		\midrule
		\multicolumn{5}{l}{Unconstrained Portfolio: No Transaction Costs (Full Sample: 2014-01 / 2019-05)}\\
		\midrule
		MON & 0.988127 & 0.987901 & 0.000000 & 0.000000\\
		CUR & 0.007469 & 0.007721 & 0.000000 & 0.000000\\
		SOV & 0.000000 & 0.000000 & 0.081421 & 0.081421\\
		COV & 0.000000 & 0.000000 & 0.710920 & 0.710920\\
		COR & 0.000000 & 0.000000 & 0.151744 & 0.151744\\
		STO & 0.003397 & 0.003259 & 0.000000 & 0.000000\\
		RES & 0.000000 & 0.000000 & 0.055916 & 0.055916\\
		LUX & 0.000000 & 0.000774 & 0.000000 & 0.000000\\
		COM & 0.001007 & 0.001016 & 0.000000 & 0.000000\\
		EWCI & 0.000000 & 0.000103 & 0.000000 & 0.000000\\
		\bottomrule
	\end{tabularx}
	\renewcommand{\thetable}{A.\arabic{table}}
	\caption{: (Optimal) Long-Only Portfolio for Case A (with and without cryptocurrencies), Source: Own Calculations.}
	\label{tab:PortfolioAllocationLongA}
\end{table}

\begin{table}[H]
	\footnotesize \noindent
	\begin{tabularx}{\textwidth}{ X  c  c c c}
	\toprule
	Asset & \multicolumn{4}{c}{Average Portfolio Weights}\\
	&\multicolumn{2}{c}{\textit{Minimum-Variance-Portfolio}} & \multicolumn{2}{c}{\textit{Tangency-Portfolio}} \\	
	\midrule
	& No EWCI & With EWCI & No EWCI & With EWCI\\
	\midrule
	\multicolumn{5}{l}{Long-Only Portfolio: No Transaction Costs}\\
	\midrule
	MON   & 0.997254 & 0.997008 & 0.192668 & 0.227877\\
	CUR   & 0.000437 & 0.000555 & 0.000000 & 0.000000\\
	SOV   & 0.000000 & 0.000000 & 0.046024 & 0.046591\\
	COV   & 0.000000 & 0.000000 & 0.291057 & 0.276957\\
	COR   & 0.000225 & 0.000274 & 0.154598 & 0.129275\\
	STO   & 0.000585 & 0.000531 & 0.000849 & 0.000849\\
	RES   & 0.000209 & 0.000201 & 0.072647 & 0.062880\\
	LUX   & 0.000006 & 0.000007 & 0.212714 & 0.215571\\
	COM   & 0.001284 & 0.001225 & 0.029441 & 0.029255\\
	EWCI  & 0.000000 & 0.000198 & 0.000000 & 0.010745\\
	\bottomrule
\end{tabularx}
\begin{tabularx}{\textwidth}{X  c  c  c  c}
	\toprule
	Measure & \multicolumn{4}{c}{Portfolio Statistics}\\
	&\multicolumn{2}{c}{\textit{Minimum-Variance-Portfolio}}&\multicolumn{2}{c}{\textit{Tangency-Portfolio}}\\
	\midrule
	& No EWCI & With EWCI & No EWCI & With EWCI\\
	\midrule
	Minimum Return & -0.002284 & -0.002324 & -0.044861 & -0.044861\\
	Mean Return & 0.000024 & 0.000025 & 0.000386 & 0.000433\\	
	Maximum Return & 0.001134 & 0.001163 & 0.035554 & 0.051399\\
	\midrule		
	Standard Deviation & 0.000412 & 0.000419 & 0.008823 & 0.010340\\
	Maximum Drawdown & 0.004555 & 0.004659 & 0.177805 & 0.199183\\
	\midrule
	Sharpe Ratio & 0.057894 & 0.059393 & 0.043771 & 0.041882\\
	Sortino Ratio & 0.078063 & 0.080314 & 0.059972 & 0.060333\\	
	\midrule
	End Value ($C_T$ in EUR) & 100.55 & 100.57 & 108.35 & 109.17\\		
	\bottomrule
\end{tabularx}
	\renewcommand{\thetable}{A.\arabic{table}}
	\caption{: Key Statistics of the Long-Only Portfolio for Case B (with and without Cryptocurrencies, 1-Year Rolling Window), Source: Own Calculations.}
	\label{tab:PortfolioStatisticsLongOnly}
\end{table}


\subsection*{A.2 Graphical Appendix}
\begin{figure}[H] 
	\begin{subfigure}{.5\textwidth}
		\centering
		\includegraphics[width=\textwidth]{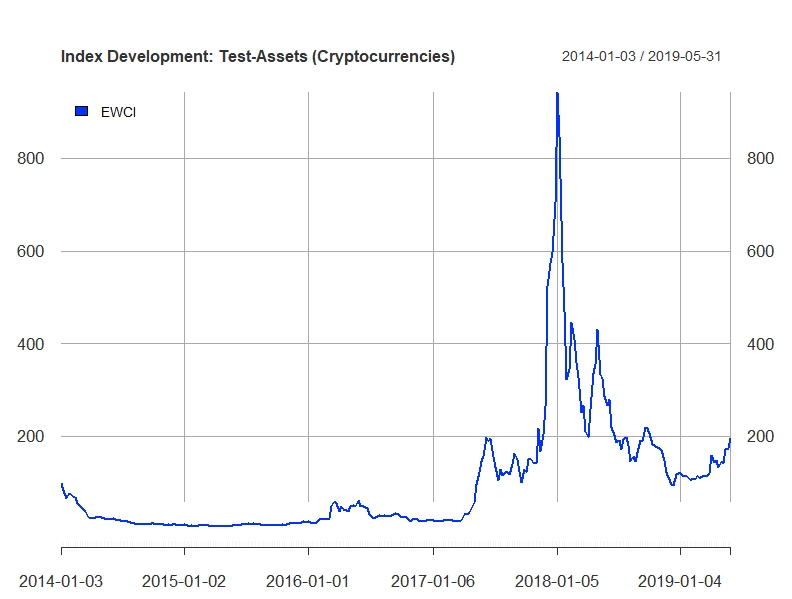}
		\caption{Index Development: Test Assets}
		\label{fig:IndexDevelopmentTest}
	\end{subfigure}
	\begin{subfigure}{.5\textwidth}
		\centering
		\includegraphics[width=\textwidth]{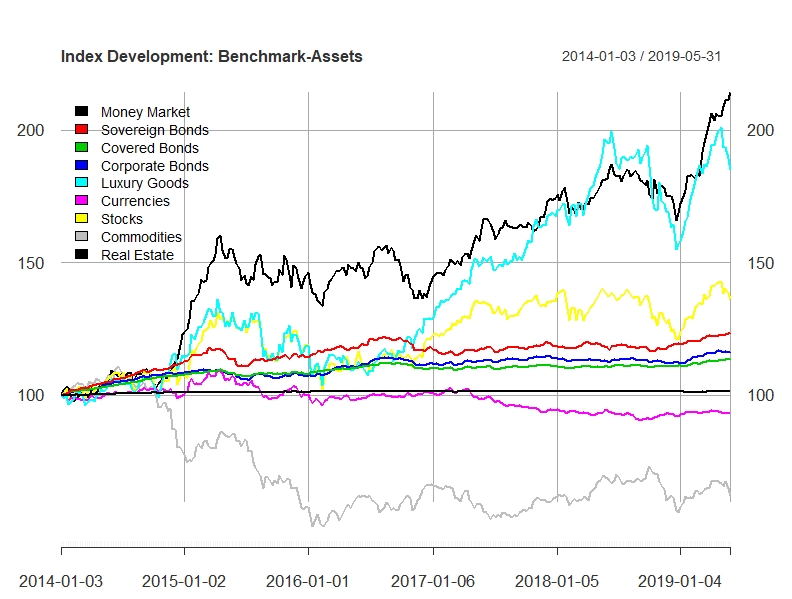}
		\caption{Index Development: Benchmark Assets}
		\label{fig:IndexDevelopmentBenchmark}
	\end{subfigure}
	\setcounter{figure}{1}
	\renewcommand{\thefigure}{A.\arabic{figure}}
	\caption{: Historical Development of the Indices considered in the Dataset, Source: Bloomberg, Thomson Reuters Eikon, CoinMarketCap.}
	\label{fig:IndexDevelopment}
\end{figure}

\begin{figure}[H]
	\begin{subfigure}{.5\textwidth}
		\centering
		\includegraphics[width=\linewidth]{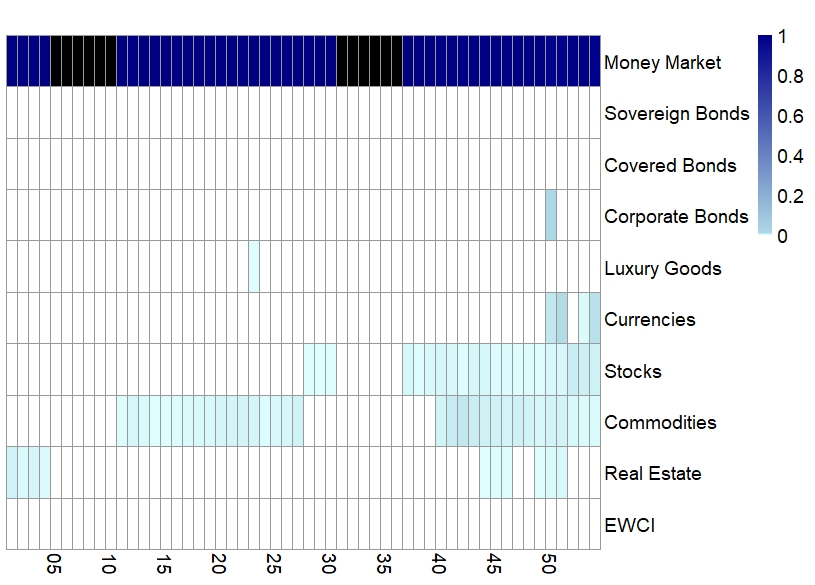}
		\caption{GMVP: Benchmark Assets only}
		\label{fig:sub2a}
		\includegraphics[width=\linewidth]{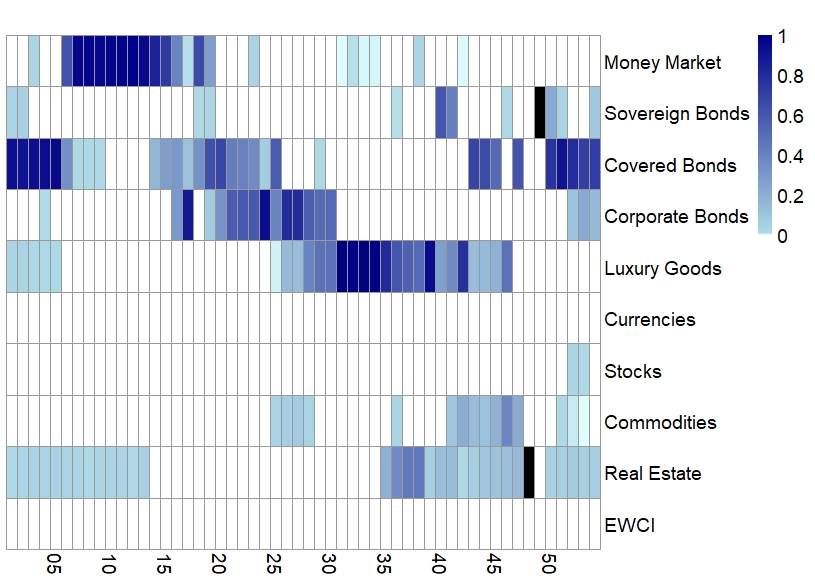}
		\caption{TP: Benchmark Assets only}
		\label{fig:sub2b}
	\end{subfigure}%
	\begin{subfigure}{0.5\textwidth}
		\centering
		\includegraphics[width=\linewidth]{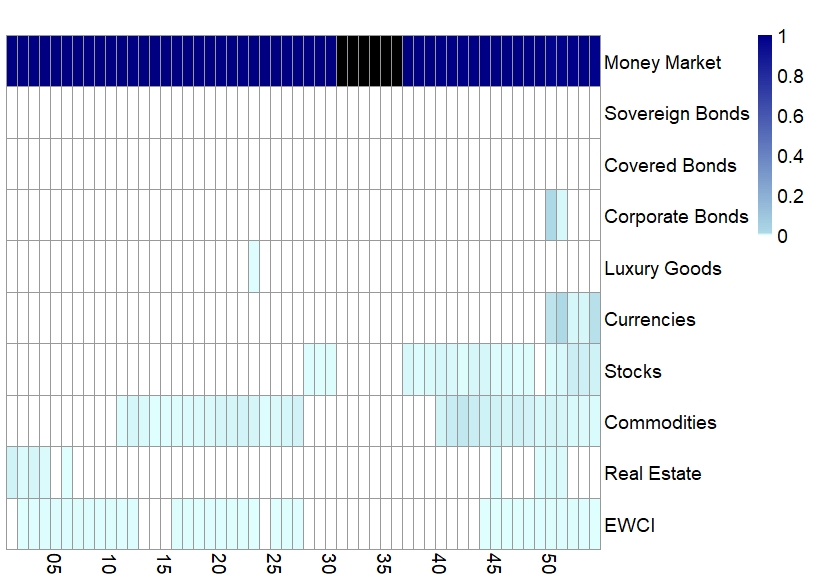}
		\caption{GMVP: Considering Cryptocurrencies}
		\label{fig:sub2c}
		\includegraphics[width=\linewidth]{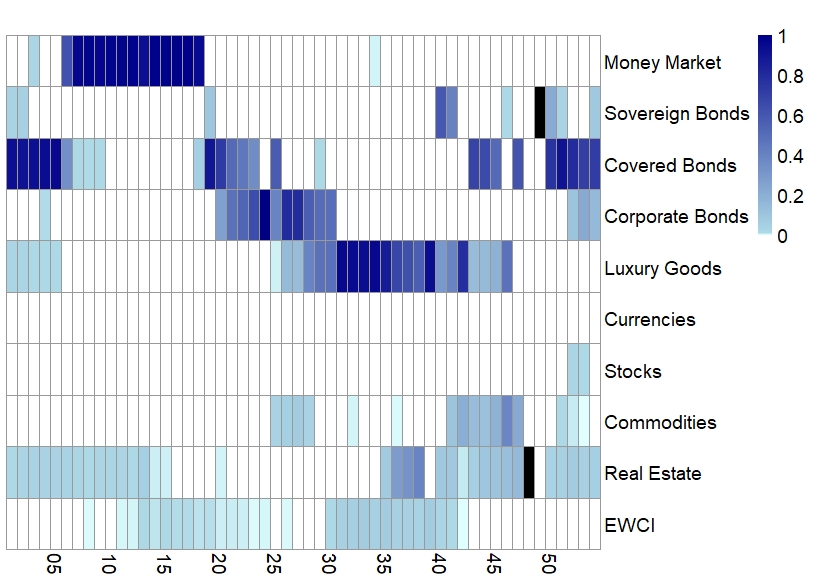}
		\caption{TP: Considering Cryptocurrencies}
		\label{fig:sub2d}
	\end{subfigure}
	\renewcommand{\thefigure}{A.\arabic{figure}}\\
	\caption{: Portfolio Weight Heatmap for Case B ($T = 54$, 1-Year Rolling Windows): Long-Only-Portfolios (No Transaction Costs), Source: Own Calculations.}
	\label{fig:HeatmapLongOnly}
\end{figure}

\subsection*{A.3 Mathematical Appendix}
\renewcommand{\theequation}{A.\arabic{equation}}
\setcounter{equation}{0}
\noindent 1.) Derivation of the transaction cost function from the Taylor series:

For small $|\Delta\omega_{i}|$, we can express a general transaction cost function $f$ as
\begin{flalign}
	f(|\Delta\omega_{i}|) = f(0) +\left. f'\right|_0 \text{ } |\Delta\omega_{i}| + \frac{1}{2}\left. f''\right|_0 \text{ } |\Delta\omega_{i}|^2 + ....
\end{flalign}
We set $f(0) = 0$. Hence, if no assets are traded, no transaction fee has to be paid. Furthermore, we also set $\left. f'\right|_0 = 0$, because transaction costs in the cryptcurrency context do not follow a linear form. Instead, the behavior of the market participants in times of historic cryptocurrency booms makes us assume, that transaction cost rise disproportional, if the trading volume exceeds a certain limit. Because the market trading volume is not equal to the individual trading volume, we have to give a short intuition, why changes in the individual portfolio allocations lead to changes in the transaction cost structure: we assume that (rational) investors follow the optimal portfolio, so that a rising number of extreme changes in the individual portfolio weights would lead to significant market changes. As a consequence, the transaction cost dynamics can better be captured by the quadratic term $\frac{1}{2}\left. f''\right|_0 |\Delta\omega_{i}|^2$ in the Taylor series. If we interpret $\left. f''\right|_0$ as $\tilde{c}_{i}^{\text{Test}}$, this also leads to Eq.\ \eqref{22b}.\\

\noindent 2.) Derivation of the parametrization of $\Psi$ in the long-only [$0 \leq \omega \leq 1$] scenario:\\

At first, we aim at equal expectation values of both cost functions leading to the equation
\begin{flalign}
\text{E}[\tau_{i}^{\text{Test (Quad.)}}(|\Delta\omega_{i}|)] \overset{!}{=} \text{E}[\tau_{i}^{\text{Test (Lin.)}}(|\Delta\omega_{i}|)].
\end{flalign}
In graphical terms, this condition can be converted into:
\begin{multline}
\int_{\text{0}}^{\text{1}} d(|\Delta\omega_{i}|)\Bigg[\frac{1}{2}\tilde{c}_{i}^{\text{Test}} \text{ } |\Delta\omega_{i}|^2 \text{ } V_{0} \text{ } \rho(|\Delta\omega_{i}|) \Bigg]\overset{!}{=} \\\ \int_{\text{0}}^{\text{1}} d(|\Delta\omega_{i}|) \Bigg[c_{i}^{\text{Test}} \text{ } |\Delta\omega_{i}|\text{ } V_{0} \text{ } \rho(|\Delta\omega_{i}|)\Bigg].
\end{multline}
For $\rho(|\Delta\omega_{i}|) = 1$ as the probability density function of the continuous uniform distribution of $|\Delta\omega_{i}|$ and $\tilde{c}_{i}^{\text{Test}} = \Psi c_{i}^{\text{Test}}$, the expression simplifies to
\begin{multline}
\int_{\text{0}}^{\text{1}} d(|\Delta\omega_{i}|)\Bigg[\frac{1}{2}\Psi \text{ } |\Delta\omega_{i}|^2 \text{ } V_{0} \Bigg]\overset{!}{=} \int_{\text{0}}^{\text{1}} d(|\Delta\omega_{i}|)\Bigg[ |\Delta\omega_{i}| \text{ } V_{0} \Bigg].
\end{multline}
Solving this equation with respect to $\Psi$ leads to
\begin{flalign}
\Leftrightarrow\hspace{0.5cm}\int_{\text{0}}^{\text{1}} d(|\Delta\omega_{i}|)\left[\frac{1}{2}\Psi \text{ } |\Delta\omega_{i}|^2 \text{ } V_{0} - |\Delta\omega_{i}| \text{ } V_{0}\right]\overset{!}{=} 0
\end{flalign}
\begin{flalign}
\Leftrightarrow\hspace{0.5cm}\left.\frac{1}{2}\Psi \text{ } \frac{|\Delta\omega_{i}|^3}{3} - \frac{|\Delta\omega_{i}|^2}{2} \right|_0^1 = 0
\end{flalign}
\begin{flalign}
\Leftrightarrow\hspace{0.5cm}\frac{1}{6}\Psi - \frac{1}{2} = 0
\end{flalign}
\begin{flalign}
\Leftrightarrow\hspace{0.5cm}\Psi = 3.
\end{flalign}

\noindent 3.) Calculation of the intersection of the linear and quadratic transaction cost functions in the long-only scenario:

To calculate the intersection of both transaction cost functions, we need to set
\begin{flalign}
\tau_{i}^{\text{Test (Quad.)}} \overset{!}{=} \tau_{i}^{\text{Test (Lin.)}}.
\end{flalign}
Inserting the respective formulas for the transaction cost functions leads to
\begin{flalign}
\frac{1}{2}\tilde{c}_{i}^{\text{Test}} \text{ } |\Delta\omega_{i}|^2 \text{ } V_{0} =  c_{i}^{\text{Test}} \text{ } |\Delta\omega_{i}| \text{ } V_{0}.
\end{flalign}
Solving this equation with respect to $|\Delta\omega_{i}|$ leads to
\begin{flalign}
\hspace{0.5cm}\frac{1}{2}\tilde{c}_{i}^{\text{Test}} \text{ } |\Delta\omega_{i}| =  c_{i}^{\text{Test}}.
\end{flalign}
Under the assumption of $\Psi = 3$, the variable $\tilde{c}_{i}^{\text{Test}}$ can now be substituted by the expression $\Psi c_{i}^{\text{Test}} = 3c _{i}^{\text{Test}}$, which leads to the final solution
\begin{flalign}
\hspace{0.5cm} |\Delta\omega_{i}| = |\Delta\omega_{i}^{*}| = \frac{2}{3},
\end{flalign}
where $|\Delta\omega_{i}^{*}|$ is the calculated amount of changes in the portfolio weights, for which both transaction cost regimes lead to the same result.


\begin{thebibliography}{100}
	\bibitem[Aggarwal et al., 2012]{AggarwalEtAl2012}Aggarwal, Raj; Kearney, Colm; Lucey, Brian M. (2012): Gravity and culture in foreign portfolio investment, in: \textit{Journal of Banking \& Finance}, Vol. 36, pp. 525-538.
	
	\bibitem[Anyfantaki et al., 2018]{Anyfantaki2018} Anyfantaki, Sofia; Arvanitis, Stelios; Topaloglou, Nikolas (2018): Diversification, integration and cryptocurrency market, Bank of Greece Working Paper, No. 244.	

	\bibitem[Balli et al., 2010]{BalliEtAl2010} Balli, Faruk; Basherb, Syed A.; Ozer-Ballia, Hatice (2010): From home bias to Euro bias: Disentangling the effects of monetary union on the European financial markets, in: \textit{Journal of Economics and Business}, Vol. 62, pp. 347–366.

	\bibitem[Baltzer et al., 2015]{BaltzerEtAl2015} Baltzer, Markus; Stolper, Oskar; Walter, Andreas (2015): Home-field advantage or a matter of ambiguity aversion? Local bias among German individual investors, in: \textit{The European Journal of Finance}, Vol. 21, No. 5, 734-754.	
	
	\bibitem[Baur et al., 2018]{BaurHongLee2018} Baur, Dirk G.; Hong, Kihoon; Lee, Adrian D. (2018): Bitcoin: Medium of exchange or speculative assets?, in: \textit{Journal of International Financial Markets, Institutions and Money}, Vol. 54, pp. 177-189.	

	\bibitem[Baur/Lucey, 2010]{BaurLucey2010} Baur, Dirk G.; Lucey, Brian M. (2010): Is gold a hedge or a safe haven? An analysis of stocks, bonds and gold, in: \textit{Financial Review}, Vol. 45, No. 2, pp. 217-229.	

	\bibitem[Bekaert/Urias, 1996]{BekaertUrias1996} Bekaert, G.; Urias, M. S. (1996): Diversification, integration and emerging market closed-end funds, in: \textit{The Journal of Finance}, Vol. 51, No. 3, pp. 835-869.

	\bibitem[Belousova/Dorfleitner, 2012]{BelousovaDorfleitner2012} Belousova, Julia; Dorfleitner, Gregor (2012): On the diversification benefits of commodities from the perspective of euro investors, in: \textit{Journal of Banking \& Finance}, Vol. 36, pp. 2455-2472.	

	\bibitem[Berndt/Savin, 1977]{BerndtSavin1977} Berndt, Ernst R.; Savin, N. Eugene (1977): Conflict among criteria for testing hypotheses in the multivariate linear regression model, in: \textit{Econometrica}, Vol. 45, pp. 1263–1278.

	\bibitem[BitcoinVisuals.com, 2020a]{BitcoinVisuals2020a} BitcoinVisuals.com (2020a): Fees Per Tx (USD), URL: https://bitcoinvisuals.com/chain-fees-tx-usd (access date: 2020-03-25).
	
	\bibitem[BitcoinVisuals.com, 2020b]{BitcoinVisuals2020b} BitcoinVisuals.com (2020b): Transactions Per Day, URL: https://bitcoinvisuals.com/chain-tx-day (access date: 2020-03-25).

	\bibitem[Börner et al., 2020]{BoernerEtAl2020} Börner, Christoph J.; Hoffmann, Ingo; Poetter, Fabian; Schmitz, Tim (2020): On Capital Allocation under Information Constraints, Working Paper.
	
	\bibitem[Borri, 2019]{Borri2019} Borri, Nicola (2019): Conditional tail-risk in cryptocurrency markets, in: \textit{Journal of Empirical Finance}, Vol. 50, pp. 1-19.

	\bibitem[Borri/Shaknov, 2018]{BorriShaknov2018} Borri, Nicola; Shakhnov, Kirill (2018): Cryptomarket discounts, SSRN Working Paper, No. 3124394.

	\bibitem[Bouri et al., 2017]{BouriEtAl2017} Bouri, Elie; Molnár, P.; Azzi, G.; Roubaud, D.; Hagfors, L. I. (2017): On the hedge and safe haven properties of Bitcoin: Is it really more than a diversifier?, in: \textit{Finance Research Letters}, Vol. 20, pp. 192-198.

	\bibitem[Brauneis/Mestel, 2019]{Brauneis2019} Brauneis, Alexander; Mestel, Roland (2019): Cryptocurrency-portfolios in a mean-variance framework, in: \textit{Finance Research Letters}, Vol. 28, pp. 259-264.

	\bibitem[Breusch, 1979]{Breusch1979} Breusch, Trevor S. (1979): Conflict among criteria for testing hypotheses: extensions and comments, in: \textit{Econometrica}, Vol. 47, pp. 203–207.

	\bibitem[Brière et al., 2015]{BriereEtAl2015} Brière, Marie; Oosterlinck, Kim; Szafarz (2015): Virtual Currency, Tangible Return: Portfolio Diversification with Bitcoin, in: \textit{Journal of Asset Management}, Vol. 16, No. 6, pp. 365-373.	

	\bibitem[Brown, 1994]{Brown1994} Brown, Patrick J. (1994): Constructing \& Calculating Bond Indices -- a Guide to the EFFAS Standardized Rules, Probus: Cambridge.	
	
	\bibitem[CCI30.com, 2020]{CCI302020} CCI30.com (2020): CCI30 - The Crypto Currencies Index, URL: https://cci30.com/ (access date: 2020-03-25).

	\bibitem[Cheah/Fry, 2015]{CheahFry2015} Cheah, Eng-Tuck; Fry, John (2015): Speculative bubbles in Bitcoin markets? An empirical investigation into the fundamental value of Bitcoin, in: \textit{Economics Letters}, Vol. 130, pp. 32-36.	
	
	\bibitem[Choi, 2018]{Choi2018} Choi, Kyoung Jin; Lehar, Alfred; Stauffer, Ryan (2018): Bitcoin Microstructure and the Kimchi Premium, SSRN Working Paper, No. 3189051.	
	
	\bibitem[Chapados, 2011]{Chopados2011} Chapados, Nicolas (2011): Portfolio Choice Problems -- An Introductory Survey of Single and Multiperiod Models, Springer: New York.	
	 
	\bibitem[Chowdhury, 2014]{Chowdhury2014} Chowdhury, Abdur (2014): Is Bitcoin the 'Paris Hilton' of the Currency World? Or Are the Early Investors onto Something That Will
	Make Them Rich?, Working Paper, Marquette University.	
	
	\bibitem[Coinmarketcap.com, 2019]{Coinmarketcap2019} Coinmarketcap.com (2019): Top 100 Cryptocurrencies by Market Capitalization, available at https://coinmarketcap.com/ (access date: 2019-06-01).
	
	\bibitem[Coinopsy.com, 2019]{Coinopsy2019} Coinopsy.com (2019): Dead Coins, available at https://www.coinopsy.com/dead-coins/ (access date: 2019-11-02).
	
	\bibitem[Corbet et al., 2018]{CorbetEtAl2018} Corbet, Shaen; Lucey, Brian M.; Peat, Maurice; Vigne, Samuel (2018): Bitcoin Futures -- What use are they?, in: \textit{Economics Letters}, Vol. 172, pp. 23–27.
	
	\bibitem[Darolles et al., 2015]{DarollesEtAl2015} Darolles, Serge; Gourieroux, Christian; Jay, Emmanuelle (2015): Robust Portfolio Allocation with Systematic Risk Contribution Restrictions, in: Jurczenko, Emmanuel (Edt.): Risk-based and Factor Investing, ISTE Press: London, pp. 123-146.
	
	\bibitem[Deadcoins.com, 2019]{Deadcoins2019} Deadcoins.com (2019): Dead Coins, available at https://deadcoins.com/ (access date: 2019-11-02).	

	\bibitem[DeSantis, 1993]{DeSantis1993} DeSantis, Giorgio (1993): Volatility Bounds for Stochastic Discount Factors -- Tests and Implications from International Financial Markets, Ph.D. Thesis, University of Chicago.

	\bibitem[Dickey/Fuller, 1979]{DickeyFuller1979} Dickey, David A.; Fuller, Wayne A. (1979): Distribution of the Estimators for Autoregressive Time Series with a Unit Root, in: \textit{Journal of the American Statistical Association}, Vol. 74, pp. 427–431.

	\bibitem[Dwyer, 2015]{Dwyer2015} Dwyer, Gerald P. (2015): The economics of Bitcoin and similar private digital currencies, in: \textit{Journal of Financial Stability}, Vol. 17, pp. 81-91.	

	\bibitem[Eisl et al., 2015]{EislEtAl2015} Eisl, Alexander; Gasser, Stephan, Weinmayer, Karl (2015): Caveat emptor: Does Bitcoin improve portfolio diversification?, Working Paper.
	
	\bibitem[ElBahrawy et al., 2018]{ElBahrawyEtAl2018} ElBahrawy, Abeer; Alessandretti, Laura; Kandler, Anne; Pastor-Satorras, Romualdo; Bronchelli, Andrea (2018): Evolutionary dynamics of the cryptocurrency market, \textit{The Royal Society Open Science}, Vol. 4: 170623.
	
	\bibitem[Engle, 2002]{Engle2002} Engle, Robert (2002): A simple class of multivariate generalized autoregressive conditional heteroskedasticity models, in: \textit{Journal of Business \& Economic Statistics}, Vol. 20, No. 3, pp. 339-350.

	\bibitem[ESMA, 2014]{ESMA2014} ESMA -- European Securities and Markets Authority (2014): Guidelines  for competent authorities and UCITS management companies -- Guidelines on ETFs and other UCITS issues, ESMA/2014/937EN, Paris.

	\bibitem[Fan et al., 2012]{FanEtAl2012} Fan, Jianqing; Zhang, Jingjin; Yu, Ke (2012): Vast Portfolio Selection With Gross-Exposure Constraints, in: \textit{ Journal of the American Statistical Association}, Vol. 107, No. 498, pp. 592-606.
	
	\bibitem[Fantazzini/Zimin, 2019]{FantazziniZimin2019} Fantazzini, Dean; Zimin, Stephan (2019): A multivariate approach for the simultaneous modelling
	of market risk and credit risk for cryptocurrencies, Working Paper.	

	\bibitem[Feder et al., 2018]{FederEtAL2018} Feder, Amir; Gandal, Neil; Hamrick, J. T.; Moore, Tyler; Vasek, Marie (2018): The Rise and Fall of Cryptocurrencies, Working Paper.	

	\bibitem[Ferson, 1995]{Ferson1995} Ferson, Wayne E. (1995): Theory and empirical testing of asset pricing models, in: Jarrow, Robert A.; Ziemba, William T.; Maksimovich, Vojislav (Eds.): \textit{Handbooks in Operational Research and Management Science}, Vol. 9, North Holland Publishers: Amsterdam, pp. 145–200.

	\bibitem[Ferson et al., 1993]{FersonEtAl1993} Ferson, Wayne E.; Foerster, Stephen R.; Keim, Donald B. (1993): General Tests of Latent Variable Models and Mean-Variance Spanning, in: \textit{Journal of Finance}, Vol. 48, No. 1, pp. 131-156.	

	\bibitem[Frijns et al., 2013]{FrijnsEtAl2013} Frijns, Bart; Gilbert, Aaron; Lehnert, Thorsten; Tourani-Rad, Alireza (2013): Uncertainty avoidance, risk tolerance and corporate takeover decisions, in: \textit{Journal of Banking \& Finance}, Vol. 37, No. 7, pp. 2457-2471.

	\bibitem[Gibbons/Hess, 1981]{GibbonsHess1981} Gibbons, Michael R; Hess, Patrick (1981): Day of the week effects and asset returns, in: \textit{Journal of Business}, Vol. 54, No. 4, pp. 579-596.

	\bibitem[Glas, 2019]{Glas2019} Glas, Tobias N. (2019): Investments in cryptocurrencies: Handle with care!, in: \textit{Journal of Alternative Investments}, Vol. 22, No. 1, pp. 96-113.

	\bibitem[Glas/Poddig, 2018]{GlasPoddig2018} Glas, Tobias N.; Poddig, Thorsten (2018): Kryptowährungen in der Asset-Allokation: Eine empirische Untersuchung auf Basis eines
	beispielhaften deutschen Multi-Asset-Portfolios, in: \textit{Vierteljahrshefte zur Wirtschaftsforschung}, Vol. 87, No. 3, pp. 107-128.	

	\bibitem[Green/Hollifield, 1992]{GreenHollifield1992} Green, Richard C.; Hollifield, Burton (1992): When Will Mean-Variance Efficient Portfolios be Well Diversified?, in: \textit{Journal of Finance}, Vol. 47, No. 5, pp. 1785-1809.		

	\bibitem[Grobys/Sapkota, 2019]{GrobysSapkota2019} Grobys, Klaus; Sapkota, Niranjan (2019): Predicting Cryptocurrency Defaults, Working Paper.		
	
	\bibitem[Härdle et al., 2018]{HaerdleEtAl2018} Härdle, Wolfgang K.; Lee, David Kuo Chuen; Nasekin, Sergey; Petukhina, Alla (2018): Tail Event Driven Asset allocation: evidence from equity and mutual funds’ markets, in: \textit{Journal of Asset Management}, Vol. 19, No. 1, pp. 49–63.
	
	\bibitem[Hale et al., 2018]{HaleEtAl2018} Hale, Galina; Krishnamurthy, Arvind; Kudlyak, Marianna; Shultz, Patrick (2018): How Futures Trading Changed Bitcoin Prices, in: \textit{FRBSF Economic Letter}, Vol. 12, pp. 1-5.

	\bibitem[Hansen, 1982]{Hansen1982} Hansen, Lars P. (1982): Large sample properties of generalized method of moments estimators, in: \textit{Econometrica}, Vol. 50, pp. 1029-1054.	

	\bibitem[Hansen/Jagannathan, 1991]{HansenJagannathan1991} Hansen, Lars P.; Jagannathan, Ravi (1991): Implications of security market data for models of dynamic economies, in: \textit{Journal of Political Economy}, Vol. 99, No. 2, pp. 225–262.

	\bibitem[He et al., 2016]{HeEtAl2016} He, Dong; Habermeier, Karl F.; Leckow, Ross B; Haksar, Vikram; Almeida, Yasmin; Kashima, Mikari; Kyriakos-Saad, Nadim; Oura, Hiroko; Sedik, Tahsin Saadi; Stetsenko, Natalia (2016): Virtual currencies and beyond: initial considerations.
	
	\bibitem[Hofstede, 2001]{Hofstede2001}Hofstede, Geert (2001): Culture’s Consequences: Comparing Values, Behaviors, Institutions and Organizations across Nations, Sage: Thousand Oaks.
	
	\bibitem[Hofstede et al., 2010]{HofstedeEtAl2010}Hofstede, Geert; Hofstede, Gert Jan;  Minkov, Michael (2010): Cultures and Organizations: Software of the Mind, $3^{rd}$ Edition, McGraw-Hill: New York. 
	
	\bibitem[Horn/Oehler, 2019]{HornOehler2019} Horn, Matthias; Oehler, Andreas (2019): Automated Portfolio Rebalancing: Autonomous Erosion of Investment Performance, Working Paper.	
	
	\bibitem[Huberman/Kandel, 1987]{HubermanKandel1987} Huberman, Gur; Kandel, Shmuel (1987): Mean-Variance-Spanning, in: \textit{Journal of Finance}, Vol. 42, No. 4, pp. 873-888.	

	\bibitem[Ince/Porter, 2006]{IncePorter2006} Ince, Ozgur S; Porter, R Burt (2006): Individual equity return data from Thomson Datastream: Handle with care!, in: \textit{Journal of Financial Research}, Vol. 29, No. 4, pp. 463-479.


	\bibitem[Jagannathan/Ma, 2003]{JagannathanMa2003} Jagannathan, Ravi; Ma, Tongshu (2003): Risk Reduction in Large Portfolios: Why Imposing the Wrong Constraints Helps, in: \textit{Journal of Finance}, Vol. 58, No. 4, pp. 1651-1683.
	
	\bibitem[Jorion, 2001]{Jorion2001} Jorion, Philippe (2001): Value at Risk: The new benchmark for managing financial risk, $2^{nd}$ Edition, McGraw Hill: New York.

	\bibitem[Kan/Zhou, 2012]{KanZou12} Kan, Raymond; Zhou, GuoFu (2012): Tests of Mean-Variance-Spanning, in: \textit{Annals of Economics and Finance}, Vol. 13, No. 1, pp. 145-193.

	\bibitem[Krückeberg/Scholz, 2018]{KrueckebergScholz2018} Krückeberg, Sinan; Scholz, Peter (2018): Cryptocurrencies as an Asset Class?, Working Paper.

	\bibitem[Mac Lane/Birkhoff, 1999]{LaneBirkhoff1999} Mac Lane, Saunders; Birkhoff, Garrett (1999): Algebra, $3^{rd}$ Edition, EDS Publications: Rhode Island.
	
	\bibitem[Ledoit/Wolf, 2004]{LedoitWolf2004} Ledoit, Olivier; Wolf, Michael (2004): A well-conditioned estimator for large-dimensional covariance matrices, in: \textit{Journal of Multivariate Analysis}, Vol. 88, No. 2, pp. 365–411.
	
	\bibitem[Lee Kuo Chuen et al., 2017]{LeeKuoChuenEtAl17} Lee Kuo Chuen, David; Guo, Li; Wang, Yu (2018): Cryptocurrency: A New Investment Opportunity?, in: \textit{Journal of Alternative Investments}, Vol. 20, No. 3, pp. 16-40.

	\bibitem[Li et al., 2019]{LiEtAl2019} Li, Tao; Shin, Donghwa; Wang, Baolian (2019): Cryptocurrency Pump-and-Dump Schemes, Working Paper.

	\bibitem[Liu, 2019]{Liu2019} Liu, Weiyi (2019): Portfolio diversification across cryptocurrencies, in: \textit{Finance Research Letters}, Vol. 29, pp. 200-205.
	
	\bibitem[Ljung/Box, 1978]{LjungBox1978} Ljung, Greta M.; Box, George E. P. (1978): On a Measure of a Lack of Fit in Time Series Models, in: \textit{Biometrika}, Vol. 65, No. 2, pp. 297–303. 
	
	\bibitem[Lobo et al., 2007]{LoboEtAl2007} Lobo, Miguel S.; Fazel, Maryam; Boyd, Stephen (2007): Portfolio optimization with linear and fixed
	transaction costs, in: \textit{Annals of Operations Research}, Vol. 152, No. 1, pp. 341-365.

	\bibitem[Malkiel, 2019]{Malkiel2019} Malkiel, Burton G. (2019): A Random Walk Down Wall Street - The Time-Tested Strategy for Successful Investing, $12^{th}$ Edition, Norton: New York.
	
	\bibitem[Markowitz, 1952]{Markowitz1952} Markowitz, Harry M. (1952): Portfolio Selection, in: \textit{Journal of Finance}, Vol. 7, No. 1, pp. 77-91.		

	\bibitem[Markowitz, 1959]{Markowitz1959} Markowitz, Harry M. (1959): Portfolio Selection, Cowles Foundation Monograph No. 16, Wiley: New York.

	\bibitem[McNeil et al., 2005]{McNeilEtAl2005} McNeil, Alexander; Frey, Rüdiger; Embrechts, Paul (2005): Quantitative Risk Management. Princeton University Press: Princeton.

	\bibitem[Morgan Stanley Capital International, 2018]{MSCI2018} Morgan Stanley Capital International (2018): MSCI Index Calculation Methodology -- Index Calculation Methodology for the MSCI Equity Indexes, January 2018.

	\bibitem[Nakamoto, 2008]{Nakamoto2008} Nakamoto, Satoshi (2008): Bitcoin: A peer-to-peer electronic cash system.
	
	\bibitem[Newey/West, 1987]{NeweyWest1987} Newey, Whitney K.; West, Kenneth D. (1987): Hypothesis testing with efficient method of moments estimation, in: \textit{International Economic Review}, Vol. 28, pp. 777-787. 
	 
	
	\bibitem[Oehler/Horn, 2019]{OehlerHorn2019} Oehler, Andreas; Horn, Matthias (2019): Does Households’ Wealth Predict the Efficiency of their Asset Mix? Empirical Evidence, in: \textit{Review of Behavioral Economics}, Vol. 6, No. 3, pp. 249-282.	

	\bibitem[Oehler et al., 2007]{OehlerEtAl2007} Oehler, Andreas; Rummer, Marco; Walker, Thomas; Wendt, Stefan (2007): Are Investors Home Biased? Evidence from Germany, in: Gregoriou, Greg N. (Ed.): Diversification and Portfolio Management of Mutual Funds, Palgrave MacMillan: Houndmills, pp. 55-77.

	\bibitem[Oehler et al., 2008]{OehlerEtAl2008} Oehler, Andreas; Rummer, Marco; Wendt, Stefan (2008): Portfolio Selection of German Investors: On the Causes of Home-biased Investment Decisions, in: \textit{The Journal of Behavioral Finance}, Vol. 9, pp. 149-162.
		
	\bibitem[Olkin/Pratt, 1958]{OlkinPratt1958} Olkin, Ingram; Pratt, John W. (1958): Unbiased Estimation of Certain Correlation Coefficients, in: \textit{The Annals of Mathematical Statistics}, Vol. 29, No. 1, pp. 201-211.
		
	\bibitem[Osterrieder et al., 2017]{OsterriederEtAl17} Osterrieder, Jörg; Lorenz, Julian; Strika, Martin (2017): Bitcoin and Cryptocurrencies - Not for the Faint-Hearted, in: \textit{International Finance and Banking}, Vol. 13, No. 1, pp. 145-193.	

	\bibitem[Rockafellar and Uryasev, 2000]{Rockafellar2000} Rockafellar, R.T.; Uryasev, S. (2000): Optimization of conditional value at risk, in: \textit{Journal of Risk}, Vol. 2, pp. 21-42.

	\bibitem[Roncalli, 2011]{Roncalli2011} Roncalli, Thierry (2011): Understanding the impact of weights constraints in portfolio theory, SSRN Working Paper, No. 1761625.
	
	\bibitem[Shahzad et al., 2019]{Shahzad2019} Shahzad, Syed J. H.; Bouri, Elie; Roubaud, D.; Kristoufek, L.; Lucey, Brian M. (2019). Is Bitcoin a better safe-haven investment than gold and commodities?, in: \textit{International Review of Financial Analysis}, Vol. 63, pp. 322-330.

	\bibitem[Sharpe, 1994]{Sharpe1994} Sharpe, William F. (1994): The sharpe ratio, in: \textit{Journal of Portfolio Management}, Vol. 21, No. 1, pp. 49-58.

	\bibitem[Sharpe, 1992]{Sharpe1992} Sharpe, William F. (1992): Asset allocation: Management style and performance measurement, in: \textit{Journal of Portfolio Management}, Vol. 18, No. 2, pp. 7-19.

	\bibitem[Sharpe, 1966]{Sharpe1966} Sharpe, William F. (1966): Mutual Fund Performance, in: \textit{The Journal of Business}, Vol. 39, No. 1, pp. 119-138.	

	\bibitem[Smales, 2019]{Smales2019} Smales, Lee A. (2019): Bitcoin as a safe haven: Is it even worth considering?, in: \textit{Finance Research Letters}, Vol. 30, No. 5, pp. 385-393.

	\bibitem[Sovbetov, 2018]{Sovbetov2018} Sovbetov, Yhlas (2018): Factors influencing cryptocurrency prices: Evidence from bitcoin, ethereum, dash, litcoin, and monero, in: \textit{Journal of Economics and Financial Analysis}, Vol. 2, No. 2, pp. 1-27.

	\bibitem[STOXX, 2018]{Stoxx2018} STOXX (edt.) (2018): iSTOXX Bond Index Guide, April 2018.
	
	\bibitem[Topaloglou/Tsomidis, 2018]{TopaloglouTsomidis2018} Topaloglou, Nikolas; Tsomidis, Georgios (2018): Investors’ Behavior in Cryptocurrency Market, Working Paper.

	\bibitem[Trimborn et al., 2018a]{TrimbornEtAl2018a} Trimborn, Simon; Härdle, Wolfgang Karl (2018a): CRIX an Index for cryptocurrencies, in: \textit{Journal of empirical finance}, Vol. 49, pp. 107-122.

	\bibitem[Trimborn et al., 2018b]{TrimbornEtAl2018b} Trimborn, Simon; Li, Mingyang; Härdle, Wolfgang K (2018b): Investing with cryptocurrencies-A liquidity constrained investment approach, Working Paper, Humboldt University Berlin.

	\bibitem[Tsay, 2014]{Tsay2014} Tsay, Ruey S. (2014): Multivariate Time Series Analysis -- With R and Financial Applications, Wiley: Hoboken.

	\bibitem[Urquhart/Zhang, 2019]{UrquhartZhang2019} Urquhart, Andrew; Zhang, Hanxiong (2019): Is Bitcoin a hedge or safe haven for currencies? An intraday analysis, in: \textit{International Review of Financial Analysis}, Vol. 63, pp. 49-57.

	\bibitem[Vitols, 2005]{Vitols2005} Vitols, Sigurt (2005): Changes in Germany's Bank-Based Financial System: implications for corporate governance, in: \textit{Corporate Governance: An International Review}, Vol. 13, No. 3, pp. 386-396.

	\bibitem[Wu/Pandey, 2014]{WuPandey2014} Wu, Chen Y.; Pandey, Vivek K. (2014): The Value of Bitcoin in Enhancing the Efficiency of an Investor's Portfolio, in: \textit{Journal of Financial Planning}, Vol. 27, No. 9, pp. 44-52.
	
	\bibitem[Yermack, 2015]{Yermack2015} Yermack, David (2015): Is Bitcoin a real currency? An economic appraisal, in: Lee, David; Chuen, Kuo (Edit.): \textit{Handbook of Digital Currency}, Academic Press: Amsterdam, pp. 31-43.			
\end{thebibliography}
\end{document}